\documentclass[10pt,aps,prd,twocolumn,superscriptaddress,nofootinbib,floatfix,longbibliography]{revtex4-2}
\usepackage{amsmath}
\usepackage{amssymb}
\usepackage{graphicx}
\usepackage{xcolor}
\usepackage{url}
\usepackage{bm}
\usepackage{mathrsfs}
\usepackage[utf8]{inputenc}
\usepackage{hyperref}
\usepackage{enumerate}
\usepackage{amsthm}
\usepackage{bbm}
\usepackage[normalem]{ulem}
\usepackage{upgreek}
\usepackage{tensor}
\usepackage{siunitx}
\usepackage{orcidlink}
\usepackage[caption=false]{subfig}
\usepackage{colortbl}

\definecolor{colour1}{HTML}{0571b0} 
\definecolor{colour2}{HTML}{92c5de} 
\definecolor{colour3}{HTML}{f4a582} 
\definecolor{colour4}{HTML}{ca0020} 
\definecolor{colour5}{HTML}{fe4a49} 
\definecolor{colour6}{HTML}{2d3092} 

\definecolor{colour99}{HTML}{0000FF} 

\hypersetup{colorlinks=true, linkcolor=colour99, citecolor=colour99,
filecolor=colour99, urlcolor=colour99}

\theoremstyle{definition}

\usepackage{cleveref}
\usepackage{multirow}
\usepackage{booktabs}
\usepackage{threeparttable}

\usepackage[export]{adjustbox}
\usepackage{times}
\usepackage{commath}
\renewcommand{\Re}{\operatorname{Re}}
\renewcommand{\Im}{\operatorname{Im}}
\usepackage{lipsum}

\crefname{equation}{Eq.}{Eqs.}
\Crefname{equation}{Equation}{Equations}

\crefname{section}{Sec.}{Secs.}
\Crefname{section}{Section}{Sections}

\crefname{table}{Table}{Tables}
\Crefname{table}{Table}{Tables}
\crefname{figure}{Figure}{Figures}
\Crefname{figure}{Figure}{Figures}

\crefname{appendix}{Appendix}{Appendix}
\providecommand{\orcidlink}[1]{}
\graphicspath{{Figures/}}

\newcommand{\HD}{\operatorname{HeunD}}

\newcommand{\HC}{\operatorname{HeunC}}

\newenvironment{qnmvalues}{\small}{}

\begin{document}
\title{Complete quasinormal modes of Type-D black holes. II. Extremal Kerr black holes}

\author{Changkai Chen\,\orcidlink{0000-0002-4023-0682}}
\affiliation{School of Physics and Astronomy, Beijing Normal University, Beijing, 100875, China}

\author{Xiaohua Zhang\,\orcidlink{0000-0002-0914-5455}}\email[Contact author: ]{zhangxiaohua@ynnu.edu.cn}
\affiliation{School of Mathematics, Yunnan Normal University, Kunming 650500, China}

\author{Zhoujian Cao\,\orcidlink{0000-0002-1932-7295} } \email[Contact author: ]{zjcao@bnu.edu.cn}
\affiliation{School of Physics and Astronomy, Beijing Normal University, Beijing, 100875, China}
\affiliation{Institute for Frontiers in Astronomy and Astrophysics, Beijing Normal University, Beijing, 102206, China}

\affiliation{School of Fundamental Physics and Mathematical Sciences, Hangzhou Institute for Advanced Study, UCAS, Hangzhou, 310024, China}

\author{Jiliang Jing\,\orcidlink{0000-0002-2803-7900} }
\affiliation{Department of Physics, Key Laboratory of Low Dimensional Quantum Structures and Quantum Control of Ministry of Education, Synergetic Innovation Center for Quantum Effects and Applications, Hunan Normal University, Changsha, 410081, Hunan, China}

\date{\today}
\begin{abstract}
We develop a double-confluent Heun method that extends our previous study of complete quasinormal-mode (QNM) spectra of Type-D black holes to extremal Kerr black holes. Radial boundary conditions and angular regularity determine the gravitational frequencies at extremality. Compared with earlier spectra restricted to the conventional Riemann sheet, analytic continuation across frequency branch cuts reveals additional QNMs on unconventional Riemann sheets. Spin continuation follows our Schwarzschild mode families toward extremality, with the survey extending through the 41st overtone on specified angular branches and Riemann sheets. We show for the first time that unconventional Class I modes with $m<0$, continued from unconventional Schwarzschild modes, form zero-damping sequences approaching the superradiant bound $\omega=m/2$ in the extremal limit. Within a fixed angular sector, finite-damping and zero-damping sequences are interspersed in the Schwarzschild overtone ordering, so a single overtone cutoff cannot separate them. Matching near the superradiant bound distinguishes nonextremal sequences accumulating at the critical frequency from isolated QNMs at exact extremality. Scattering-amplitude calculations verify representative roots and their sheet assignments. The expanded spectrum clarifies how Schwarzschild mode families reorganize at extremality, with completeness established by contour counting within a bounded frequency domain on a specified angular branch and Riemann sheet.
\end{abstract}
\maketitle
\raggedbottom

\section{Introduction}
A perturbed black hole emits gravitational waves whose ringdown is dominated by damped oscillations. Quasinormal modes (QNMs) satisfy purely ingoing boundary conditions at the event horizon and purely outgoing boundary conditions at infinity. With time dependence ${\rm e}^{-i\omega t}$, the real part of the complex frequency gives the oscillation frequency, and the decay rate of a damped mode is $-\operatorname{Im}(\omega)$. For Kerr black holes, these frequencies depend on the mass and spin, so measurements of multiple modes provide a means of testing the Kerr description of astrophysical black holes~\cite{Dreyer:2003bv,Berti:2005ys,Berti:2009kk,Berti:2025hly}. Numerical relativity studies have demonstrated the importance of overtones in modeling early ringdown~\cite{Giesler:2019uxc}. Simulation-based inference is also being developed for efficient gravitational-wave parameter estimation~\cite{wanghe}.

Extremal Kerr black holes have maximal spin $a=M$, where $M$ is the mass and $a$ the angular momentum per unit mass; at this spin, the inner and outer horizons coincide and the surface gravity vanishes. For azimuthal index $m$, as $a/M\to1$, zero-damping modes (ZDMs) accumulate at the superradiant bound frequency $\omega_{\rm SR}=m/2$, with vanishing damping and diverging decay times~\cite{Casals:2019vdb}. Damped modes (DMs), by contrast, approach frequencies with finite negative imaginary parts and remain exponentially damped in the extremal Kerr spacetime.

Detweiler~\cite{Detweiler:1980gk} identified the long-lived resonances of Kerr black holes with nearly extremal spin. Hod~\cite{Hod:2008zz,Hod:2009td} derived analytical expressions for the near-extremal resonance spectrum and examined their relation to the geometric optics approximation. Yang et al.~\cite{Yang:2012he,Yang:2012pj,Yang:2013uba} established the ZDM--DM distinction and studied the associated spectrum bifurcation and power-law ringdown. The geometric optics analysis explains much of this structure, but it does not determine which low-multipole overtones retain finite damping in the extremal Kerr limit. Berti emphasized this overtone dependence of the extremal limit in his discussion of open spectral problems~\cite{Berti:2004md}.

For nonextremal Kerr black holes, Cook and Zalutskiy~\cite{Cook:2014cta} provided high-precision gravitational frequencies, including branches at nearly extremal spin and near the Schwarzschild algebraically special (AS) frequency. The reported spectrum is incomplete: our subsequent study obtained the complete spectrum within the investigated mode ranges~\cite{Chen:2025sbz}, and its structure near the algebraically special frequency was independently confirmed by Kubota and Motohashi~\cite{Kubota:2026hdv}. Cook and Zalutskiy also studied purely imaginary Kerr frequencies, distinguishing QNMs from total-transmission and other polynomial solutions~\cite{Cook:2016fge}. Direct computations of extremal Kerr QNMs are needed alongside these finite-spin calculations and approximations for Kerr black holes with nearly extremal spin. The merged horizon is an irregular singular point of the radial perturbation equation, so the usual horizon expansion in Leaver's method~\cite{Leaver:1985ax} must be modified. Richartz~\cite{Richartz:2015saa} constructed a continued-fraction method for extremal Kerr black holes. Carneiro da Cunha and Cavalcante~\cite{daCunha:2021jkm} also obtained extremal frequencies through an isomonodromic construction involving the Painlev\'e III tau function. Casals and Longo Micchi~\cite{Casals:2019vdb} used the Mano--Suzuki--Takasugi (MST) representation~\cite{Mano:1996vt} to calculate scattering amplitudes and QNM frequencies for both extremal Kerr black holes and Kerr black holes with nearly extremal spin. In particular, their gravitational DMs with angular multipoles $\ell=m=2$ and $3$ demonstrate that finite damping can coexist with ZDM accumulation in the same angular sector. The QNM spectrum must therefore be determined from the full radial boundary conditions, rather than inferred solely from the leading approximation near the superradiant bound~\cite{Zimmerman:2015rua,Hod:2015swa}.

Recent calculations have advanced several complementary approaches. Hatsuda and Shiga~\cite{Hatsuda:2026ghx} derived exact Wentzel--Kramers--Brillouin (WKB) quantization conditions and evaluated Borel--Pad\'e-resummed quantum periods for selected scalar sectors of extremal Kerr and Reissner--Nordstr\"om black holes. For gravitational Kerr QNMs, Hu et al.~\cite{Hu:2026cyq} developed high-order WKB resummations about a potential maximum that improve DM frequencies. This local expansion loses accuracy for ZDM branches as the spin approaches extremality. Ferrer-S\'anchez et al.~\cite{FerrerSanchez:2026Hyperboloidal} used physics-informed neural networks in a two-dimensional hyperboloidal formulation of the gravitational Teukolsky equation, following six selected branches to $a/M=0.9998$ without separating the radial and angular variables. These calculations resolve selected sectors and finite-spin branches; they do not supply a complete gravitational spectrum at $a=M$.

Heun connection methods provide a direct way to impose both QNM boundary conditions. Related approaches express Kerr scalar scattering coefficients through isomonodromic deformations~\cite{CarneirodaCunha:2015hzd} and formulate Schwarzschild and Kerr spectral quantization using Seiberg--Witten theory~\cite{Aminov:2020yma}. The separated angular and radial Teukolsky equations~\cite{Teukolsky:1973ha,Teukolsky:1974yv} admit Heun representations, and the connection coefficients relate local solutions to scattering amplitudes~\cite{Motohashi:2021zyv,Bonelli:2021uvf}. Furugori et al.~\cite{Furugori:2026bqu} derived connection conditions and controlled spectral expansions for a separable spin-weighted equation on Kerr--Newman--de Sitter backgrounds; this formal spin equation does not describe the full coupled gravitational and electromagnetic perturbations of a charged rotating black hole. Rossi et al.~\cite{Rossi:2026kds} independently compared Heun and hypergeometric series calculations of Kerr--de Sitter excitation factors and studied exceptional points. These studies illustrate why the full connection amplitudes matter. The asymptotically flat extremal Kerr gravitational problem considered here instead requires a double-confluent radial equation.

The analytic structure of this radial problem is also part of the spectral description. For extremal Kerr black holes, the frequency-domain Green function has branch points in the complex frequency plane at $\omega=0$ and $\omega=m/2$~\cite{Casals:2018eev,Casals:2019vdb}. The associated cut contributions are distinct from isolated QNM poles. For scalar perturbations, the superradiant branch point is associated with the extremal horizon instability~\cite{Casals:2016mel}; transient near-horizon growth also occurs for scalar, electromagnetic, and gravitational perturbations of Kerr black holes with nearly extremal spin~\cite{Gralla:2016sxp}. In particular, the real frequency reached by a ZDM sequence must not be counted as an additional isolated extremal Kerr QNM~\cite{Richartz:2017qep}. Spectral branches may also continue onto unconventional Riemann sheets. Leung et al.~\cite{Leung:2003eq} identified such Schwarzschild poles near the AS frequency. At the AS frequency itself, the Regge--Wheeler and Zerilli equations have different mode content, and total-transmission solutions require separate treatment~\cite{MaassenvandenBrink:2000iwh}. More recently, Kubota and Motohashi~\cite{Kubota:2026hdv} tracked poles and zeros across Riemann sheets and related the near-AS Kerr anomalies to avoided crossings, resonant excitation, and pole skipping. The pole skipping of an inverse incident amplitude must be distinguished from cancellation in the full Green function: homogeneous solutions and connection coefficients can have singular factors that cancel only after they are combined~\cite{Motohashi:2026mbn}. Thus a continued frequency and its excitation in a physical response require separate tests.

Our earlier work developed exact Heun solutions and scattering amplitudes for Type-D perturbations~\cite{Chen:2023ese,Chen:2026bym}, and used analytic continuation of the confluent Heun (HeunC) eigenvalue equation to construct complete Schwarzschild and Kerr QNM spectra over the reported mode ranges~\cite{Chen:2025sbz}. This continuation recovers branches crossing or lying on the negative imaginary axis and connects them continuously to their Schwarzschild origins. The gravitational $\ell=2$ unconventional and AS modes define distinct parent sequences, assigned the Schwarzschild overtone labels $\hat n=8$ and $9$ in our convention. These labels follow the continued roots and do not change when their damping order changes. Our subsequent study of the complete spectrum of Kerr total-transmission modes (TTMs)~\cite{Chen:2026ane} further demonstrated the importance of distinguishing globally continued families and their frequency symmetries. The next step is to extend the QNM construction to $a=M$, where the radial equation changes from HeunC to double-confluent Heun (HeunD) form.

In this paper, we solve the gravitational QNM problem directly for extremal Kerr black holes. We initialize the horizon-ingoing and infinity-outgoing solutions from their asymptotic series and propagate their values and derivatives to a common ordinary point using HeunD functions. The radial Wronskian and the original angular HeunC regularity condition are then solved simultaneously. The continuation retains the frequency logarithms associated with both radial branch points. Independent endpoint constructions and scattering-amplitude calculations provide complementary checks of the resulting frequencies and their Riemann sheet assignments. We present the computed Class I gravitational spectrum for $\ell=2,3$, including roots with negative real parts on unconventional Riemann sheets. Class I is one of the two symmetry-related mode families; its label is retained along each continued sequence, including across the imaginary-frequency axis.

To relate these extremal Kerr frequencies to our Schwarzschild spectrum, we compare them with Kerr trajectories that retain their Schwarzschild parent labels. This comparison distinguishes finite-damping limits from ZDM accumulation without identifying an overtone by its instantaneous damping order. In the present extremal calculation, references to completeness are restricted to the surveyed sectors $2\le\ell\le3$, $|m|\le\ell$, parent labels $0\le\hat n\le41$, and the specified angular branches and radial frequency sheets. The cutoff includes the identified low-multipole families crossing the frequency cuts, notably the unconventional $(3,-3,40)$ sequence and the $(3,-2,41)$ AS family. It therefore covers both zero-damping accumulation and finite-damping endpoints of these families; it is not a damping threshold or a claim to exhaust all Riemann sheets. A certified contour establishes exactly one simple DM in the stated $(3,3)$ domain, while critical matching gives local accumulation and exclusion results near the branch points. These certified domains and the broader parent survey have distinct coverage, specified in Sec.~\ref{sec:connection-counting}.

Section~\ref{sec:method} presents the HeunD solver and its scattering-amplitude checks. Section~\ref{sec:complete-extremal-results} gives the spectrum, parent identifications, and continuation from nearly extremal spins. Section~\ref{sec:spectral-completeness} examines bounded completeness, critical frequencies, and their physical interpretation. The three appendices collect the formulas for the separated solutions and their connections, the analytic conditions and error bounds for counting, and the asymptotic derivations near frequency branch points. We use $c=G=1$ and express frequencies and spins in black-hole mass units, setting $M=1$. Explicit mass factors are retained where needed in analytical formulas.

\section{HeunD method for solving and verifying QNMs}\label{sec:method}
\subsection{Teukolsky equations for extremal Kerr black holes}
The Teukolsky master variable $\psi$ of spin weight $s$ separates in Boyer--Lindquist coordinates $(t,r,\theta,\phi)$ as~\cite{Teukolsky:1973ha}
\begin{equation}\label{eq:separation}
 \psi(t,r,\theta,\phi)={\rm e}^{-i\omega t}{\rm e}^{im\phi}
 S_{\ell m}(\Theta)R_{\ell m}(r).
\end{equation}
Here $m\in\mathbb Z$ is the azimuthal index with $|m|\le\ell$, $\Theta=\cos\theta$, and $S_{\ell m}$ and $R_{\ell m}$ are the angular and radial functions, and $\ell\ge\max(|m|,|s|)$. With this Fourier convention, damped modes have $\operatorname{Im}(\omega)<0$. The outer and inner horizon radii are denoted by $r_+$ and $r_-$; at extremality $r_+=r_-=M$. The separation constant $\lambda$ and the angular eigenvalue $ {}_sA_{\ell m}(a\omega)$ are related by
\begin{equation}\label{eq:lambda}
 \lambda={}_s A_{\ell m}(a\omega)+a^2\omega^2-2ma\omega.
\end{equation}
The Schwarzschild limit is $ {}_sA_{\ell m}(0)=\ell(\ell+1)-s(s+1)$.
The angular equation is
\begin{align}\label{eq:angular}
 &\frac{{\rm d}}{{\rm d}\Theta}\!\left[(1-\Theta^2)\frac{{\rm d}S_{\ell m}}{{\rm d}\Theta}\right]
 +{}_s A_{\ell m}(a\omega)S_{\ell m}
\nonumber\\
 &+\left[a^2\omega^2\Theta^2-2sa\omega\Theta+s
 -\frac{(m+s\Theta)^2}{1-\Theta^2}\right]S_{\ell m}=0.
\end{align}
A solution of the angular equation~\eqref{eq:angular} regular at the north pole and one regular at the south pole must be linearly dependent in their common domain; equivalently, their angular Wronskian $W_\theta$, formed with derivatives with respect to $\Theta$, vanishes. We evaluate this Wronskian at $\Theta=0$ using their HeunC representations~\cite{Chen:2025sbz,Chen:2026ane}; Appendix~\ref{app:angular-regularity} gives the explicit solutions and matching formula.

The extremal radial Teukolsky equation in the Boyer--Lindquist coordinate $r$ is
\begin{align}\label{eq:radial}
 &\frac{{\rm d}^2R_{\ell m}}{{\rm d}r^2}+\frac{2(s+1)}{r-M}\frac{{\rm d}R_{\ell m}}{{\rm d}r}
 +\bigg[\omega^2+\frac{4M\omega^2+2is\omega}{r-M}
\nonumber\\
 &\,+\frac{7M^2\omega^2-{}_s A_{\ell m}(M\omega)}{(r-M)^2}
 +\frac{M(4M\omega-2is)(2M\omega-m)}{(r-M)^3}
\nonumber\\
 &\,+\frac{M^2(2M\omega-m)^2}{(r-M)^4}\bigg]R_{\ell m}=0.
\end{align}
The merged horizon and infinity are rank-one irregular singularities when $\omega\ne 0$ and $2M\omega\ne m$.

\subsection{Double-confluent Heun equation}
The M\"obius transformation maps the horizon and infinity to the two HeunD singularities:
\begin{equation}\label{eq:mobius}
 x=\frac{r-M-\varpi}{r-M+\varpi},\qquad
 \varpi^2=-\frac{M(2M\omega-m)}{\omega}.
\end{equation}
The square root follows a continuous frequency path. Here $r=M$ maps to $x=-1$ and $r\to\infty$ to $x=1$, with inverse $r-M=\varpi\frac{1+x}{1-x}$.

We write the S-homotopic transformation as
\begin{equation}\label{eq:gauge}
 R_{\ell m}(r)=\mathcal S_r(x)\mathcal H(x),
\end{equation}
where the radial prefactor is
\begin{equation}\label{eq:radial-prefactor}
 \mathcal S_r(x)=(r-M)^{-s-\frac12}
 {\rm e}^{i\omega(r-M)-\frac{i\omega\varpi^2}{r-M}}.
\end{equation}
Here $r$ is understood as the inverse of Eq.~\eqref{eq:mobius}; the dependence of $\mathcal S_r$ on $s,M,m,\omega$ is implicit. Its powers retain the chosen continuation path.
The essential horizon exponential replaces the powers associated with the separate horizons in the nonextremal solution. The resulting double-confluent Heun equation is
\begin{align}\label{eq:mapled}
 &\frac{{\rm d}^2\mathcal H}{{\rm d}x^2}
 +\left[\frac{2x}{x^2-1}-\frac{\alpha(x^2+1)}{(x^2-1)^2}\right]\frac{{\rm d}\mathcal H}{{\rm d}x}
\nonumber\\
 &\hspace{8mm}+\frac{\beta x^2+(\gamma+2\alpha)x+\delta}{(x^2-1)^3}\mathcal H=0.
\end{align}
We use Maple's symmetric four-parameter form $\HD(\alpha,\beta,\gamma,\delta;x)$, following the parameter convention of our earlier Heun formulations~\cite{Chen:2025sbz,Chen:2026ane}. This choice places the horizon and infinity on equal footing at $x=-1$ and $x=1$. The radial equation is encoded by the HeunD parameter quadruple
\begin{subequations}\label{eq:parameters}
\begin{equation}\label{eq:heund-parameters}
 \begin{aligned}
 \alpha&=-8i\omega\varpi,\quad\quad\beta=4\mathcal{V} - 16is\omega\varpi,\\
 \gamma&=-64M\omega^2\varpi,\quad\delta=-4\mathcal{V} - 16is\omega\varpi.
 \end{aligned}
\end{equation}
\begin{equation}\label{eq:heund-v}
 \mathcal V\equiv(\beta-\delta)/8=4M\omega(3M\omega-m)-\lambda-(s+\tfrac12)^2.
\end{equation}
\end{subequations}
The HeunD parameter quadruple has the coordinate symmetries
\begin{equation}\label{eq:heund-coordinate-symmetries}
 \begin{aligned}
 x\mapsto-x:\quad&
 (\alpha,\beta,\gamma,\delta)
 \mapsto(-\alpha,\beta,-\gamma,\delta),\\
 x\mapsto x^{-1}:\quad&
 (\alpha,\beta,\gamma,\delta)
 \mapsto(-\alpha,-\delta,-\gamma,-\beta).
 \end{aligned}
\end{equation}
The corresponding solution transformations and conventions for evaluating the HeunD functions are summarized in Appendix~\ref{app:map}.
Changing the square-root sign $\varpi\mapsto-\varpi$ in Eq.~\eqref{eq:mobius} realizes the $x\mapsto x^{-1}$ symmetry in Eq.~\eqref{eq:heund-coordinate-symmetries}. For noncritical frequencies, the combinations $\gamma/\alpha=-8iM\omega$ and $\beta-\delta=8\mathcal V$ are invariant, so this transformation preserves the physical frequency and separation constant. The radial solutions and their continued logarithms must be transported with the coordinate change.
The HeunD parameter quadruple is also the input to the MST calculation for extremal Kerr black holes in Appendix~\ref{app:mst}. The recurrence coefficients are written directly in these parameters in Eq.~\eqref{eq:mst-maple-direct-recurrence}; no separate MST frequency or spin parameters are introduced.

For gravitational perturbations, $\beta=4\mathcal V-4\alpha$ and $\delta=-4\mathcal V-4\alpha$. The map degenerates at $\omega=0$ or $2M\omega=m$, where separate limiting equations are required.

Further confluence relates the nonextremal and extremal radial equations, but the HeunC and HeunD parameters and normalizations at ordinary points are different. They cannot be identified by renaming the arguments.

\subsection{Radial solutions and QNM conditions}
Substitution of the two independent HeunD functions into the S-homotopic transformation~\eqref{eq:gauge} gives two particular solutions of the radial Teukolsky equation:
\begin{subequations}\label{eq:heund-local-solutions}
\begin{align}
 R_1(r)&=\mathcal S_r(x)\HD(\alpha,\beta,\gamma,\delta;x),
 \label{eq:heund-alg}\\
 R_2(r)&=\mathcal S_r(x){\rm e}^{-\frac{\alpha x}{x^2-1}}
 \HD(-\alpha,\beta,\gamma,\delta;x).
 \label{eq:heund-exp}
\end{align}
\end{subequations}
The exponential multiplying the second HeunD function changes $\alpha$ to $-\alpha$ while leaving $\beta,\gamma,\delta$ unchanged. For $\alpha\ne0$, these functions form a linearly independent radial basis; their ordinary-point normalization is given in Appendix~\ref{app:map}, and their endpoint expansions and connection coefficients in Appendix~\ref{app:endpoint-basis}. The QNM boundary conditions are imposed on the complete radial solutions below.

The asymptotic solutions of the radial Teukolsky equation~\cite{Teukolsky:1973ha,Teukolsky:1974yv} can be written as
\begin{equation}\label{eq:RTE-asympt}
 R_{\ell m}(r)\to
 \begin{cases}
 Z_{\mathrm{in}}^M R_{\mathrm{in}}^M(r)
 +Z_{\mathrm{up}}^M R_{\mathrm{up}}^M(r),&r\to M,\\
 Z_{\mathrm{in}}^{\infty} R_{\mathrm{in}}^{\infty}(r)
 +Z_{\mathrm{up}}^{\infty} R_{\mathrm{up}}^{\infty}(r),&r\to\infty.
 \end{cases}
\end{equation}
Here $Z_{\mathrm{in}}^M,Z_{\mathrm{up}}^M$ are the complex amplitudes of the ingoing and outgoing waves at the extremal horizon, and $Z_{\mathrm{in}}^{\infty},Z_{\mathrm{up}}^{\infty}$ are the corresponding amplitudes at infinity. The superscripts $M$ and $\infty$ denote the horizon $r=M$ and spatial infinity, respectively. The leading wave factors in Eq.~\eqref{eq:RTE-asympt} are defined by
\begin{subequations}\label{eq:physicalbc}
\begin{align}
 R_{\mathrm{in}}^M(r)&=
 (r-M)^{-2s-2iM\omega}
 {\rm e}^{\frac{iM(2M\omega-m)}{r-M}},\label{eq:bc-in}\\
 R_{\mathrm{up}}^M(r)&=
 (r-M)^{2iM\omega}
 {\rm e}^{-\frac{iM(2M\omega-m)}{r-M}},\label{eq:bc-out-h}\\
 R_{\mathrm{in}}^\infty(r)&=
 (r-M)^{-1-2iM\omega}
 {\rm e}^{-i\omega(r-M)},\label{eq:bc-inc}\\
 R_{\mathrm{up}}^\infty(r)&=
 (r-M)^{-1-2s+2iM\omega}
 {\rm e}^{i\omega(r-M)}.\label{eq:bc-up}
\end{align}
\end{subequations}
 The complete radial solutions $R_{\ell m}^{\mathrm{in}}$ and $R_{\ell m}^{\mathrm{up}}$, selected below by these boundary conditions, each contain only one wave at their defining endpoint and generally both waves at the other.

\subsubsection{Radial matching and the QNM equation}\label{sec:two-radial-frameworks}
We select the complete radial solutions $R_{\ell m}^{\mathrm{in}}$ and $R_{\ell m}^{\mathrm{up}}$ by normalizing the ingoing amplitude at the horizon and the outgoing amplitude at infinity, respectively. Their endpoint behavior is
\begin{equation}\label{eq:mst-amplitudes}
 R_{\ell m}^{\mathrm{in}}(r)\to
 \begin{cases}
 R_{\mathrm{in}}^M(r),&r\to M,\\
 B^{\mathrm{inc}}_{\ell m}R_{\mathrm{in}}^\infty(r)
 +B^{\mathrm{ref}}_{\ell m}R_{\mathrm{up}}^\infty(r),&r\to\infty.
 \end{cases}
\end{equation}
\begin{equation}\label{eq:infinity-framework-connection}
 R_{\ell m}^{\mathrm{up}}(r)\to
 \begin{cases}
 C^{\mathrm{ref}}_{\ell m}R_{\mathrm{in}}^M(r)
 +C^{\mathrm{up}}_{\ell m}R_{\mathrm{up}}^M(r),&r\to M,\\
 R_{\mathrm{up}}^\infty(r),&r\to\infty.
 \end{cases}
\end{equation}
Here $B^{\mathrm{inc}}_{\ell m}$ and $B^{\mathrm{ref}}_{\ell m}$ are the two amplitudes of the horizon-ingoing solution at infinity, while $C^{\mathrm{ref}}_{\ell m}$ and $C^{\mathrm{up}}_{\ell m}$ are the two amplitudes of the infinity-outgoing solution at the horizon. For $R_{\ell m}^{\mathrm{in}}$ with the normalization in Eq.~\eqref{eq:mst-amplitudes}, the coefficients in Eq.~\eqref{eq:RTE-asympt} satisfy $Z_{\mathrm{in}}^{\infty}=B^{\mathrm{inc}}_{\ell m}$ and $Z_{\mathrm{up}}^{\infty}=B^{\mathrm{ref}}_{\ell m}$. The factors in Eq.~\eqref{eq:physicalbc} are their leading endpoint waves; $R_{\ell m}^{\mathrm{in}}$ and $R_{\ell m}^{\mathrm{up}}$ denote the complete solutions.

For a QNM, these two solutions $R_{\ell m}^{\mathrm{in/up}}$ must be linearly dependent. At an ordinary matching radius $r^\star$, define the radial Wronskian by $W_r[f,g]\equiv f\,{\rm d}g/{\rm d}r-g\,{\rm d}f/{\rm d}r$. The frequency and separation constant therefore satisfy
\begin{equation}\label{eq:radial-matching-qnm}
 \begin{aligned}
 W_r[R_{\ell m}^{\mathrm{in}},R_{\ell m}^{\mathrm{up}}]
 \big|_{r=r^\star}&=0,\\
 W_\theta(\omega,\lambda)&=0.
 \end{aligned}
\end{equation}
This is the boundary-value problem for the extremal Kerr QNM spectrum.

We take the matching point to be
\begin{equation}
\label{eq:matching-point}
 r^\star=2M,\qquad r^\star-M=M.
\end{equation}
This ordinary point balances the propagation distances from the horizon and from infinity after inversion when measured in units of $M$. It is not an additional boundary condition: the conserved weighted Wronskian makes the exact root independent of $r^\star$. Varying the matching point is therefore a numerical convergence test.

The scattering interpretation follows from the conserved quantity $(r-M)^{2(s+1)}W_r$. With the endpoint normalizations above,
\begin{subequations}\label{eq:dual-amplitude-projections}
\begin{align}
 B^{\mathrm{inc}}_{\ell m}
 &=\frac{(r-M)^{2(s+1)}}{2i\omega}
 W_r[R_{\ell m}^{\mathrm{in}},R_{\ell m}^{\mathrm{up}}],\\
 C^{\mathrm{up}}_{\ell m}
 &=\frac{(r-M)^{2(s+1)}}{2iM(2M\omega-m)}
 W_r[R_{\ell m}^{\mathrm{in}},R_{\ell m}^{\mathrm{up}}].
\end{align}
\end{subequations}
\begin{align}\label{eq:dual-amplitude-identity}
 &(r-M)^{2(s+1)}
 W_r[R_{\ell m}^{\mathrm{in}},R_{\ell m}^{\mathrm{up}}]
\nonumber\\
 &\qquad=2i\omega B^{\mathrm{inc}}_{\ell m}
 =2iM(2M\omega-m)C^{\mathrm{up}}_{\ell m}.
\end{align}
Consequently, away from $\omega=0$ and $2M\omega=m$, Eq.~\eqref{eq:radial-matching-qnm} is equivalent to either amplitude condition,
\begin{subequations}\label{eq:QNM}
\begin{align}
 B^{\mathrm{inc}}_{\ell m}&=0,\qquad W_\theta(\omega,\lambda)=0,
 \label{eq:qnm-horizon}\\
 C^{\mathrm{up}}_{\ell m}&=0,\qquad W_\theta(\omega,\lambda)=0.
 \label{eq:qnm-infinity}
\end{align}
\end{subequations}

Equations~\eqref{eq:qnm-horizon} and~\eqref{eq:qnm-infinity} are alternative forms of the same radial condition, not additional independent constraints. The coupled iteration uses the radial Wronskian and angular regularity; the separately evaluated amplitudes provide checks.

\subsubsection{Analytic continuation across the frequency cuts}
The radial boundary-value problem is multivalued in frequency. We use $\operatorname{Log}$ for the principal logarithm and $\operatorname{Log}_{\mathfrak b}$ for its value continued along the specified path; $\log$ of a positive real quantity is the real logarithm. Frequency logarithms written without a sheet subscript inherit the prescribed continuation. The endpoint powers and logarithms involve both $\omega$ and $2M\omega-m$, while the HeunD map also contains the square root $\varpi^2=-M(2M\omega-m)/\omega$. Consequently, continuing a solution around either branch point can change its frequency sheet and its connection amplitudes. We therefore specify the cuts and continuation path before identifying modes on different sheets.
The endpoint solutions used in the frequency-domain Green function construction are selected by continuation from the upper half of the frequency plane. We place the frequency branch cuts along $\omega=-i\sigma$ and $\omega=\omega_{\rm SR}-i\sigma$, with $\sigma\ge0$~\cite{Casals:2018eev,Casals:2019vdb}. A frequency path retains the logarithms of both frequencies and the branch of $\varpi$; a principal-branch restart after a cut crossing changes the connection problem.

We use $\mathfrak b$ to record the continued angular branch and both radial frequency logarithms. The sheet integers $(\mathfrak b_W,\mathfrak b_K)$ count additional full turns of $\log\omega$ and $\log(2M\omega-m)$ relative to the conventional sheet $(0,0)$; $\arg_{\mathfrak b}$ denotes the corresponding continuously tracked argument. It is a continuation specification, not a continuous physical parameter. The canonical radial Wronskian is
\begin{equation}\label{eq:Evans}
 \mathcal W_{\mathfrak b}(\omega)
 =(r-M)^{2(s+1)}W_r[R_{\ell m}^{\mathrm{in}},R_{\ell m}^{\mathrm{up}}].
\end{equation}
It is independent of $r$. A change of Riemann sheet can change its zeros; moving the displayed cut while retaining the same continued solution cannot. For $m=0$, the two radial branch points coincide. The sign of $\operatorname{Re}(\omega)$ alone does not identify a sheet: modes on the conventional Riemann sheet also have symmetry partners with negative real frequencies. Sheet assignments refer to the continued logarithms and boundary solutions~\cite{Kubota:2026hdv}.

Suitable rays within the decay sectors satisfy
\begin{subequations}\label{eq:radial-rays}
\begin{align}
 \arg(r_H-M)&=\arg(2M\omega-m)-\frac{\pi}{2}+\epsilon_H,\label{eq:horizon-ray}\\
 \arg(r_\infty-M)
 &=\frac{\pi}{2}-\arg\omega+\epsilon_\infty.\label{eq:infinity-ray}
\end{align}
\end{subequations}
Here $r_H$ and $r_\infty$ are finite starting points near the horizon and infinity, and $\epsilon_H,\epsilon_\infty$ are real phase offsets within the chosen Stokes sectors. Varying them tests the same radial solution; it does not determine the frequency sheet. We continue HeunD solutions between the endpoints and solve the angular and radial conditions jointly. Precision, endpoint distance, matching radius, and spatial phases supply independent convergence tests.

\subsubsection{Numerical solution by HeunD propagation}\label{sec:numerical-protocol}
 For each trial $(\omega,\lambda)$, we initialize the horizon and infinity solutions from their asymptotic expansions and propagate their values and derivatives to $r^\star$. Their matching condition and the angular regularity condition then determine the next frequency and separation constant iterates. The angular branch is initialized from its eigenvalue in the Schwarzschild limit and followed continuously as a function of $a\omega$. The radial sheet integers specify frequency logarithms; they do not independently select an angular eigenvalue. Along the trajectories this branch selection is numerical continuation, whereas the angular isolation used in the bounded count is established throughout $D_{33}$.

The ingoing horizon and outgoing infinity solutions have the sectorial expansions
\begin{subequations}\label{eq:numerical-endpoint-series}
\begin{align}
 R_{\ell m}^{\mathrm{in}}(r)&\underset{r\to M}{\sim}
 R_{\mathrm{in}}^M(r)\sum_{n=0}^{\infty}a_n^M(r-M)^n,\\
 R_{\ell m}^{\mathrm{up}}(r)&\underset{r\to\infty}{\sim}
 R_{\mathrm{up}}^{\infty}(r)\sum_{n=0}^{\infty}a_n^{\infty}(r-M)^{-n}.
\end{align}
\end{subequations}
 The coefficients $a_n^M$ and $a_n^{\infty}$ follow recursively from the radial equation, with $a_0^M=a_0^{\infty}=1$. At finite starting points $r_H$ and $r_\infty$ on the rays~\eqref{eq:radial-rays}, we truncate near the smallest term and differentiate the truncated expression. Thus the endpoint data are the states $(R,R')$, where $R'={\rm d}R/{\rm d}r$, at starting radii set by a positive dimensionless endpoint scale $L$:
\begin{equation}\label{eq:numerical-endpoint-radii}
 |r_H-M|=\frac{|M(2M\omega-m)|}{L},\qquad
 |r_\infty-M|=\frac{L}{|\omega|}.
\end{equation}
 Each endpoint state therefore contains the subleading terms of the radial asymptotic expansions. The recurrence and truncation errors are described in Appendix~\ref{app:sectorial-states}.

The endpoint states are then transported to the matching point. We remove the first derivative in the radial equation by setting
\begin{equation}\label{eq:numerical-radial-state}
 \begin{aligned}
 \Psi_{\ell m}(r)&=(r-M)^{s+1}R_{\ell m}(r),\\
 \mathbf v(r)&=\begin{pmatrix}\Psi_{\ell m}(r)\\ \frac{\rm d}{{\rm d}r}\Psi_{\ell m}\end{pmatrix}.
 \end{aligned}
\end{equation}
 For each propagation step, the two local HeunD solutions~\eqref{eq:heund-local-solutions} are a computational basis, not the two endpoint state vectors. Set $\Psi_j=(r-M)^{s+1}R_j$ and define
\begin{equation}\label{eq:heund-transfer-matrix}
 \begin{aligned}
  \mathsf F(r)&=\begin{pmatrix}\Psi_1(r)&\Psi_2(r)\\
  \Psi_1'(r)&\Psi_2'(r)\end{pmatrix},\\
  \mathbf c(r_0)&=\mathsf F(r_0)^{-1}\mathbf v(r_0),\\
  \mathbf v(r_1)&=\mathsf F(r_1)\mathbf c(r_0)\\
  &=\mathsf F(r_1)\mathsf F(r_0)^{-1}\mathbf v(r_0).
 \end{aligned}
\end{equation}
 Here $r_0,r_1$ are neighboring ordinary points on the radial path, $\mathbf c=(c_1,c_2)^{\mathsf T}$ is obtained by a $2\times2$ linear solve, and $\mathsf T$ denotes transpose. Explicitly, with $\Delta_F=\det\mathsf F(r_0)$,
 \begin{equation}
  c_1=\frac{\Psi_2'\Psi-\Psi_2\Psi'}{\Delta_F},\qquad
  c_2=\frac{-\Psi_1'\Psi+\Psi_1\Psi'}{\Delta_F}.
 \end{equation}
 The coefficients are local coordinates and are recomputed in the next local basis; they are not QNM amplitudes. The implementation rescales the local radial coordinate at each step and evaluates a HeunD basis at an ordinary point using the convention in Appendix~\ref{app:map}. Changing that basis leaves the transfer matrix unchanged.

The horizon state is propagated outward. For the infinity state, radial inversion converts infinity to a small-coordinate endpoint; after propagation, the chain rule returns its value and radial derivative at $r^\star$. Both states are then expressed in the same radial coordinate and can be compared directly. Their frequency logarithms and radial powers retain the prescribed continuation paths throughout.

Once both transported states are expressed in the same radial coordinate at the matching point, write the unscaled determinant as
 \begin{equation}
  \mathcal W_{\mathfrak b}=\det[\mathbf v_{\mathrm{in}},\mathbf v_{\mathrm{up}}]
  =W_r[\Psi_{\ell m}^{\mathrm{in}},\Psi_{\ell m}^{\mathrm{up}}].
 \end{equation}
 Since the infinity expansion gives $R_{\ell m}^{\mathrm{in}}=B_{\ell m}^{\mathrm{inc}}R_{\mathrm{in}}^\infty+B_{\ell m}^{\mathrm{ref}}R_{\mathrm{up}}^\infty$,
 \begin{equation}
  B_{\ell m}^{\mathrm{inc}}=\frac{\mathcal W_{\mathfrak b}}{2i\omega},
  \qquad
  C_{\ell m}^{\mathrm{up}}=\frac{\mathcal W_{\mathfrak b}}{2iM(2M\omega-m)}.
 \end{equation}
 For numerical conditioning, each state is normalized by its Euclidean norm:
\[
 \widehat{\mathbf v}=\frac{\mathbf v}{\|\mathbf v\|_2}.
\]
The residual used in the iteration is
\begin{equation}\label{eq:numerical-radial-residual}
 \mathcal F_r(\omega,\lambda)
 =\det\big[\widehat{\mathbf v}_{\mathrm{in}},
 \widehat{\mathbf v}_{\mathrm{up}}\big].
\end{equation}
Overall nonzero factors in the initialized states and the subsequent positive rescalings preserve the determinant zeros. Before these rescalings, the normal-form Wronskian obeys
\begin{equation}\label{eq:normal-radial-wronskian}
 W_r[\Psi_{\ell m}^{\mathrm{in}},\Psi_{\ell m}^{\mathrm{up}}]
 =(r-M)^{2(s+1)}W_r[R_{\ell m}^{\mathrm{in}},R_{\ell m}^{\mathrm{up}}].
\end{equation}
 The positive rescaling changes $\mathcal W_{\mathfrak b}$ by a nonzero factor but preserves its zeros. Thus $\mathcal F_r=0$ implements the radial condition in Eq.~\eqref{eq:radial-matching-qnm}. Because the normalization involves absolute values, we solve four real equations,
\begin{equation}\label{eq:numerical-coupled-root}
 \operatorname{Re}\mathcal F_r=\operatorname{Im}\mathcal F_r
 =\operatorname{Re}W_\theta=\operatorname{Im}W_\theta=0,
\end{equation}
for $\operatorname{Re}(\omega)$, $\operatorname{Im}(\omega)$, $\operatorname{Re}(\lambda)$, and $\operatorname{Im}(\lambda)$. A fixed positive scale balances the radial and angular residuals during each solve. In this implementation, every iteration recomputes the endpoint data, HeunD transfers, and angular residual at the trial values. For supplementary roots, the same radial connection condition is also evaluated by local Taylor continuation of the radial equation. Numerical accuracy is assessed from the residuals and changes under variations of precision, endpoint data, matching radius, and spatial phases. For three representative roots, these variations change the frequencies by less than $7\times10^{-39}$; the independently evaluated incoming-to-reflected amplitude ratios for the QNMs in Table~\ref{tab:extremal-validation} are below $10^{-34}$. The normalized residual is used for locating zeros; canonical amplitudes and contour counts require their own normalization. At $\omega=0$ or $2M\omega=m$, the limiting radial equations replace the degenerate generic HeunD map.

\subsection{Scattering amplitudes in HeunD parameters}\label{sec:dual-mst-check}
For a noncritical frequency with $\alpha\gamma\ne0$, the scattering amplitudes can be written in terms of the HeunD parameters and the Wronskian of the horizon and infinity solutions. With the endpoint-normalized radial solutions continued on the chosen frequency sheet $\mathfrak b$, the unwanted amplitudes are
\begin{subequations}\label{eq:mst-heund-amplitudes-main}
\begin{align}
 B^{\mathrm{inc}}_{\ell m}[\alpha,\beta,\gamma,\delta;\mathfrak b]
 &=-\frac{4\alpha M}{\gamma}(r-M)^{2\left(1+\frac{\beta+\delta}{4\alpha}\right)}
\nonumber\\[-2pt]
 &\quad\times W_r[R_{\ell m}^{\mathrm{in}},R_{\ell m}^{\mathrm{up}}],\\
 C^{\mathrm{up}}_{\ell m}[\alpha,\beta,\gamma,\delta;\mathfrak b]
 &=\frac{4\gamma}{M\alpha^3}(r-M)^{2\left(1+\frac{\beta+\delta}{4\alpha}\right)}
\nonumber\\[-2pt]
 &\quad\times W_r[R_{\ell m}^{\mathrm{in}},R_{\ell m}^{\mathrm{up}}] .
\end{align}
\end{subequations}
The bracketed arguments specify the HeunD parameters and the continuation choice. For the amplitude check, we construct the normalized solutions using bilateral MST $U$-series rather than local HeunD transfer matrices. This provides a different radial representation and numerical evaluation path for the same differential equation, angular eigenvalue, HeunD parameters, and frequency sheet. The independence of this check concerns the radial representation, not these shared inputs. Details and normalization are given in Appendix~\ref{app:mst}. At a noncritical QNM, these amplitudes satisfy the zero condition in Eq.~\eqref{eq:QNM}.
To check the matching roots independently, we solve the amplitude conditions~\eqref{eq:QNM} through two separate endpoint projections. The horizon construction propagates three states and projects $R_{\ell m}^{\mathrm{in}}$ onto the two infinity waves. The infinity construction propagates a separate set of three states and projects $R_{\ell m}^{\mathrm{up}}$ onto the two horizon waves. Its infinity data are initialized after radial inversion. Each construction solves its own coupled equations~\eqref{eq:QNM}. The frequency logarithms, radial powers, and angular branch follow the same continuation path throughout. Propagation uses the full solution and derivative.

To express these amplitudes in HeunD parameters, let $\mathbf h\equiv(\alpha,\beta,\gamma,\delta)$ denote the HeunD parameter quadruple in the Maple convention. The two endpoint projections are explicitly the same continued connection:
\begin{equation}\label{eq:dual-heund-mst-result}
 \begin{aligned}
 -\frac{\gamma}{4\alpha M}B^{\mathrm{inc}}_{\ell m}[\mathbf h;\mathfrak b]
 &=\mathcal W_{\mathfrak b}\\
 &=\frac{M\alpha^3}{4\gamma}C^{\mathrm{up}}_{\ell m}[\mathbf h;\mathfrak b] .
 \end{aligned}
\end{equation}
At a fixed HeunD frequency, the separately constructed amplitudes test the QNM boundary conditions. For $\alpha\gamma\ne 0$, the two unwanted amplitudes vanish at the same connection zero.
Root verification and parent identification are separate tasks. The next section compares the extremal frequencies with finite-spin trajectories and follows selected roots backward to their Schwarzschild limits.

\section{Extremal Kerr spectrum and overtone continuation}\label{sec:complete-extremal-results}

\subsection{Finite-damping spectrum}\label{sec:dm-results}
Following Ref.~\cite{Chen:2025sbz}, Class I ($\omega_{\mathrm I}$) and Class II ($\omega_{\mathrm{II}}$) denote the two families continued from the corresponding Schwarzschild roots. Their labels follow the continued boundary solutions, including their angular branch and frequency sheet, rather than the sign of $\Re(\omega)$.

The traditional positive-frequency and negative-frequency labels classify roots solely by the sign of their real parts. A QNM sequence that crosses the negative imaginary axis can contain roots of both signs, so these labels would divide a single continued sequence into two categories. In particular, our Class I sequences include unconventional roots with $\Re(\omega)<0$; these remain Class I rather than being renamed negative-frequency modes. The Class label is therefore retained along each continued mode sequence, including at purely imaginary frequencies.

The Kerr frequency symmetry relates the corresponding families by
\begin{equation}\label{eq:qnm-class-symmetry}
 \omega_{\mathrm{II},\ell m}(a)=-\omega_{\mathrm I,\ell,-m}^{*}(a).
\end{equation}
The star denotes complex conjugation. The transformation interchanges $m$ and $-m$ and acts on the continued boundary solutions and their frequency logarithms. A mode is therefore specified by its family, angular branch, frequency sheet, and frequency, not by $\omega$ alone. In the Schwarzschild limit, azimuthal degeneracy reduces this relation to $\omega_{\mathrm{II}}=-\omega_{\mathrm I}^{*}$.

In the earlier HeunC calculation, the two classes were selected using the phase condition for the radial contour
\begin{equation}\label{eq:qnm-class-phase}
 \arg r=2k_r\pi+\frac{\pi}{2}(1+\varepsilon)-\arg\omega,
\end{equation}
where $k_r\in\mathbb Z$ and $-1<\varepsilon<1$. The choices $\varepsilon<0$ and $\varepsilon>0$ yield Class I and Class II, respectively; $\varepsilon$ tilts the complex radial contour to avoid the branch cut. We retain the same root labels in the extremal HeunD calculation. The particular frequency sheet is specified separately by $(\mathfrak b_W,\mathfrak b_K)$.

\begin{figure*}[tb]
\centering
\subfloat[$\ell=2$\label{fig:native-spectrum-l2}]{\includegraphics[width=.49\textwidth]{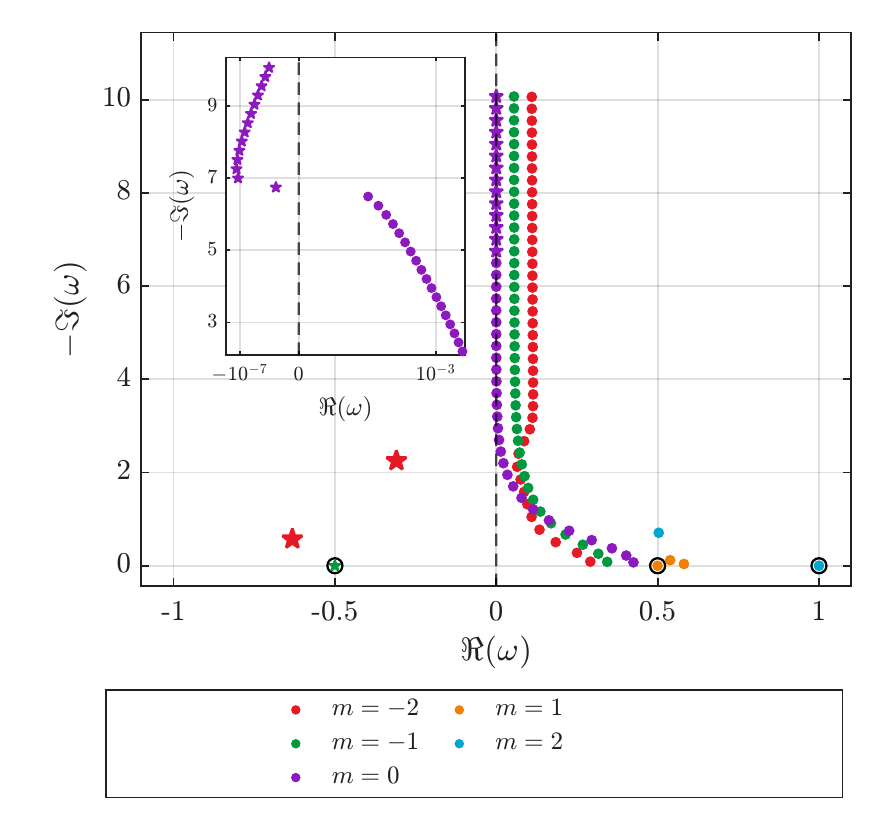}}\hfill
\subfloat[$\ell=3$\label{fig:native-spectrum-l3}]{\includegraphics[width=.49\textwidth]{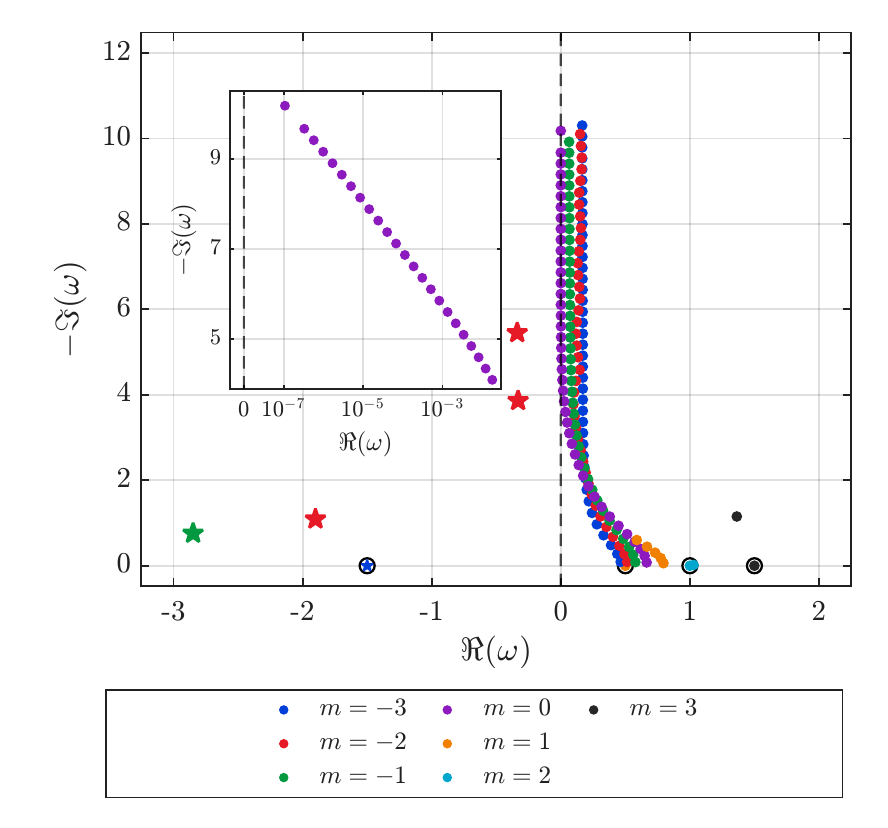}}
\caption{Class I extremal Kerr gravitational roots for $\ell=2,3$ and additional computed $m=0$ roots; panel (a) includes 42 distinct $m=0$ numerical roots. Colors distinguish $m$ consistently across both panels; five-pointed stars outside the black rings identify roots on unconventional Riemann sheets and retain the corresponding $m$ colors. Ordinary dots mark roots on the conventional sheet. The insets resolve the $m=0$ roots with a symmetric logarithmic horizontal scale, linear for $|\Re(\omega)|\le10^{-9}$ in panel (a) and $|\Re(\omega)|\le10^{-7}$ in panel (b). Black rings mark the positive-$m$ superradiant bounds and verified negative-$m$ accumulation frequencies $\omega=m/2$. Their colored centers are dots for positive $m$ and stars for the verified unconventional negative-$m$ sequences. These symbols mark zero-damping accumulation, not additional isolated extremal QNMs. Black dashed lines denote $\Re(\omega)=0$.}\label{fig:native-spectrum}
\end{figure*}

Figure~\ref{fig:native-spectrum} displays the computed extremal gravitational roots for $s=-2$ and $\ell=2,3$. Their distribution is organized by the sign of $m$; stars outside rings identify roots on unconventional sheets, while ringed stars mark the accumulation limits of unconventional negative-$m$ sequences. For $m>0$, the isolated finite-damping roots shown lie in the right half-plane. Panel (a) contains two such roots for $m=1$ and one for $m=2$, while panel (b) contains five for $m=1$ and one each for $m=2,3$. Many nonextremal sequences instead approach the common superradiant frequencies $\omega=m/2$ with vanishing damping. Their limits are represented by the rings, so a small number of isolated extremal points does not imply a small number of continued parent sequences. These rings are not additional isolated extremal QNMs.

For $m=0$, each panel contains 42 numerical roots. At larger damping the roots lie close to the imaginary axis, and the insets resolve their small real parts. The $\ell=2$ sequence extends across the axis: 28 roots have positive real parts, whereas the 14 higher-damping roots have small negative real parts on $(\mathfrak b_W,\mathfrak b_K)=(-1,-1)$. By contrast, all displayed $\ell=3$, $m=0$ roots remain in the right half-plane throughout the computed range; their smallest real part is approximately $1.08\times10^{-7}$. Thus the apparent coincidence with the axis in the main panels hides different behavior for the two multipoles.

For $m<0$, the conventional-sheet roots form long finite-damping sequences in the right half-plane. At high damping these sequences form approximately vertical columns, with larger positive real offsets for larger $|m|$. The unconventional-sheet stars show an additional part of the spectrum in the left half-plane, in the $(2,-2)$, $(3,-1)$, and $(3,-2)$ sectors. Negative-$m$ accumulation also occurs: the unconventional Class I sequences $(\ell,m,\hat n)=(2,-1,8)$ and $(3,-3,40)$ approach $\omega=-1/2$ and $-3/2$, respectively, as $a/M\to1$. Negative-frequency zero-damping limits have previously been discussed~\cite{Yang:2013uba,Richartz:2017qep}; the two sequences identified here retain their Class I Schwarzschild-parent labels and differ from the Class II partners obtained by reflecting the usual positive-$m$ modes [Eq.~\eqref{eq:qnm-class-symmetry}].

Figure~\ref{fig:native-spectrum} therefore combines three patterns: a small set of positive-$m$ isolated roots alongside superradiant accumulation, nearly imaginary axisymmetric sequences with a multipole-dependent change of sign, and extended negative-$m$ sequences supplemented by unconventional-sheet roots and negative-frequency accumulation. In particular, the sign of $m$ does not determine the sign of $\Re(\omega)$.

The frequency, Riemann sheet, and Schwarzschild parent must be specified together to identify a mode. The integers $(\mathfrak b_W,\mathfrak b_K)$ record the extra windings of $\log\omega$ and $\log(2\omega-m)$ relative to the conventional sheet; $(0,0)$ denotes zero additional turns. Table~\ref{tab:extremal-validation} lists these quantities and the independently evaluated amplitude ratios $|B^{\mathrm{inc}}/B^{\mathrm{ref}}|$ for selected roots. The ratios are at most $9.34\times10^{-35}$ for the listed QNMs, demonstrating that the high-precision frequencies satisfy the outgoing boundary condition to high numerical accuracy. The five $(3,1)$ entries are included here because their parent labels distinguish the finite-damping sequence from the intervening ZDM parents. Frequencies are rounded only for display; the amplitude tests use the stored full-precision values. The parent labels are assigned by continuation of the mode sequences, rather than by the size of an amplitude residual. For the unconventional $(2,-2)$ root, $\hat n=8^\dagger$ denotes the parent association supported by the stored finite-spin trajectory and endpoint convergence; its extremal frequency sheet is verified separately.

\begin{table*}[tb]
\centering\small
\caption{Frequencies, Schwarzschild parents, Riemann sheets, and checks of the QNM boundary conditions for selected extremal Kerr QNMs. The conventional sheet is $(0,0)$; $(-1,0)$ denotes continuation across the $\omega=0$ branch cut. The superscript $\dagger$ qualifies only the numerical parent assignment $\hat n=8$, for which continuous sheet transport along the full spin path has not been certified; the extremal root itself is independently verified on its stated sheet. The last row is a conventional-sheet nonroot control, without a parent assignment.}\label{tab:extremal-validation}
\setlength{\tabcolsep}{8pt}
\begin{tabular}{ccccc}
\toprule
$(\ell,m)$ & $\hat n$ & $\omega$ & $(\mathfrak b_W,\mathfrak b_K)$ & $\left|B^{\mathrm{inc}}/B^{\mathrm{ref}}\right|$\\
\midrule
$(2,-2)$ & $0$ & $0.291553464366-0.088025837319i$ & $(0,0)$ & $7.28\times10^{-43}$\\
$(2,0)$ & $0$ & $0.425145109123-0.071806183967i$ & $(0,0)$ & $1.07\times10^{-43}$\\
$(2,2)$ & $5$ & $0.503459950790-0.707398820291i$ & $(0,0)$ & $2.77\times10^{-43}$\\
$(3,1)$ & $0$ & $0.795283356124-0.058965041864i$ & $(0,0)$ & $3.75\times10^{-44}$\\
$(3,1)$ & $1$ & $0.773674497444-0.179000338686i$ & $(0,0)$ & $2.90\times10^{-35}$\\
$(3,1)$ & $2$ & $0.730708008605-0.305803994055i$ & $(0,0)$ & $4.37\times10^{-35}$\\
$(3,1)$ & $3$ & $0.667856253312-0.444854368338i$ & $(0,0)$ & $6.56\times10^{-35}$\\
$(3,1)$ & $6$ & $0.588738959735-0.601671595031i$ & $(0,0)$ & $9.34\times10^{-35}$\\
$(3,2)$ & $0$ & $1.028553339198-0.018571648863i$ & $(0,0)$ & $1.06\times10^{-43}$\\
$(3,3)$ & $34$ & $1.363326725335-1.152382740908i$ & $(0,0)$ & $2.98\times10^{-53}$\\
$(2,-2)$ & $8^\dagger$ & $-0.631844938475-0.571639143992i$ & $(-1,0)$ & $2.41\times10^{-53}$\\
\midrule
\multicolumn{2}{c}{Nonroot control $(2,-2)$} & $-0.631844938475-0.571639143992i$ & $(0,0)$ & $9.26\times10^{-4}$\\
\bottomrule
\end{tabular}
\end{table*}

For the $(2,-2,\hat n=0)$, $(3,3,\hat n=34)$, and unconventional $(2,-2,\hat n=8)$ roots, the two endpoint constructions were also solved with independent numerical settings. The largest differences between their frequencies are $1.08\times10^{-38}$, $2.4\times10^{-38}$, and $6.9\times10^{-40}$, respectively. Simultaneously varying precision, endpoint data, matching radius, and spatial phases changes the corresponding roots by $4.61\times10^{-39}$, $6.26\times10^{-39}$, and $3.36\times10^{-40}$. These are numerical convergence comparisons, rather than certified frequency error bounds. The independently evaluated incoming-to-reflected amplitude ratios in Table~\ref{tab:extremal-validation}, together with finite nonzero outgoing amplitudes, provide a separate test of the QNM boundary conditions.
The sheet dependence is explicit in the final two rows of the table. At the unconventional root on $(\mathfrak b_W,\mathfrak b_K)=(-1,0)$, the amplitude ratio is $2.41\times10^{-53}$. Keeping that frequency fixed but evaluating it on $(0,0)$ gives $9.26\times10^{-4}$. The final row is therefore a nonroot control on the conventional Riemann sheet, not a poorly converged QNM or an additional overtone. It is excluded from the mode count. Small residuals establish neither a Schwarzschild parent nor completeness of the unsampled spectrum; these require the continuation and counting arguments that follow.

\subsection{Continuity from nonextremal to extremal Kerr}\label{sec:parent-anchor}
We establish continuity with nonextremal Kerr by following the coupled angular and radial roots as the spin varies. We retain the Schwarzschild parent label $\hat n$ along a sequence and, where the necessary history is available, track its angular branch and frequency sheet. The label is not reassigned by the instantaneous damping order or the sign of $\Re(\omega)$. A sequence may be followed forward from a Schwarzschild root or backward from a near-extremal HeunC frequency. When the angular branch and frequency sheet are tracked along the entire path, this continuation identifies the parent independently of the scattering residual. Associations supported by stored samples are discussed separately below. The continuation equation and supporting parent checks are given in Appendix~\ref{app:continuation-details}.

\subsubsection{Mode trajectories and Schwarzschild parents}\label{sec:damped-number}
\begin{figure*}[tb]
\centering
\subfloat[$\ell=2,\ m=1$\label{fig:transition-l2m1}]{\includegraphics[width=.32\textwidth]{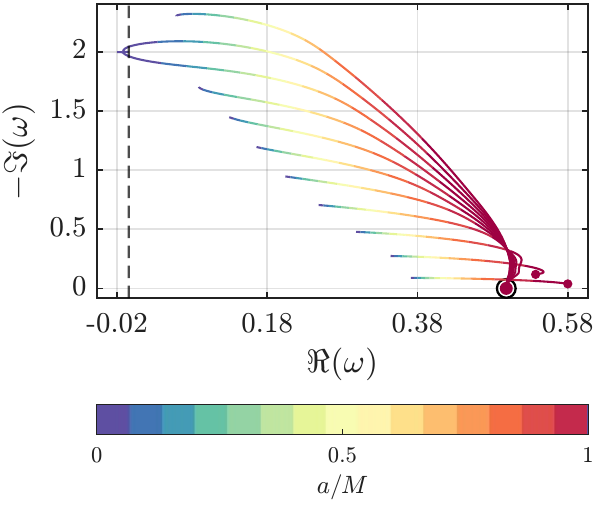}}\hfill
\subfloat[$\ell=m=2$, $\hat n\ne5$\label{fig:transition-zdm}]{\includegraphics[width=.32\textwidth]{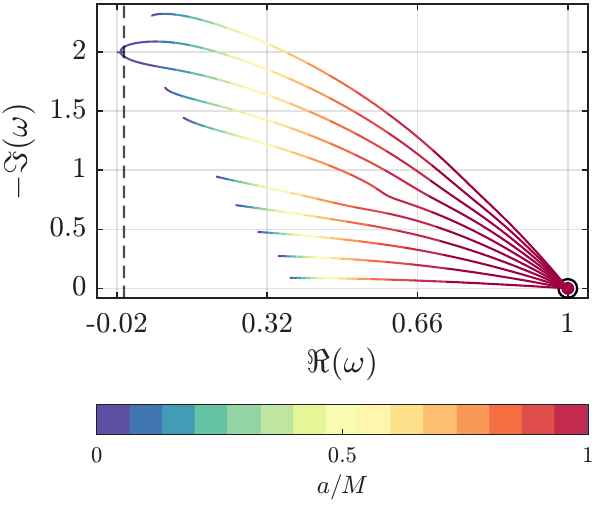}}\hfill
\subfloat[$\ell=m=2,\ \hat n=5$: DM\label{fig:transition-dm}]{\includegraphics[width=.32\textwidth]{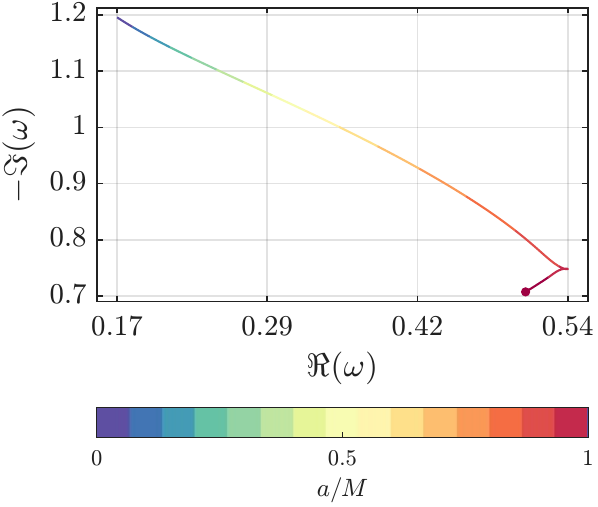}}
\caption{Kerr QNM trajectories with $0\le\hat n\le10$ over the available spin range $0\le a/M<1$, with independently computed extremal Kerr frequencies shown as endpoint dots. Black circles mark superradiant critical frequencies, with filled centers indicating the sequence limits. The common color scale is $0\le a/M\le1$. \protect\subref{fig:transition-l2m1}: $\ell=2,m=1$, with all plotted parents continued to $a/M=0.99999999$. \protect\subref{fig:transition-zdm}: $\ell=m=2$ trajectories with $\hat n\ne5$. \protect\subref{fig:transition-dm}: the $\hat n=5$ trajectory approaching a finite-damping mode. At the Schwarzschild starting points, the trajectories lie farther from the real axis as $\hat n$ increases. Parent labels follow our complete Schwarzschild spectrum~\cite{Chen:2025sbz}.}\label{fig:transition}
\end{figure*}

\begin{figure*}[tb]
\centering
\subfloat[$\ell=2,\ m=0$\label{fig:l2-all-m-continuation-m0}]{\includegraphics[width=.32\textwidth]{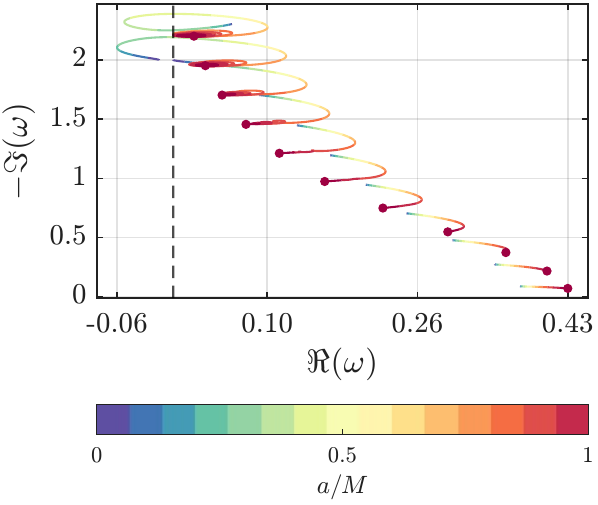}}\hfill
\subfloat[$\ell=2,\ m=-1$\label{fig:l2-all-m-continuation-mm1}]{\includegraphics[width=.32\textwidth]{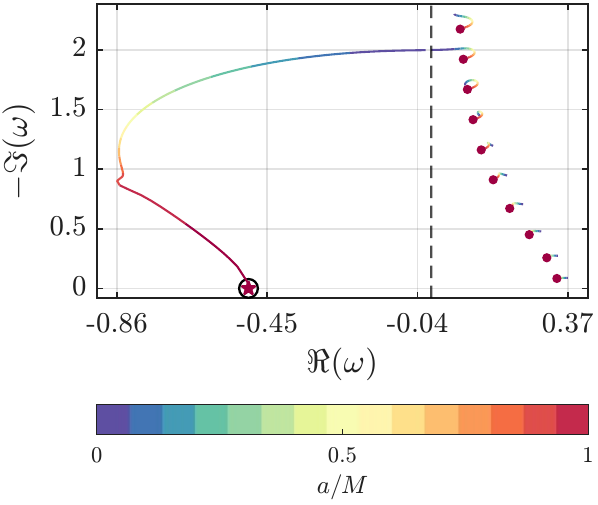}}\hfill
\subfloat[$\ell=2,\ m=-2$\label{fig:l2-all-m-continuation-mm2}]{\includegraphics[width=.32\textwidth]{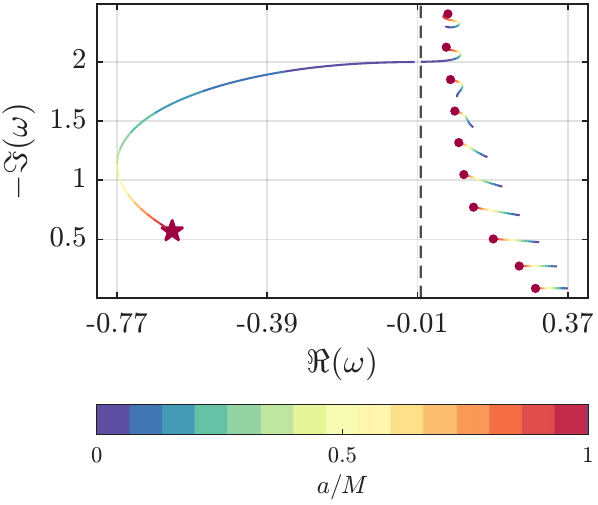}}
\caption{Kerr trajectories for $\ell=2$, $-2\le m\le0$, and $0\le\hat n\le10$. Colors indicate $a/M$; outside rings, dots and filled stars show independently computed isolated extremal Kerr QNMs on conventional and unconventional sheets, respectively. The stars use the final color of the spin scale. In panel \protect\subref{fig:l2-all-m-continuation-mm1}, the unconventional $\hat n=8$ sequence approaches $\omega=m/2=-1/2$ with vanishing damping as $a/M\to1$. A black ring enclosing a filled star marks this unconventional sequence's accumulation limit, not an additional isolated extremal QNM. At the Schwarzschild starting points, the trajectories lie farther from the real axis as $\hat n$ increases. The black dashed line denotes the imaginary axis, $\Re(\omega)=0$.}\label{fig:l2-all-m-continuation}
\end{figure*}

We first compare the nonextremal $\ell=2$ trajectories for $0\le\hat n\le10$ and all $-2\le m\le2$ with the independently computed extremal frequencies in Figs.~\ref{fig:transition} and~\ref{fig:l2-all-m-continuation}. The independently computed extremal frequencies are shown as dots on the conventional sheet and filled stars outside rings on unconventional Riemann sheets. The black circles mark the superradiant critical frequencies. The trajectories distinguish two types of limit: DMs retain finite damping and approach isolated extremal roots, whereas ZDM sequences approach $\omega=m/2$ with vanishing damping. This distinction depends on the continued parent sequence. In particular, the neighboring $\ell=m=2$ parents in Figs.~\ref{fig:transition-zdm} and~\ref{fig:transition-dm} can have different extremal outcomes. The $\hat n=9,10$ trajectories are extended by solving the radial and angular conditions from their stored finite-spin anchors; the corresponding $\ell=m=2$ sequences are included in the near-extremal scaling comparison. Higher-overtone results will be included in our updated data set.

\begin{figure*}[tb]
\centering
\subfloat[$9\le\hat n\le14$\label{fig:parent-identification-low}]{\includegraphics[width=.49\textwidth]{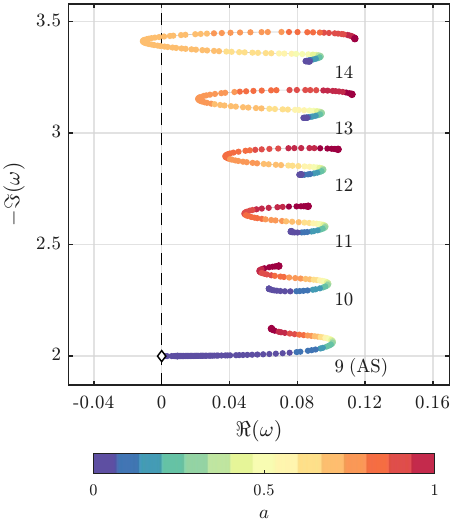}}\hfill
\subfloat[$16\le\hat n\le22$\label{fig:parent-identification-high}]{\includegraphics[width=.49\textwidth]{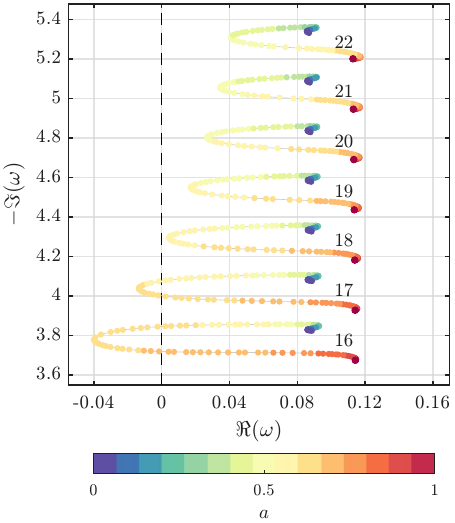}}
\caption{Backward continuation of the $(\ell,m)=(2,-2)$ sequences from the HeunC frequencies at $a=0.999999$. The trajectories are computed with an independent scattering-amplitude representation and colored by $a$; the labels indicate Schwarzschild parents $\hat n$. Endpoint dots at $a/M=1$ mark independently computed extremal HeunD frequencies. The diamond marks the algebraically special limit $\omega=-2i$ of the $\hat n=9$ sequence. Black dashed lines denote $\Re(\omega)=0$.}\label{fig:parent-identification}
\end{figure*}

For the more strongly damped $(\ell,m)=(2,-2)$ sequences, Fig.~\ref{fig:parent-identification} illustrates backward continuation toward the Schwarzschild limit. Twelve trajectories reach the Schwarzschild parents $10\le\hat n\le22$, $\hat n\ne15$. The remaining sequence approaches the algebraically special frequency $\omega=-2i$ as $a\to0$ and carries the label $\hat n=9$ in our convention; its limit is marked by a diamond in Fig.~\ref{fig:parent-identification-low}. The $a/M=1$ endpoint dots in Fig.~\ref{fig:parent-identification-high} mark the independently solved extremal roots, including the endpoint associated with $\hat n=22$. Table~\ref{tab:cook-l2-parent-candidates} lists the Schwarzschild, near-extremal, and extremal frequencies, making these connections explicit. The $\hat n=15$ branch is absent from this conventional-sheet comparison because its verified extremal endpoint lies on $(-1,0)$, among the unconventional roots in Fig.~\ref{fig:native-spectrum-l2}.

\begin{table*}[tb]
\caption{Frequencies $\omega$ for $(\ell,m)=(2,-2)$ at the indicated spins, labeled by Schwarzschild parent $\hat n$. Nonextremal frequencies are recomputed with HeunC at $a=0.999999$; $\hat n=9$ denotes the algebraically special limit.}\label{tab:cook-l2-parent-candidates}
\centering\small
\begin{tabular}{@{}c@{\hspace{1em}}c@{\hspace{1em}}c@{\hspace{1em}}c@{}}
\toprule
$\hat n$ & $a=0$ & $a=0.999999$ & $a=1$\\
\midrule
9 & $-2i$ & $0.0647832760-2.1233474924i$ & $0.0647832733-2.1233476885i$\\
10 & $0.0632635051-2.3026447652i$ & $0.0691163690-2.4041154241i$ & $0.0691165535-2.4041155612i$\\
11 & $0.0765534629-2.5608266174i$ & $0.0864419792-2.6722106619i$ & $0.0864421825-2.6722106625i$\\
12 & $0.0825981446-2.8154423726i$ & $0.1040840577-2.9275985028i$ & $0.1040842276-2.9275983681i$\\
13 & $0.0857279221-3.0686947439i$ & $0.1122856549-3.1742669446i$ & $0.1122856867-3.1742668087i$\\
14 & $0.0873942624-3.3212302384i$ & $0.1139756845-3.4239679694i$ & $0.1139756822-3.4239678730i$\\
16 & $0.0885903871-3.8251052081i$ & $0.1142712502-3.6759318540i$ & $0.1142712362-3.6759317799i$\\
17 & $0.0886329025-4.0766645887i$ & $0.1141591886-3.9290978216i$ & $0.1141591692-3.9290977619i$\\
18 & $0.0884767066-4.3280502096i$ & $0.1139171109-4.1829744402i$ & $0.1139170886-4.1829743906i$\\
19 & $0.0881908481-4.5792969569i$ & $0.1136426785-4.4373067812i$ & $0.1136426542-4.4373067394i$\\
20 & $0.0878201241-4.8304297491i$ & $0.1133720815-4.6919500391i$ & $0.1133720559-4.6919500037i$\\
21 & $0.0873943883-5.0814670995i$ & $0.1131176023-4.9468166670i$ & $0.1131175756-4.9468166369i$\\
22 & $0.0869339503-5.3324232006i$ & $0.1128819190-5.2018511749i$ & $0.1128818913-5.2018511496i$\\
\bottomrule
\end{tabular}
\end{table*}

\begin{figure*}[tb]
\centering
\subfloat[$\ell=3,\ m=-3$\label{fig:l3-all-m-continuation-mm3}]{\includegraphics[width=.47\textwidth]{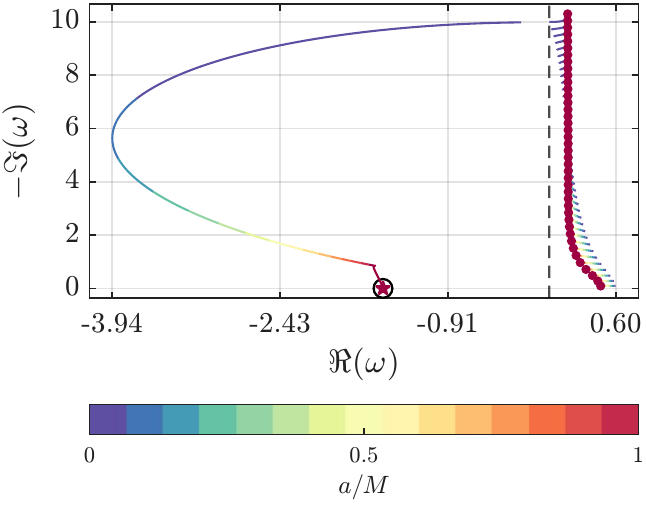}}\hfill
\subfloat[$\ell=3,\ m=-2$\label{fig:l3-all-m-continuation-mm2}]{\includegraphics[width=.47\textwidth]{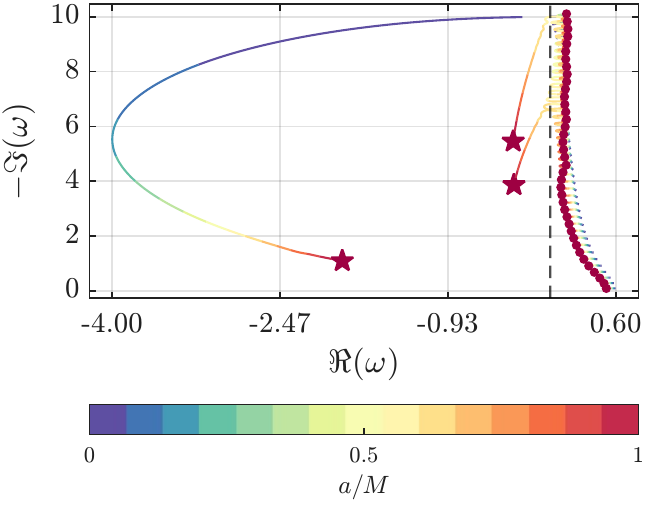}}\\[1ex]
\subfloat[$\ell=3,\ m=-1$\label{fig:l3-all-m-continuation-mm1}]{\includegraphics[width=.47\textwidth]{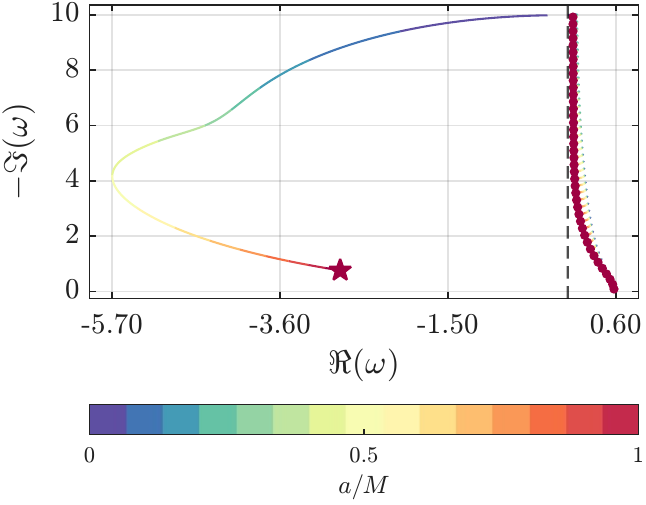}}\hfill
\subfloat[$\ell=3,\ m=0$\label{fig:l3-all-m-continuation-m0}]{\includegraphics[width=.47\textwidth]{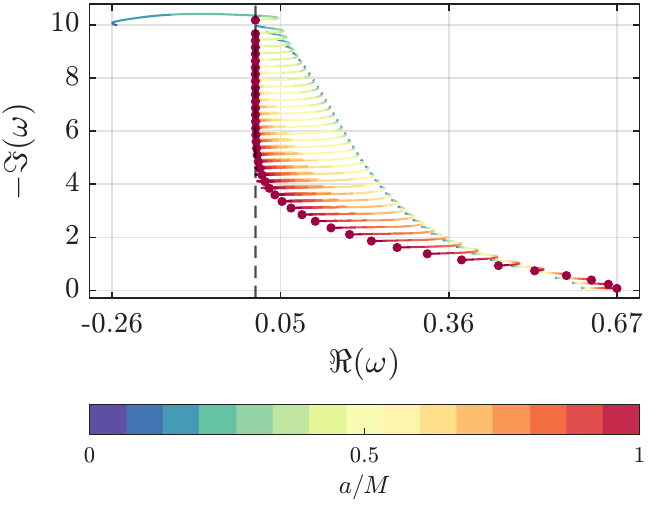}}
\caption{Kerr QNM trajectories for $\ell=3$, $-3\le m\le0$, and $0\le\hat n\le41$. Colors indicate $a/M$ on the common scale $0\le a/M\le1$; outside rings, dots and filled stars show independently computed isolated extremal Kerr QNMs on conventional and unconventional sheets, respectively. The stars use the final color of the spin scale. In panel \protect\subref{fig:l3-all-m-continuation-mm3}, the unconventional $m=-3,\hat n=40$ sequence approaches $\omega=m/2=-3/2$ with vanishing damping as $a/M\to1$. A black ring enclosing a filled star marks this unconventional sequence's accumulation limit, not an additional isolated extremal QNM.}\label{fig:l3-all-m-continuation}
\end{figure*}

\begin{figure*}[tb]
\centering
\subfloat[$\ell=3,\ m=1$\label{fig:l3-positive-m-continuation-m1}]{\includegraphics[width=.32\textwidth]{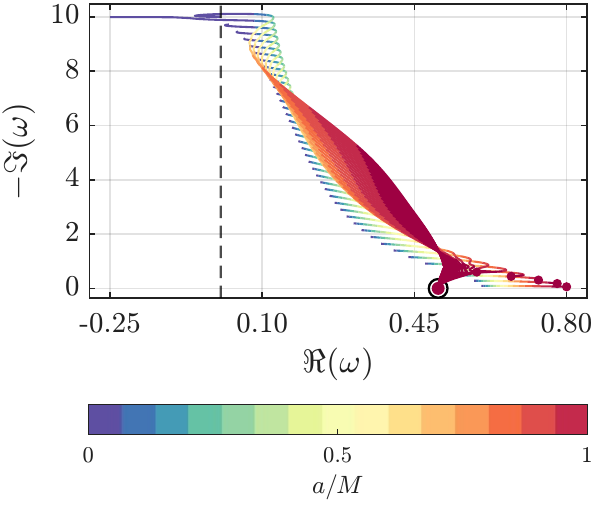}}\hfill
\subfloat[$\ell=3,\ m=2$\label{fig:l3-positive-m-continuation-m2}]{\includegraphics[width=.32\textwidth]{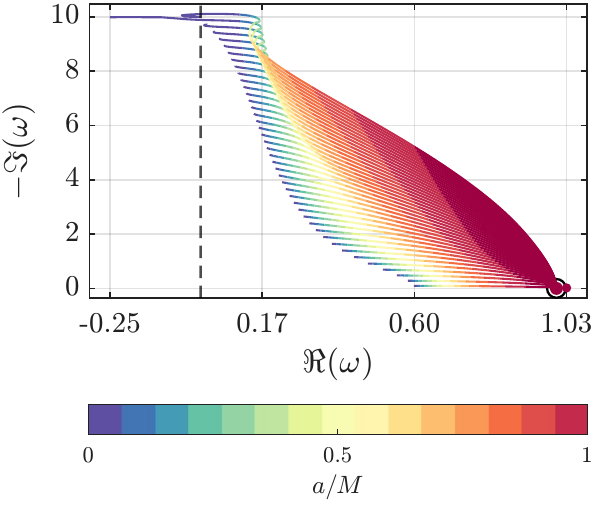}}\hfill
\subfloat[$\ell=3,\ m=3$\label{fig:l3-positive-m-continuation-m3}]{\includegraphics[width=.32\textwidth]{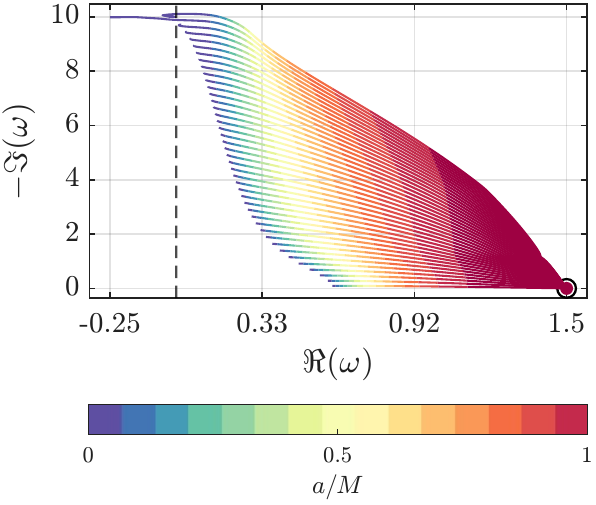}}
\caption{Class I Kerr QNM trajectories for $\ell=3$, $1\le m\le3$, and $0\le\hat n\le41$. Colors and endpoint markers follow Fig.~\ref{fig:l3-all-m-continuation}; black circles mark $\omega=m/2$.}\label{fig:l3-positive-m-continuation}
\end{figure*}

Figures~\ref{fig:l3-all-m-continuation} and~\ref{fig:l3-positive-m-continuation} extend this comparison to all $-3\le m\le3$ and $0\le\hat n\le41$ for $\ell=3$. The positive-$m$ panels show that superradiant accumulation can coexist with isolated finite-damping endpoints in each sector. Their parent assignments require following the trajectories, rather than sorting the roots by damping at one spin. The continuations for $m\le0$ also include Class I sequences on unconventional Riemann sheets.

\subsubsection{Near-extremal scaling of the $\ell=2$ modes}\label{sec:scaling}
Following the near-extremal matching analysis~\cite{Yang:2012pj,Yang:2013uba}, we evaluate the separation exponent at the superradiant frequency:
\begin{equation}\label{eq:deltasr}
 \delta_{\rm SR}^{\,2}=\frac{7m^2}{4}-(s+\frac{1}{2})^2-{}_s A_{\ell m}(\frac{m}{2}).
\end{equation}
The principal square root may be real or purely imaginary. With surface gravity $\kappa=\frac{r_+-r_-}{2(r_+^2+a^2)}$ and horizon angular velocity $\Omega_H=\frac{a}{r_+^2+a^2}$, the analytical near-extremal resonance formula~\cite{Hod:2008zz,Hod:2009td} can be written in the present notation as~\cite{Yang:2012pj,Yang:2013uba,Casals:2019vdb}
\begin{equation}\label{eq:ZDM}
 \omega_j\simeq m\Omega_H+\kappa[m-\delta_{\rm SR}-i(j+\frac{1}{2})].
\end{equation}
The nonnegative integer $j$ labels the local ZDM ladder and need not coincide with the Schwarzschild parent label $\hat n$. For the $m>0$ sequences considered here, we evaluate the extremal-map expression $\alpha^2=64M\omega(2M\omega-m)$ at the finite-spin frequencies as a scaling diagnostic. Equation~\eqref{eq:ZDM} then gives
\begin{equation}\label{eq:alpha-zdm-index}
 \frac{\alpha^2}{64mM\kappa}\simeq-\delta_{\rm SR}-i(j+\frac{1}{2}).
\end{equation}
Both the real and imaginary parts of this rate are needed to test ZDM scaling. For real $\delta_{\rm SR}$, the leading matching relation has small corrections that may oscillate with $\log(1-a/M)$ and need not vanish after division by $\kappa$~\cite{Yang:2013uba}; the displayed rates are therefore leading matching values. For $\delta_{\rm SR}^{\,2}<0$, the correction decays as a power of $1-a/M$. The QNM boundary conditions select the roots, while this scaled parameter describes their approach to the critical frequency.

\begin{figure*}[tb]
\centering
\subfloat[Scaling of the real part.\label{fig:scaling-real}]{\includegraphics[width=.32\textwidth]{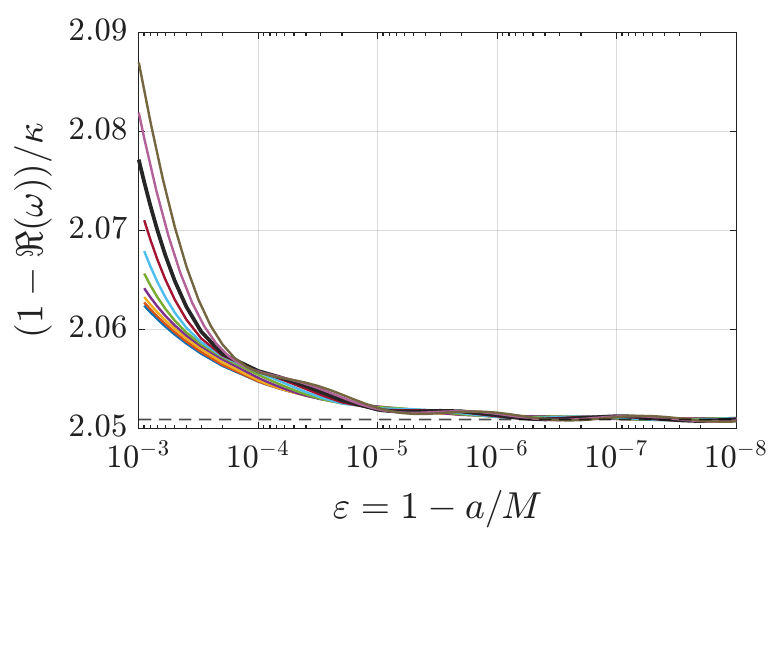}}\hfill
\subfloat[ZDM damping ladder.\label{fig:scaling-damping}]{\includegraphics[width=.32\textwidth]{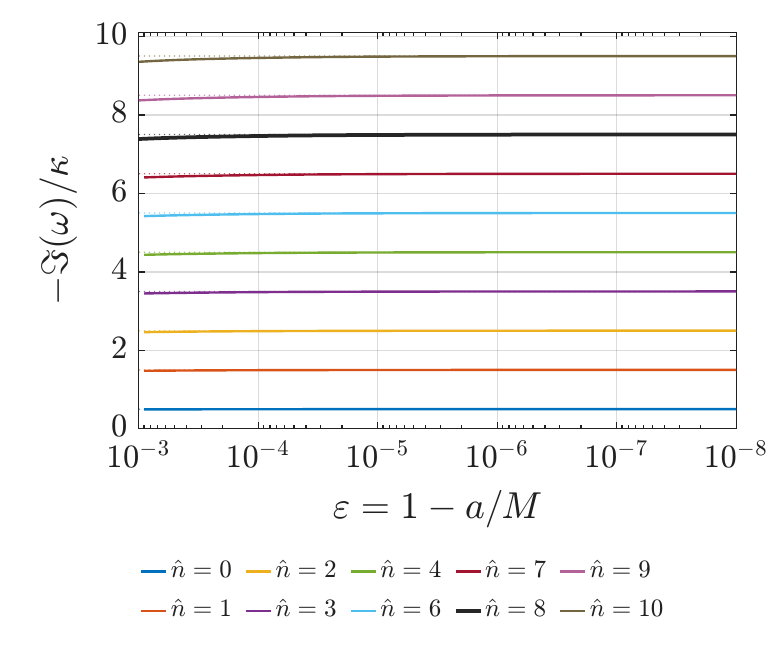}}\hfill
\subfloat[Finite-damping convergence.\label{fig:scaling-dm}]{\includegraphics[width=.32\textwidth]{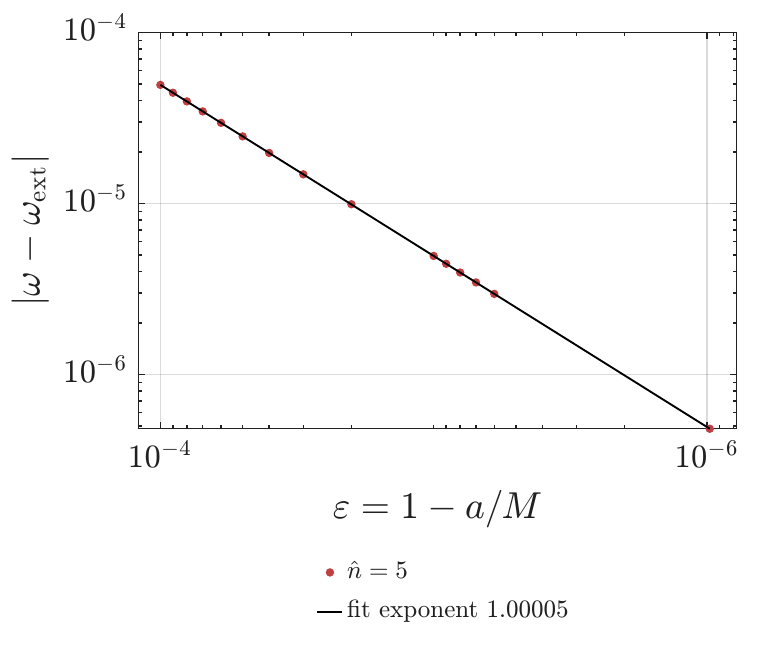}}
\caption{Near-extremal scaling of the $\ell=m=2$ trajectories. Panels~\protect\subref{fig:scaling-real} and~\protect\subref{fig:scaling-damping}: The scaled frequency shift $[1-\Re(\omega)]/\kappa$ and damping $-\Im(\omega)/\kappa$ are compared with the leading matching values $\delta_{\rm SR}$ and $j+\frac{1}{2}$, respectively; these are not exact constant limits. Curve colors distinguish the parent labels $\hat n$; in panel~\protect\subref{fig:scaling-damping}, the curves are ordered from bottom to top by increasing local index $j$. The ZDM panels include $0\le\hat n\le10$, $\hat n\ne5$, including the unconventional $\hat n=8$ sequence; the new $\hat n=9,10$ trajectories reach $1-a/M=10^{-8}$. \protect\subref{fig:scaling-dm} The $\hat n=5$ distance from the extremal DM has fitted exponent $1.00005$.}\label{fig:scaling}
\end{figure*}

The contrasting $\ell=m=2$ limits are resolved in Figs.~\ref{fig:scaling-real}--\ref{fig:scaling-dm}. The ten parents with $0\le\hat n\le10$ and $\hat n\ne5$ approach $\omega=1$, while the $\hat n=5$ trajectory approaches
\begin{qnmvalues}
\begin{equation}\label{eq:dm-frequency}
 \omega_{\mathrm{ext}}=0.5034599507901126-0.7073988202908459i.
\end{equation}
\end{qnmvalues}
The ZDM indices are $j=\hat n$ for $0\le\hat n\le4$ and $j=\hat n-1$ for $6\le\hat n\le10$: the DM interrupts this finite parent list before the remaining sequences are indexed on the critical ladder. Figures~\ref{fig:scaling-real} and~\ref{fig:scaling-damping} show the corresponding square-root approach in $1-a/M$, with fitted exponents close to $1/2$. In contrast, Fig.~\ref{fig:scaling-dm} gives exponent $1.00005$ for the distance to the $\hat n=5$ DM. The unconventional $\hat n=8$ parent belongs to the former group, with $j\simeq6.99724$ at the closest stored spin, consistent with $j=7$ in the leading approximation. The $\hat n=5$ rate instead diverges because its limiting $\alpha^2$ is nonzero. Thus a local index inferred from damping alone would misclassify this DM. The $m=1$ fundamental and first overtone in Fig.~\ref{fig:transition-l2m1} likewise approach the damped frequencies $0.581433202454-0.038255455258i$ and $0.538854664198-0.118627904339i$, respectively. In the same panel, the remaining displayed parents, $2\le\hat n\le10$, approach $\omega=1/2$ with decreasing damping.

\subsubsection{Damped and zero-damping $\ell=3$ modes}\label{sec:l3-damped}
Figure~\ref{fig:l3-parent-classification-trajectory} reveals a second interruption of the ZDM parent ordering for $(\ell,m)=(3,1)$. The parents with $0\le\hat n\le3$ or $\hat n=6$ retain finite damping, while $4\le\hat n\le20$, $\hat n\ne6$, approach $\omega=1/2$. We associate the $\hat n=6$ sequence with the independently computed HeunD root
\begin{qnmvalues}
\begin{equation}\label{eq:l3m1-parent6}
 \omega_{\mathrm{ext}}=0.5887389597350677-0.6016715950314355i.
\end{equation}
\end{qnmvalues}
Its high-precision amplitude check is included in Table~\ref{tab:extremal-validation}. Here $\delta_{\rm SR}=3.17492296964616i$, so its imaginary part contributes to the scaled damping in Eq.~\eqref{eq:alpha-zdm-index}. The ZDM parents have $j=\hat n-4$ for $4\le\hat n\le5$ and $j=\hat n-5$ for $7\le\hat n\le20$. For the three examples in Fig.~\ref{fig:l3-parent-classification-zdm}, the final samples give $j\simeq0.000160$, $1.000204$, and $2.000248$ for $\hat n=4,5,7$, respectively, with scaled real parts tending to zero. The diverging rate of $\hat n=6$ in Fig.~\ref{fig:l3-parent-classification-dm} separates its finite-damping endpoint from these neighboring ZDMs. A single threshold in the Schwarzschild overtone therefore does not describe the DM/ZDM division.

\begin{figure*}[tb]
\centering
\subfloat[Kerr QNM trajectories\label{fig:l3-parent-classification-trajectory}]{\includegraphics[width=.32\textwidth]{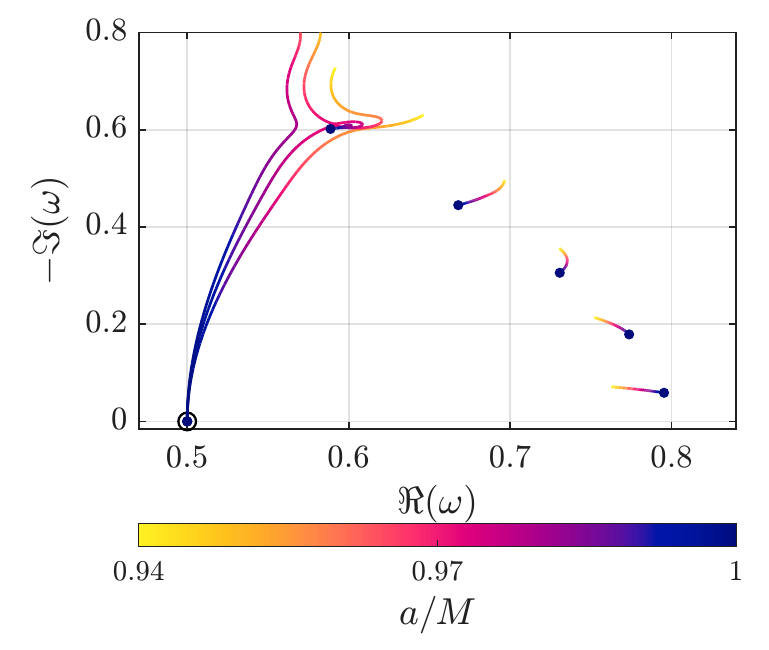}}\hfill
\subfloat[ZDMs\label{fig:l3-parent-classification-zdm}]{\includegraphics[width=.32\textwidth]{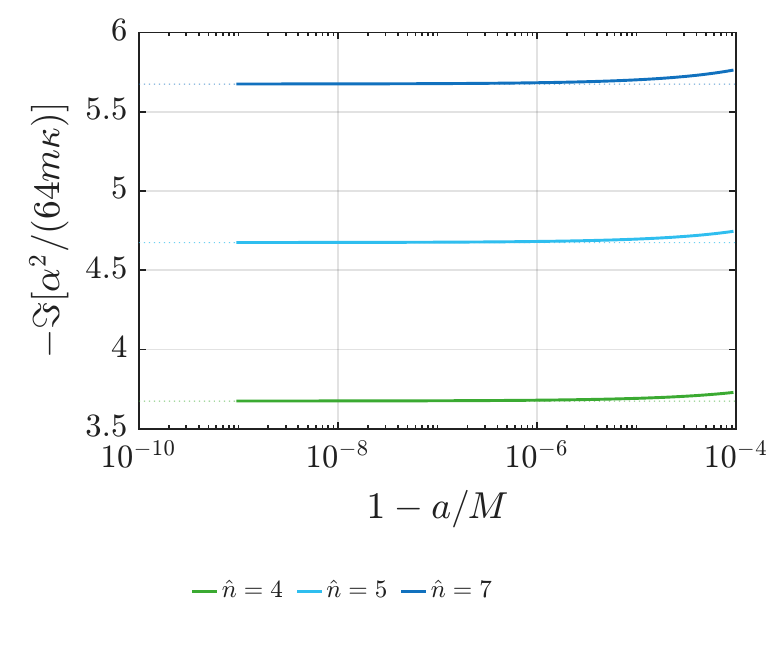}}\hfill
\subfloat[DM: $\hat n=6$\label{fig:l3-parent-classification-dm}]{\includegraphics[width=.32\textwidth]{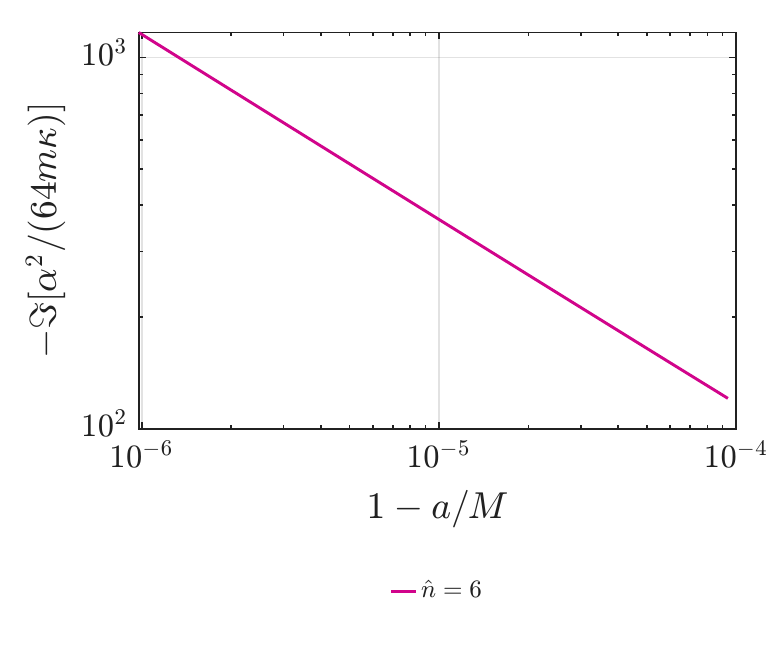}}
\caption{The $(\ell,m)=(3,1)$ trajectories and their limiting behavior. \protect\subref{fig:l3-parent-classification-trajectory} Comparison trajectories from Ref.~\cite{Cook:2014cta}, assigned Schwarzschild parent labels using our HeunC calculation, are colored by $a/M$; endpoint dots mark the five finite-damping HeunD frequencies and the black circle marks $\omega=1/2$. \protect\subref{fig:l3-parent-classification-zdm} The HeunD parameter rates for the ZDM parents $4\le\hat n\le7$, $\hat n\ne6$, with dotted lines indicating their limiting values. \protect\subref{fig:l3-parent-classification-dm} The corresponding rate for the DM parent $\hat n=6$.}\label{fig:l3-parent-classification}
\end{figure*}

The remaining corotating sectors provide complementary examples in Figs.~\ref{fig:l3-positive-m-continuation-m2} and~\ref{fig:l3-positive-m-continuation-m3}. For $(3,2)$, the fundamental approaches the finite-damping frequency $1.028553339198-0.018571648863i$ in Table~\ref{tab:extremal-validation}, while the next twenty parents follow critical scaling with $j=\hat n-1$. For $(3,3)$, the parents with $0\le\hat n\le20$ follow the ZDM ladder with $j=\hat n$. The isolated extremal DM previously found from scattering amplitudes~\cite{Casals:2019vdb} nevertheless remains in the same sector. Our paired continuation associates this root with the Schwarzschild parent $\hat n=34$, while its neighboring parents $\hat n=33$ and $35$ approach the critical frequency with local ZDM indices $j=33$ and $34$. Its finite limiting damping implies $-\Im(\omega)/\kappa\to\infty$, in contrast to the finite scaled rates of the neighboring ZDMs. Thus the DM lies between ZDM parents in this higher-overtone set, as does $\hat n=6$ in the $(3,1)$ sector. These results establish individual parent associations and show why a single overtone cutoff cannot describe the DM/ZDM division.

\subsection{QNMs across the frequency branch cuts}
The continuation also retains Class I roots on Riemann sheets reached across the frequency branch cuts. Our earlier HeunC calculation~\cite{Chen:2025sbz} followed the Kerr branches originating from the unconventional $\ell=2$, $\hat n=8$ Schwarzschild mode. At $a=0$, all $m$ with $-2\le m\le2$ share the frequency
\begin{qnmvalues}
\begin{equation}
 \omega\simeq-0.01532450476070-1.99841184185905i,
\end{equation}
\end{qnmvalues}
on the same unconventional sheet. The $m=1,2$ sequences in Figs.~\ref{fig:transition-l2m1} and~\ref{fig:transition-zdm} approach their respective superradiant frequencies, even though their common Schwarzschild parent has $\Re(\omega)<0$. Their Class I labels are retained along the continuation.

For $m<0$, unconventional Class I QNM sequences also reach the superradiant critical frequency $\omega=m/2$ in the extremal limit $a/M\to1$, with vanishing damping. In Fig.~\ref{fig:l2-all-m-continuation-mm1}, the $(\ell,m,\hat n)=(2,-1,8)$ sequence reaches $1-a/M=10^{-20}$ with
\begin{qnmvalues}
\begin{equation}\label{eq:negative-m-zdm-l2}
 \omega+\frac12\simeq -3.23807\times10^{-20} - 1.710541\times10^{-10}i.
\end{equation}
\end{qnmvalues}
Its damping scales as $(1-a/M)^{1/2}$, with $-\Im(\omega)/\kappa$ approaching $2.419070586$, while its real part approaches $-1/2$. The $(3,-3,40)$ sequence in Fig.~\ref{fig:l3-all-m-continuation-mm3} reaches $1-a/M=10^{-14}$ with
\begin{qnmvalues}
\begin{equation}\label{eq:negative-m-zdm-l3}
 \omega+\frac32\simeq -1.102377\times10^{-7} - 7.708572\times10^{-7}i.
\end{equation}
\end{qnmvalues}
The two sequences therefore reach $\omega=m/2=-1/2$ and $-3/2$, respectively, in the extremal limit. They belong to Class I, whereas the negative-$m$ partners obtained by applying Kerr symmetry to the usual positive-$m$ Class I ZDMs belong to Class II~\cite{Yang:2013uba,Richartz:2017qep}.

The $m=-2$ sequence associated with the unconventional $\ell=2$, $\hat n=8$ Schwarzschild mode has a different limit. Figure~\ref{fig:unconventional-endpoint-trajectory} shows a finite-damping endpoint in the left half of the frequency plane. Solving the extremal HeunD connection on $(\mathfrak b_W,\mathfrak b_K)=(-1,0)$ gives
\begin{subequations}\label{eq:unconventional-root}
\begin{qnmvalues}
\begin{align}
 \omega_{\mathrm{ext}}&=-0.6318449384746158-0.5716391439917598i,\label{eq:unconventional-frequency}\\
 \lambda_{\mathrm{ext}}&=-0.1738342624017932-3.6490524069801881i.\label{eq:unconventional-lambda}
\end{align}
\end{qnmvalues}
\end{subequations}
Here the continued argument of $\omega$ differs by $-2\pi$ from the conventional value, while the horizon-frequency sheet is unchanged. Table~\ref{tab:extremal-validation} verifies the outgoing condition on this sheet and shows that the same frequency is a nonroot on $(0,0)$. Figure~\ref{fig:unconventional-endpoint-distance} resolves the approach to the extremal root, and Fig.~\ref{fig:unconventional-endpoint-extrapolation} compares its extrapolation from finite spin. Increasing the polynomial degree from two to five reduces the difference from $2.56\times10^{-8}$ to $1.11\times10^{-15}$. Together with the stored trajectory, this convergence supports the numerical $\hat n=8$ association marked by a dagger in Table~\ref{tab:extremal-validation}. The endpoint sheet test and the parent association use separate evidence: the former evaluates the extremal boundary conditions, while the latter compares the continued finite-spin sequence with that endpoint.

\begin{figure*}[tb]
\centering
\subfloat[HeunC-Chen method\label{fig:unconventional-endpoint-trajectory}]{\includegraphics[width=.32\textwidth]{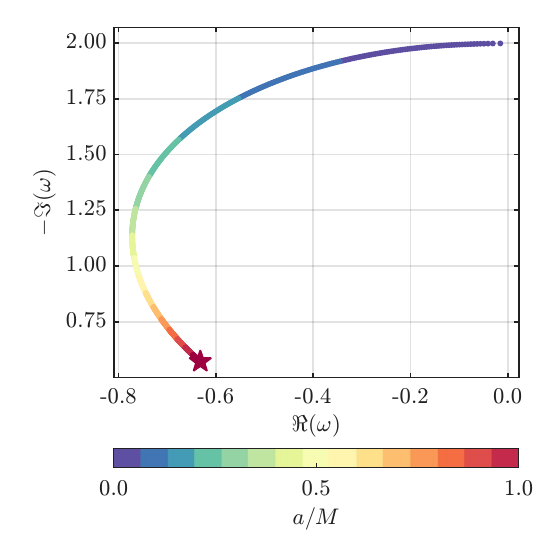}}\hfill
\subfloat[Distance from the HeunD frequency\label{fig:unconventional-endpoint-distance}]{\includegraphics[width=.32\textwidth]{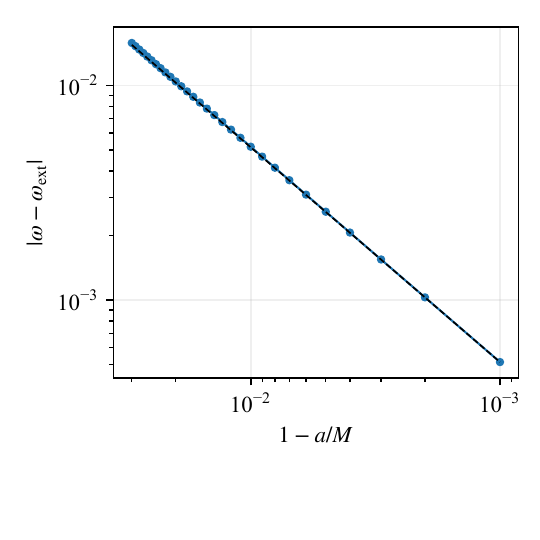}}\hfill
\subfloat[Extrapolation comparison\label{fig:unconventional-endpoint-extrapolation}]{\includegraphics[width=.32\textwidth]{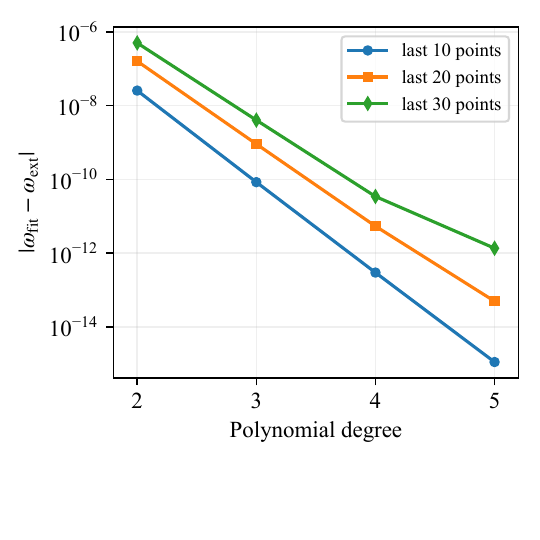}}
\caption{Continuation of the unconventional $(\ell,m)=(2,-2)$ sequence. \protect\subref{fig:unconventional-endpoint-trajectory} Stored HeunC frequencies, colored by $a/M$, and the independently solved extremal HeunD root on an unconventional Riemann sheet (filled star in the final color of the spin scale). \protect\subref{fig:unconventional-endpoint-distance} Distance of the final samples from that root. \protect\subref{fig:unconventional-endpoint-extrapolation} Polynomial extrapolation differences using the final 10, 20, and 30 samples. These differences measure convergence to the independently solved endpoint and support the numerical parent association.}\label{fig:unconventional-endpoint}
\end{figure*}

This QNM is close to, but distinct from, a total-transmission mode. Our earlier study identified the close approach of the $n_\infty=3$ TTM family and the unconventional QNM sequence, with $n_\infty$ denoting the family index used there (Sec.~III B.3 and Fig.~13 of Ref.~\cite{Chen:2026ane}). At extremality, solving the vanishing Teukolsky--Starobinsky constant and angular regularity for $(\ell,m)=(2,-2)$ gives the algebraically special frequency
\begin{qnmvalues}
\begin{equation}
\omega_{\rm AS}^{\rm Kerr}\simeq-0.6318448883309036-0.5716391237271185i,
\end{equation}
\end{qnmvalues}
which differs from Eq.~\eqref{eq:unconventional-frequency} by $5.40837\times10^{-8}$. An independent amplitude evaluation gives $|B^{\mathrm{inc}}/B^{\mathrm{ref}}|=4.40\times10^{-36}$ at the QNM and $2.23\times10^{-3}$ at the AS frequency, with finite nonzero outgoing amplitudes. Thus agreement with the rounded AS frequency reported in Appendix A.3 of Ref.~\cite{Casals:2019vdb} does not identify the two roots. For $(\ell,m)=(3,-3)$, the unconventional Class I QNM sequence with $\hat n=40$ in Fig.~\ref{fig:l3-all-m-continuation-mm3} also passes close to the $n_\infty=3$ TTM sequence at nonextremal spins. At $a/M=0.966$, their frequency separation is $|\omega_{\rm QNM}-\omega_{\rm TTM}|\simeq3.17\times10^{-6}$. At higher sampled spins the two frequencies move apart. As $a/M\to1$, the QNM approaches $\omega=-3/2$ with vanishing damping, whereas the TTM approaches $\omega\simeq-1.57115828-0.84760488i$; their limiting separation is approximately $0.851$.

The negative-real-part filled stars in Figs.~\ref{fig:l3-all-m-continuation-mm1} and~\ref{fig:l3-all-m-continuation-mm2} include additional finite-damping extremal roots on $(-1,0)$,
\begin{qnmvalues}
\begin{equation}\label{eq:additional-negative-heund}
\begin{aligned}
 \omega_{3,-1}&=-2.84861967615503-0.759022659262235i,\\
 \omega_{3,-2}&=-1.90047866015572-1.09119899403845i.
\end{aligned}
\end{equation}
\end{qnmvalues}
The $(3,-2)$ trajectories labeled $\hat n=26$ and $41$ in Fig.~\ref{fig:l3-all-m-continuation-mm2} also retain finite damping. An independent extremal scattering calculation gives
\begin{qnmvalues}
\begin{equation}\label{eq:mminus2-extremal-mst}
\begin{aligned}
 \omega_{\hat n=26}&=-0.330815005095-3.863733721882i,\\
 \omega_{\hat n=41}&=-0.337614470615-5.452506277776i.
\end{aligned}
\end{equation}
\end{qnmvalues}
These Class I endpoints are verified independently and marked by filled stars. The $\hat n=41$ sequence retains its AS-limit anchor at $a=0$. The branch-cut continuations therefore connect unconventional Class I parents to both superradiant accumulation and isolated finite-damping roots; the frequency sheet and the continued parent label remain essential in either case.

\section{Spectral completeness and critical frequencies}\label{sec:spectral-completeness}
The preceding section determines frequencies and their parent associations. We now test whether roots have been missed in specified frequency regions and distinguish isolated QNMs from accumulation near the branch points. We first establish bounded connection counts, then examine the critical frequencies and their physical interpretation.

\subsection{QNM completeness in bounded frequency domains}\label{sec:connection-counting}
\begin{figure*}[tb]
\centering
\subfloat[Frequency domain\label{fig:connection-count-domain}]{\includegraphics[width=.49\textwidth]{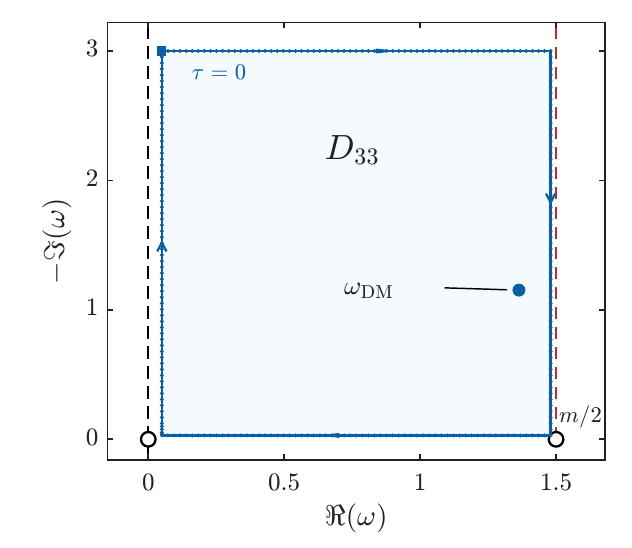}}\hfill
\subfloat[Accumulated connection phase\label{fig:connection-count-phase}]{\includegraphics[width=.49\textwidth]{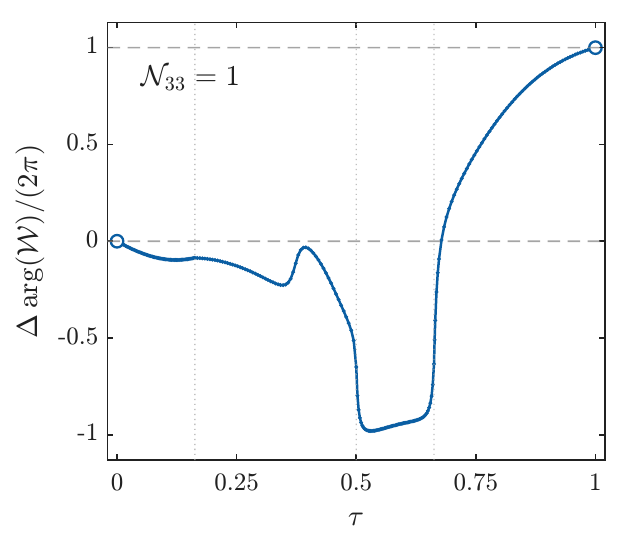}}
\caption{Counting QNMs for $(\ell,m)=(3,3)$ on the conventional Riemann sheet. \protect\subref{fig:connection-count-domain} The domain $D_{33}$ and the known DM (dot); open circles mark the critical frequencies, and dashed lines show the downward frequency cuts. \protect\subref{fig:connection-count-phase} Accumulated phase of $\mathcal W_{\mathfrak b}$ along the positively oriented boundary, starting at $\omega=0.05-3i$, with $\tau$ denoting normalized contour arclength. The net phase change $2\pi$ gives a total count of one QNM, namely the known DM in panel (a), confirming that no additional roots have been missed in $D_{33}$ on the specified angular branch and Riemann sheet. Positive orientation appears clockwise because the vertical axis is $-\Im(\omega)$.}\label{fig:connection-count}
\end{figure*}

To check for missed QNMs, the contour in Fig.~\ref{fig:connection-count-domain} encloses the already known $(\ell,m)=(3,3)$ DM from Fig.~\ref{fig:l3-positive-m-continuation-m3}. The search domain $D_{33}$ lies on the conventional Riemann sheet and is defined by $0.05<\Re(\omega)<1.48$ and $-3<\Im(\omega)<-0.03$. The dot marks the known DM, while the two critical frequencies and their branch cuts lie outside the rectangle.

The number of enclosed modes follows from the phase change of the canonical radial Wronskian $\mathcal W_{\mathfrak b}=\det[\mathbf v_{\mathrm{in}},\mathbf v_{\mathrm{up}}]$, where the two state vectors represent the canonically normalized horizon-ingoing and infinity-outgoing solutions. The label $\mathfrak b$ specifies the continued angular branch and radial frequency logarithms. For a bounded domain $D$ with positively oriented boundary $\partial D$, the argument principle gives
\begin{equation}\label{eq:count}
 N_{\rm zero}-N_{\rm pole}=\frac{1}{2\pi}\Delta_{\partial D}\arg\mathcal W_{\mathfrak b},
\end{equation}
provided $\mathcal W_{\mathfrak b}$ is meromorphic in $D$ and has no zeros or poles on $\partial D$. Here $N_{\rm zero}$ and $N_{\rm pole}$ count its zeros and poles in $D$, including multiplicity, and $\Delta_{\partial D}\arg\mathcal W_{\mathfrak b}$ is its total continuously tracked phase change around the boundary. With the canonical normalization, the connection has no poles in $D_{33}$.

Figure~\ref{fig:connection-count-phase} shows a net phase increase of $2\pi$ after one circuit. The angular branch remains isolated throughout $D_{33}$, and the boundary errors and phase variation between samples are controlled as described in Appendix~\ref{app:holomorphic-radiation}. The resulting QNM count is
\begin{equation}\label{eq:certified-D33-count}
 \mathcal N_{33}(D_{33})=1,
\end{equation}
where $\mathcal N_{33}$ is the total root count on the specified angular branch, including multiplicity. The single counted root is the already known DM marked in Fig.~\ref{fig:connection-count-domain}. It is therefore simple, and no additional QNMs have been missed in $D_{33}$ on this branch and Riemann sheet. For $\ell=m=2$, numerical contour checks around the DM in Fig.~\ref{fig:transition-dm} also give winding one in three nested domains. These supporting tests, including their convergence checks and angular continuation conditions, are collected in Appendix~\ref{app:bounded-count-details}.

The same contour procedure can be applied to other bounded domains on a specified angular branch and Riemann sheet. The present survey extends to Schwarzschild parent labels $\hat n=41$ in the stated angular sectors, including the identified unconventional and AS families. The parent inventory is not uniformly resolved across this entire range: for $(\ell,m)=(2,0)$, the associations of the candidates with $28\le\hat n\le41$ remain unresolved. They are not included as established parent assignments in Fig.~\ref{fig:native-spectrum}. Accordingly, completeness is established here in the certified bounded domain and stated local exclusion neighborhoods; an exhaustive count for every surveyed parent and every continued sheet would require further continuation and counting. Families with $\hat n>41$ are outside the parent survey.

\subsection{Critical frequencies and Riemann sheets}\label{sec:critical}
The positive-$m$ ZDM sequences in Figs.~\ref{fig:transition-l2m1}--\ref{fig:transition-zdm}, \ref{fig:l3-positive-m-continuation}, and~\ref{fig:l3-parent-classification-trajectory} approach $\omega_{\rm SR}=m/2$ with vanishing damping. Figures~\ref{fig:scaling-real}--\ref{fig:scaling-damping} and~\ref{fig:l3-parent-classification-zdm} resolve the near-extremal scaling of representative sequences. The negative-$m$ Class I sequences in Figs.~\ref{fig:l2-all-m-continuation-mm1} and~\ref{fig:l3-all-m-continuation-mm3} have the same zero-damping limit at negative $\omega_{\rm SR}$; the positive- and negative-$m$ accumulation frequencies are marked by black rings in Fig.~\ref{fig:native-spectrum}. For either sign of $m$, $\omega_{\rm SR}=m\Omega_H$ is where the horizon-frame frequency $\omega-m\Omega_H$ vanishes. The negative bounds here describe accumulation of the continued Class I sequences. These plots establish where the nonextremal sequences accumulate as $a\to M$. The physical issue is how this accumulation appears in the spectrum at exactly $a=M$: an isolated QNM is a pole of the radial Green function, whereas an accumulation can give rise to a branch point with a different contribution to the response.

At exact extremality, $\omega_{\rm SR}$ is a branch point of the radial problem; $\omega=0$ is another critical frequency, treated separately below. The generic HeunD map degenerates at these frequencies, so the local analysis must use the limiting radial solutions while retaining the ingoing horizon and outgoing infinity boundary conditions. Their frequency dependence contains logarithms whose values change upon continuation around a branch point. A Riemann sheet records this continuation, and the existence of isolated poles near a critical frequency must therefore be assessed on a specified sheet. The common endpoint of the nonextremal trajectories alone does not identify an additional isolated extremal pole. Appendix~\ref{app:uniform-critical} gives the limiting equations and matching needed to determine this local spectral structure.

For $m>0$ and real positive $\delta_{\rm SR}$, defined in Eq.~\eqref{eq:deltasr}, the leading local matching condition near the superradiant bound is~\cite{Gralla:2017lto,Casals:2018eev}
\begin{equation}\label{eq:critical-sheet-model}
\begin{aligned}
 {\rm e}^{2i\delta_{\rm SR}L_{\mathfrak b}}&=C_{\ell m},\\
 L_{\mathfrak b}&=\operatorname{Log}_{\mathfrak b}[-2i(2M\omega-m)]\\
 &\quad+\operatorname{Log}_{\mathfrak b}(-2iM\omega).
\end{aligned}
\end{equation}
Here $C_{\ell m}$ is the matching coefficient at the critical angular eigenvalue, given in Appendix~\ref{app:critical-angular-isolation}; $\operatorname{Log}_{\mathfrak b}$ denotes the logarithm continued along the chosen frequency path. These logarithms retain the Riemann sheet of the radial solutions. Table~\ref{tab:critical-ray-arguments} gives the limiting arguments of $\Delta\omega=\omega-\omega_{\rm SR}$. For $\ell=m=2,3$, they lie outside the conventional-sheet interval $[-\pi/2,3\pi/2)$ and on the adjacent unconventional sheet reached across the cut from $\omega_{\rm SR}$.

For these two diagonal sectors, with real $\delta_{\rm SR}>0$, uniform matching confirms an infinite sequence of simple poles on one fixed adjacent horizon-frequency sheet, with successive detunings shrinking by $\exp(-\pi/\delta_{\rm SR})$. Their detunings approach zero at a fixed limiting argument, so the sequence does not require successively visiting new sheets. For $\delta_{\rm SR}^{2}\le0$, the geometric-sequence conclusion does not apply; each fixed finite sheet instead has a sufficiently small zero-free punctured critical neighborhood. On the conventional sheet, the exact extremal problem has no isolated QNMs in the punctured disks
\begin{equation}\label{eq:critical-exclusion-domain}
 0<|\omega-\omega_{\rm SR}|\le\frac{\rho}{2M},
\end{equation}
with $\rho=0.018$ for $(2,2)$ and $0.040$ for $(3,3)$; the bounds are derived in Appendix~\ref{app:explicit-conventional-sheet-collar}. The finite-damping endpoints in Figs.~\ref{fig:transition-dm} and~\ref{fig:l3-positive-m-continuation-m3} lie outside these critical disks. Local exclusion near the superradiant bound is therefore compatible with the DMs shown in those panels. The remaining corotating sectors and the zero-frequency limit are treated in Appendices~\ref{app:negative-critical} and~\ref{app:origin-collar}.

\begin{table}[!htbp]
\centering\small
\caption{Critical exponents, limiting rays, and conventional-sheet exclusion radii. The exponents use the regular angular eigenvalue at $a\omega=m/2$ in Eq.~\eqref{eq:deltasr} and are rounded for display. Continued arguments are in radians; $\rho$ defines the disk in Eq.~\eqref{eq:critical-exclusion-domain}. A dash denotes an inapplicable limiting ray; n.q. denotes an exclusion radius not quantified here.}\label{tab:critical-ray-arguments}
\setlength{\tabcolsep}{4pt}
\begin{tabular}{crrr}
\toprule
$(\ell,m)$ & $\delta_{\rm SR}^{2}$ & $\arg_{\mathfrak b}\Delta\omega$ & $\rho$\\
\midrule
$(2,1)$ & $-3.682831914$ & --- & $\mathrm{n.q.}$\\
$(2,2)$ & $4.206309624$ & $6.072088$ & $0.018$\\
$(3,1)$ & $-10.080135863$ & --- & $\mathrm{n.q.}$\\
$(3,2)$ & $-3.501190916$ & --- & $\mathrm{n.q.}$\\
$(3,3)$ & $7.804277648$ & $6.239925$ & $0.040$\\
\bottomrule
\end{tabular}
\end{table}

\subsection{Physical interpretation and overtone counting}\label{sec:physics}
Figures~\ref{fig:transition-zdm}, \ref{fig:transition-dm}, and~\ref{fig:l3-positive-m-continuation-m3} show two distinct limits as $a\to M$. The ZDM sequences approach the real superradiant frequency and their damping times grow without bound, while the DMs approach complex frequencies with finite damping times. Both behaviors occur for $\ell=m=2$ and $3$: the accumulation of ZDMs does not exhaust the extremal spectrum. Near-horizon symmetry constrains the critical radial scaling~\cite{Bardeen:1999px}, but the usual near-horizon DM criterion does not predict the DMs found here~\cite{Casals:2019vdb}. Their frequencies follow from the full boundary-value problem, with an ingoing radial solution at the horizon, an outgoing solution at infinity, and a regular angular solution, as in Sec.~\ref{sec:method} and Appendix~\ref{sec:accessory}.

The overtone label $\hat n$ identifies the Schwarzschild mode from which a Kerr trajectory starts~\cite{Berti:2004md}. Following that trajectory determines whether the mode becomes a DM or belongs to a ZDM sequence. In Fig.~\ref{fig:l3-parent-classification-trajectory}, the $(3,1)$ mode with $\hat n=6$ remains a DM, while the neighboring parents with $4\le\hat n\le7$, $\hat n\ne6$, belong to ZDM sequences. Likewise, the $(3,3)$ DM in Fig.~\ref{fig:l3-positive-m-continuation-m3} has parent $\hat n=34$, while its immediate neighbors belong to ZDM sequences. Thus no single overtone threshold separates DMs from ZDMs. Figure~\ref{fig:connection-count-domain} checks a different aspect: the chosen rectangle contains exactly one root, the identified $(3,3)$ DM, on the specified branch and sheet. This count determines the number of roots in that domain; the trajectory determines the Schwarzschild parent.

To relate these modes to ringdown, the radial Green function describes the response to a perturbing source. For a conventional-sheet QNM with a simple pole, excitation by the source produces an exponentially damped oscillation, with frequency and decay rate set by the real and imaginary parts of the QNM frequency. For the $(2,2)$ DM in Fig.~\ref{fig:transition-dm}, our numerical checks support a simple pole: the connection function has a nonzero derivative at its zero, while the Green function numerator remains nonzero. The accumulating ZDMs can produce transient power-law ringdown near extremality~\cite{Yang:2013uba}; at exact extremality, the superradiant branch cut contributes to the late-time response~\cite{Casals:2018eev}. For unconventional-sheet poles, including the mode in Fig.~\ref{fig:unconventional-endpoint-trajectory}, a waveform contribution additionally requires the pole residue, source coupling, and the appropriate continuation of the Green function. The present calculation determines their frequencies and sheets; residues and source excitation are outside its scope.

\section{Conclusions}
We have extended our confluent Heun construction of Type-D QNM spectra to extremal Kerr black holes. The HeunD radial equation treats the merged horizon directly, and matching the horizon-ingoing and infinity-outgoing solutions together with angular regularity determines the frequencies at $a=M$. Retaining the frequency logarithms under analytic continuation resolves QNMs on unconventional Riemann sheets as well as on the conventional sheet. Independent scattering-amplitude and endpoint calculations verify representative roots and their sheet assignments.

The gravitational results for $\ell=2,3$ connect this extremal spectrum to the continued Schwarzschild mode families. In particular, the unconventional Class I sequences $(\ell,m,\hat n)=(2,-1,8)$ and $(3,-3,40)$ approach $\omega=m/2$ with vanishing damping. Their Class I identities are retained along continuation, distinguishing them from the Class II partners obtained by reflecting the usual positive-$m$ ZDMs through Kerr symmetry. Their limiting real frequencies are critical accumulation points of the nonextremal spectrum. The finite-damping unconventional $(2,-2)$ root has a numerical $\hat n=8$ association supported by its finite-spin trajectory and endpoint convergence, complementing the two zero-damping sequences. Its extremal sheet is independently verified, and it remains distinct from a nearby algebraically special TTM. For $(3,-3)$, the QNM and TTM also pass close at intermediate spin but separate again toward extremality.

The parent associations reveal a nonmonotonic DM/ZDM ordering. The known $\ell=m=2$ and $3$ damped frequencies are reproduced, while continuation identifies their Schwarzschild parents as $\hat n=5$ and $34$, respectively. The $(3,1)$ parent $\hat n=6$ provides another finite-damping endpoint between zero-damping parents. These examples show that neither instantaneous damping order nor a single threshold in the Schwarzschild overtone determines which modes retain finite damping. Figures~\ref{fig:transition}, \ref{fig:l3-positive-m-continuation}, and~\ref{fig:l3-parent-classification} display these connections.

The connection count supplies a complementary test of spectral coverage: the stated $(3,3)$ rectangle contains exactly one simple QNM on the specified angular branch and conventional Riemann sheet. For $\ell=m=2,3$, real positive critical exponents yield an infinite pole sequence on a fixed adjacent horizon sheet, together with explicit zero-free neighborhoods on the conventional sheet. These local results distinguish isolated extremal modes from critical accumulation. The parent survey extends to $\hat n=41$ to include the identified cut-crossing and AS families in the low-multipole sectors. Completeness is established in the stated bounded domain and local exclusion neighborhoods on their specified angular branches and Riemann sheets; extending those counts and resolving the remaining parent associations would test the coverage of the broader truncated survey.

\section*{Data availability}\label{sec:data-availability}
The numerical results reported here will be included in an update of our QNM dataset accompanying Ref.~\cite{Chen:2025sbz}.

\begin{acknowledgments}
This work was supported by the National Key Research and Development Program of China (No.~2021YFC2203001) and the National Natural Science Foundation of China (No.~12475049).
\end{acknowledgments}

\appendix
\section{Separated solutions and connection formulas}\label{app:separated-derivations}
This appendix explains how the separated solutions implement the boundary conditions used in the main text. We first match the angular solutions regular at the two poles, then construct radial solutions with the required endpoint behavior. An alternative radial representation supplies scattering-amplitude checks, and continuation of the coupled angular and radial conditions connects the frequencies to their Schwarzschild parents.

\subsection{Angular regularity condition}\label{app:angular-regularity}
The HeunC solutions regular at the north and south poles are written in the coordinates $ {}_{\theta}x_{\pm1}=(1\mp\Theta)/2$, with parameters~\cite{Chen:2025sbz,Chen:2026ane}
\begin{subequations}\label{eq:angular-heunc-parameters}
\begin{align}
 \alpha_{\pm1}&=\mp4a\omega,\qquad
 \beta_{\pm1}=|m\pm s|,\\
 \gamma_{\pm1}&=|m\mp s|,\qquad
 \delta_{\pm1}=\mp4sa\omega,\\
 \eta_{\pm1}&=\frac{m^2-s^2}{2}-s-2ma\omega
 \pm2sa\omega-\lambda.
\end{align}
\end{subequations}
Let $\HC_{\pm1}(z)$ denote the corresponding normalized HeunC functions. The regular solutions are
\begin{align}\label{eq:angular-pole-regular}
 S_{\ell m}^{(\pm1)}({}_{\theta}x_{\pm1})={}&
 {\rm e}^{\alpha_{\pm1}{}_{\theta}x_{\pm1}/2}
 {}_{\theta}x_{\pm1}^{\beta_{\pm1}/2}
 (1-{}_{\theta}x_{\pm1})^{\gamma_{\pm1}/2}
\nonumber\\[-2pt]
 &\times\HC_{\pm1}({}_{\theta}x_{\pm1}).
\end{align}
The superscript $(\pm1)$ labels the pole at which the local solution is regular. The normalization $\HC_{\pm1}(0)=1$ selects the regular Frobenius branch; a logarithmic solution is excluded when a pole exponent vanishes. Each coordinate vanishes at its associated pole. A globally regular angular solution exists when the two local solutions are linearly dependent in their common domain, so their Wronskian $W_\theta$ must vanish. Abel's identity shows that $(1-\Theta^2)W_\theta$ is constant, and the condition can be evaluated at the equator, where both HeunC arguments equal $1/2$. The two prefactors in Eq.~\eqref{eq:angular-pole-regular} differ only by a constant as functions of $\Theta$, so their derivative contributions cancel in the Wronskian. The opposite signs of ${\rm d}{}_{\theta}x_{\pm1}/{\rm d}\Theta=\mp1/2$ then give
\begin{align}\label{eq:angular-heunc-wronskian}
 0={}&\HC_{-1}(\tfrac12)\HC_{+1}'(\tfrac12)
\nonumber\\
 &+\HC_{+1}(\tfrac12)\HC_{-1}'(\tfrac12),
\end{align}
where primes denote derivatives with respect to the HeunC argument. This condition determines $\lambda$ in the coupled frequency solve. The label $\ell$ follows the eigenvalue continued from $ {}_sA_{\ell m}(0)=\ell(\ell+1)-s(s+1)$. At extremality, $a=M$.

\subsection{HeunD transformations and normalization}\label{app:map}
For the symmetric equation~\eqref{eq:mapled}, the coordinate transformations act as
\begin{equation}
\begin{aligned}
 x\mapsto-x &: (\alpha,\beta,\gamma,\delta)\mapsto(-\alpha,\beta,-\gamma,\delta),\\
 x\mapsto x^{-1} &: (\alpha,\beta,\gamma,\delta)\mapsto(-\alpha,-\delta,-\gamma,-\beta).
\end{aligned}
\end{equation}
The HeunD function used here is normalized by $\mathcal H(0)=1$ and $\mathcal H'(0)=0$. Uniqueness then gives
\begin{align}\label{eq:heund-normalized-reflection}
 &\HD(\alpha,\beta,\gamma,\delta,-x)
\nonumber\\
 &\qquad=\HD(-\alpha,\beta,-\gamma,\delta,x),
\end{align}
with both sides continued along the transformed path. Inversion instead exchanges the ordinary points $0$ and infinity, so it relates solution spaces with transported initial data rather than independently normalized functions at $x=0$. In the Kerr map, it is induced by $\varpi\mapsto-\varpi$ and preserves $\omega$, $\mathcal V$, and $\lambda$. The same physical radial problem is recovered when the boundary factors, normalizations, and frequency sheet are transported with the solution. A change of coordinate alone does not select the required boundary conditions.

\subsection{Connection coefficients and radial solutions}\label{app:endpoint-basis}
For the ordinary-point basis $R_j$ in Eq.~\eqref{eq:heund-local-solutions}, define the endpoint connection coefficients by
\begin{equation}\label{eq:heund-asymptotic}
\begin{aligned}
 R_j&\sim D_{1j}^{M}R_{\mathrm{in}}^M+D_{2j}^{M}R_{\mathrm{up}}^M, &&r\to M,\\
 R_j&\sim D_{1j}^{\infty}R_{\mathrm{up}}^\infty+D_{2j}^{\infty}R_{\mathrm{in}}^\infty, &&r\to\infty.
\end{aligned}
\end{equation}
The coefficients $D_{\sigma j}^{e}$ multiply the full radial waves after the factor $\mathcal S_r$ has been restored. Here the wave factors are those of Eq.~\eqref{eq:physicalbc}, and $\sigma,j=1,2$. Thus the first row of each connection matrix contains the desired horizon-ingoing or infinity-outgoing amplitudes, and the second row contains the opposite waves. The expansions hold in the selected Stokes sectors with the continued logarithms of the physical problem. To obtain an endpoint-normalized solution, take $R=c_1R_1+c_2R_2$ and impose unit amplitude for the desired wave and zero amplitude for the unwanted wave. At either endpoint $e=M,\infty$, this gives
\begin{equation}
\begin{aligned}
 D_{11}^{e}c_1+D_{12}^{e}c_2&=1,\\
 D_{21}^{e}c_1+D_{22}^{e}c_2&=0.
\end{aligned}
\end{equation}
Put $\Delta_e=D_{11}^{e}D_{22}^{e}-D_{12}^{e}D_{21}^{e}$. At noncritical frequencies the basis is independent, and solving these two equations gives
\begin{subequations}
\begin{align}\label{eq:rin-construction}
 R_{\ell m}^{\mathrm{in}}&=\frac{D_{22}^{M}R_1-D_{21}^{M}R_2}{\Delta_M},\\
 R_{\ell m}^{\mathrm{up}}&=\frac{D_{22}^{\infty}R_1-D_{21}^{\infty}R_2}{\Delta_\infty}.\label{eq:rup-construction}
\end{align}
\end{subequations}
The scattering amplitudes therefore satisfy
\begin{equation}\label{eq:heund-infinity-amplitude-map}
 \begin{aligned}
 B^{\mathrm{ref}}_{\ell m}
 &=\frac{D_{22}^{M}D_{11}^{\infty}-D_{21}^{M}D_{12}^{\infty}}
 {D_{11}^{M}D_{22}^{M}-D_{12}^{M}D_{21}^{M}},\\
 B^{\mathrm{inc}}_{\ell m}
 &=\frac{D_{22}^{M}D_{21}^{\infty}-D_{21}^{M}D_{22}^{\infty}}
 {D_{11}^{M}D_{22}^{M}-D_{12}^{M}D_{21}^{M}},
 \end{aligned}
\end{equation}
and
\begin{equation}\label{eq:heund-horizon-amplitude-map}
 \begin{aligned}
 C^{\mathrm{ref}}_{\ell m}
 &=\frac{D_{22}^{\infty}D_{11}^{M}-D_{21}^{\infty}D_{12}^{M}}
 {D_{11}^{\infty}D_{22}^{\infty}-D_{12}^{\infty}D_{21}^{\infty}},\\
 C^{\mathrm{up}}_{\ell m}
 &=\frac{D_{22}^{\infty}D_{21}^{M}-D_{21}^{\infty}D_{22}^{M}}
 {D_{11}^{\infty}D_{22}^{\infty}-D_{12}^{\infty}D_{21}^{\infty}}.
 \end{aligned}
\end{equation}
The first pair is obtained by substituting the horizon-normalized solution into its infinity expansion; the second follows by substituting the infinity-normalized solution into its horizon expansion. In particular, $B^{\mathrm{inc}}=0$ removes the incoming wave at infinity from the horizon-ingoing solution, while $C^{\mathrm{up}}=0$ removes the outgoing wave at the horizon from the infinity-outgoing solution. The numerical solver propagates the endpoint solutions directly; these formulas express the same boundary conditions in the HeunD basis.

\subsection{Radial accessory parameter representation}\label{sec:accessory}
The change of variables $\mathcal Y_{\ell m}=(r-M)^{s+1/2}R_{\ell m}$ and
\begin{subequations}\label{eq:log-coordinate}
\begin{align}
 \chi&=\log\frac{r-M}{\varpi}=\log\frac{1+x}{1-x},\label{eq:chi-coordinate}\\
 \mathcal Y_{\ell m}(\chi)&={\rm e}^{-\frac{\alpha}{4}\sinh\chi}
 \mathcal H\!\left(\tanh\frac{\chi}{2}\right),\label{eq:accessory-gauge}
\end{align}
\end{subequations}
puts the radial equation in the form
\begin{align}\label{eq:accessory-operator}
 -\frac{{\rm d}^2\mathcal Y_{\ell m}}{{\rm d}\chi^2}&+\left(\frac{\alpha^2}{16}\cosh^2\chi
 +\frac{\gamma}{8}\sinh\chi\right)\mathcal Y_{\ell m}
\nonumber\\
 &+\frac{s\alpha}{2}\cosh\chi\,\mathcal Y_{\ell m}=\mathcal V \mathcal Y_{\ell m},
\end{align}
with $\mathcal V=(\beta-\delta)/8$ and $\beta+\delta=4s\alpha$. The endpoint powers are
\begin{subequations}\label{eq:powers}
\begin{align}
 p_+&=\frac{\phantom{-}\beta+\gamma+\delta+2\alpha}{4\alpha}
      =\phantom{-}s+\frac{1}{2}-2iM\omega,\\
 p_-&=\frac{-\beta+\gamma-\delta+2\alpha}{4\alpha}
      =-s+\frac{1}{2}-2iM\omega,
\end{align}
\end{subequations}
and the corresponding boundary conditions are
\begin{subequations}\label{eq:accessory-boundary}
\begin{align}
 \mathcal Y_{\mathrm{in}}^M&\sim\varpi^{p_-}{\rm e}^{\frac{\alpha}{8}{\rm e}^{-\chi}+p_-\chi},
 &\operatorname{Re}\chi&\to-\infty,\label{eq:accessory-in}\\
 \mathcal Y_{\mathrm{up}}^\infty&\sim\varpi^{-p_+}{\rm e}^{-\frac{\alpha}{8}{\rm e}^{\chi}-p_+\chi},
 &\operatorname{Re}\chi&\to+\infty.\label{eq:accessory-up}
\end{align}
\end{subequations}
The logarithm and contour follow the selected Stokes sectors. For fixed nonzero $\alpha$, fixed $\gamma$, and sheet $\mathfrak b$, the radial condition is the zero of
\begin{equation}\label{eq:accessory-determinant}
 \mathcal D_{\mathfrak b}(\alpha,\gamma;\mathcal V)
 =W_\chi[\mathcal Y_{\mathrm{in}}^M,\mathcal Y_{\mathrm{up}}^\infty].
\end{equation}
With canonical endpoint normalizations this equals $\mathcal W_{\mathfrak b}$ in Eq.~\eqref{eq:Evans}. Angular regularity supplies the spectral intersection
\begin{align}\label{eq:accessory-intersection}
 \mathcal V(\alpha(\omega),\gamma(\omega);\mathfrak b)
 ={}&11M^2\omega^2-2mM\omega
\nonumber\\
 &-{}_s A_{\ell m}(M\omega)-(s+\frac{1}{2})^2.
\end{align}
The logarithmic coordinate sends the horizon and infinity to opposite ends of the $\chi$ plane. For fixed $\alpha$, $\gamma$, and sheet, the Wronskian condition selects accessory values $\mathcal V$ for which one solution obeys both endpoint conditions. The Kerr frequency must also satisfy the angular relation in Eq.~\eqref{eq:accessory-intersection}; an accessory zero by itself is therefore not yet a coupled QNM frequency. This is a global connection condition in the complex radial plane. Its accessory values do not assign Schwarzschild parent labels by a local ordering.

\subsection{Series and amplitudes in HeunD parameters}\label{app:mst}
Scattering amplitudes are evaluated from the MST radial series expressed in HeunD parameters~\cite{Mano:1996vt,Casals:2019vdb}.
For $\alpha\gamma\ne 0$, set $q_n=n+\nu+1$, where $n\in\mathbb Z$ and the renormalized angular momentum $\nu$ is fixed by the condition for a bilateral minimal sequence. The recurrence is written only in the Maple HeunD parameters:
\begin{equation}\label{eq:mst-maple-direct-recurrence}
\begin{aligned}
 \mathsf a_n&=\frac{\alpha^2}{32}
 \frac{\left(q_n+\frac{\beta+\delta+\gamma}{4\alpha}\right)
       \left(q_n-\frac{\gamma-\beta-\delta}{4\alpha}\right)}{q_n(2q_n+1)},\\
 \mathsf b_n&=(q_n-1)q_n-\left(\frac{\alpha^2}{32}-\frac{\beta-\delta}{8}-\frac14\right)\\
 &\quad-\frac{\frac{\alpha^2}{32}
 \frac{\beta+\delta+\gamma}{4\alpha}
 \frac{\gamma-\beta-\delta}{4\alpha}}{(q_n-1)q_n},\\
 \mathsf c_n&=\frac{\alpha^2}{32}
 \frac{\left(q_n-1-\frac{\beta+\delta+\gamma}{4\alpha}\right)
       \left(q_n-1+\frac{\gamma-\beta-\delta}{4\alpha}\right)}{(q_n-1)(2q_n-3)} .
\end{aligned}
\end{equation}
The radial-series coefficients $a_n^\nu$ satisfy
\begin{equation}\label{eq:mst-recurrence}
 \mathsf a_n a_{n+1}^\nu+\mathsf b_n a_n^\nu
 +\mathsf c_n a_{n-1}^\nu=0.
\end{equation}
The renormalized angular momentum $\nu$ selects the common minimal sequence in both directions: the solution minimal as $n\to+\infty$ and that minimal as $n\to-\infty$ must satisfy the same recurrence at a common index. In practice their coefficient ratios can be obtained from continued fractions and matched there; $a_0^\nu=1$ fixes the remaining overall scale. The HeunD parameters and the chosen Riemann sheet are held fixed when this auxiliary condition is solved; the frequency equation remains the HeunD connection condition.

Put $\rho_r=(r-M)/(2M)$. For the three bilateral series used to construct the radial solutions, define
\begin{equation}
 g_n=\frac{\Gamma\!\left(q_n+\frac{\beta+\delta+\gamma}{4\alpha}\right)}
 {\Gamma\!\left(q_n-\frac{\beta+\delta+\gamma}{4\alpha}\right)},\qquad a_0^\nu=1.
\end{equation}
The two-sided series for the three required radial solutions are
\begin{subequations}\label{eq:mst-series}
\begin{align}
 F_{\mathrm{in}}^M={}&\rho_r^{-\frac{\beta+\delta}{4\alpha}-\nu-1}
 {\rm e}^{-\frac{\gamma}{4\alpha}\rho_r+\frac{\alpha^3}{16\gamma\rho_r}}
\nonumber\\
 &\times\sum_{n=-\infty}^\infty g_n a_n^\nu
 \left(-\frac{\alpha^3}{8\gamma\rho_r}\right)^n
\nonumber\\
 &\hspace{3mm}\times U\!\left(q_n+\frac{\gamma-\beta-\delta}{4\alpha},2q_n,
 -\frac{\alpha^3}{8\gamma\rho_r}\right),\label{eq:mst-in-series}\\
 F_{\mathrm{up}}^\infty={}&\rho_r^{-\frac{\beta+\delta}{4\alpha}+\nu}
 {\rm e}^{-\frac{\gamma}{4\alpha}\rho_r+\frac{\alpha^3}{16\gamma\rho_r}}
\nonumber\\
 &\times\sum_{n=-\infty}^\infty g_n a_n^\nu
 \left(\frac{\gamma\rho_r}{2\alpha}\right)^n
\nonumber\\
 &\hspace{3mm}\times U\!\left(q_n+\frac{\beta+\delta+\gamma}{4\alpha},2q_n,
 \frac{\gamma\rho_r}{2\alpha}\right),\label{eq:mst-up-series}\\
 F_{\mathrm{in}}^\infty={}&\rho_r^{-\frac{\beta+\delta}{4\alpha}+\nu}
 {\rm e}^{\frac{\gamma}{4\alpha}\rho_r+\frac{\alpha^3}{16\gamma\rho_r}}
\nonumber\\
 &\times\sum_{n=-\infty}^\infty a_n^\nu
 \left(\frac{\gamma\rho_r}{2\alpha}\right)^n
\nonumber\\
 &\hspace{3mm}\times U\!\left(q_n-\frac{\beta+\delta+\gamma}{4\alpha},2q_n,
 -\frac{\gamma\rho_r}{2\alpha}\right).\label{eq:mst-inc-series}
\end{align}
\end{subequations}
Here $U(a,b,u)$ is Tricomi's confluent hypergeometric function, distinct from the radial function $\mathcal Y$ in Appendix~\ref{sec:accessory}. The first series supplies a horizon-ingoing solution, and the other two supply outgoing and incoming solutions at infinity. They solve the same radial equation, but their leading amplitudes have not yet been normalized to unity.
The horizon-ingoing and infinity-outgoing solutions with normalized leading amplitudes follow by dividing out the leading coefficients of these series:
\begin{subequations}\label{eq:mst-normalized-bases}
\begin{align}
 R_{\ell m}^{\mathrm{in}}
 &= (2M)^{-\frac{\beta+\delta}{2\alpha}+\frac{\gamma}{4\alpha}}
 \frac{\left(-\frac{\alpha^3}{8\gamma}\right)^{\nu+1+\frac{\gamma-\beta-\delta}{4\alpha}}
 F_{\mathrm{in}}^M}{\sum_n g_n a_n^\nu},\\
 R_{\ell m}^{\mathrm{up}}
 &= (2M)^{-\frac{\beta+\delta}{2\alpha}-1-\frac{\gamma}{4\alpha}}
 \frac{\left(\frac{\gamma}{2\alpha}\right)^{\nu+1+\frac{\beta+\delta+\gamma}{4\alpha}}
 F_{\mathrm{up}}^\infty}{\sum_n g_n a_n^\nu}.
\end{align}
\end{subequations}
All powers and hypergeometric functions follow the same frequency and radial continuation paths. A zero of an explicit series denominator requires a regularized basis before an amplitude ratio or pole count can be interpreted.

At an ordinary matching radius, the infinity amplitudes are
\begin{subequations}\label{eq:mst-wronskian-amplitudes}
\begin{align}
 B^{\mathrm{inc}}_{\ell m}
 &=-\frac{4\alpha M}{\gamma}(r-M)^{2\left(1+\frac{\beta+\delta}{4\alpha}\right)}
 W_r[R_{\ell m}^{\mathrm{in}},R_{\ell m}^{\mathrm{up}}],\\
 B^{\mathrm{ref}}_{\ell m}
 &=-\frac{4\alpha M}{\gamma}(r-M)^{2\left(1+\frac{\beta+\delta}{4\alpha}\right)}
 (2M)^{-1+\frac{\gamma}{4\alpha}}
\nonumber\\
 &\quad\times\frac{\left(-\frac{\gamma}{2\alpha}\right)^{\nu+1-\frac{\beta+\delta+\gamma}{4\alpha}}}
 {\sum_n(-1)^n a_n^\nu}
 W_r[F_{\mathrm{in}}^\infty,R_{\ell m}^{\mathrm{in}}].
\end{align}
\end{subequations}
The second prefactor removes the leading incoming amplitude of $F_{\mathrm{in}}^\infty$. To evaluate the Wronskians, the series may be differentiated term by term using ${\rm d}U(a,b,u)/{\rm d}u=-aU(a+1,b+1,u)$ together with the chain rule for $u(r)$. Both amplitudes are independent of matching radius wherever the continued solutions are regular.

The frequency and radial logarithms fix the continued branches of all special functions. Critical frequencies require the limiting equations of Sec.~\ref{sec:critical} rather than a direct specialization of the noncritical formulas.

\subsection{Direct series continuation}\label{app:fast-connection}
The transfer method of Sec.~\ref{sec:numerical-protocol} propagates an endpoint state using a local two-solution HeunD basis. A complementary implementation instead uses Taylor expansions of the radial equation to propagate the endpoint-selected states. Near the horizon it takes $z=r-M$ and expands the normal-form solution $\Psi_{\ell m}$; at infinity it uses the inverted coordinate $z=(r-M)^{-1}$ and expands the corresponding transformed solution $y(z)=z\Psi_{\ell m}(M+z^{-1})$. At each ordinary point $z_0$, the expansion variable is $\tau=(z-z_0)/z_0$. The radial equation supplies the coefficients, and the series gives the solution and its derivative at the next point. After propagation, the inverted-coordinate state is converted to the common physical state $(\Psi_{\ell m},{\rm d}\Psi_{\ell m}/{\rm d}r)$ at the matching radius. Rescaling either state does not change the zero of their Wronskian. Both calculations impose the same physical endpoint conditions and track the same Riemann sheet, so they test the same connection-zero condition; the endpoint scales, spatial paths, and matching radius may be varied as convergence checks.

When neither solution vanishes at the matching point, one may also divide the Wronskian condition by their product and use the equivalent local condition
\begin{equation}
 \left.\frac{(\Psi_{\ell m}^{\mathrm{up}})'}{\Psi_{\ell m}^{\mathrm{up}}}\right|_{r^\star}
 -\left.\frac{(\Psi_{\ell m}^{\mathrm{in}})'}{\Psi_{\ell m}^{\mathrm{in}}}\right|_{r^\star}=0.
\end{equation}
Primes here denote radial derivatives. This logarithmic-derivative form is used only where both denominators are nonzero; the Wronskian remains the appropriate condition when either vanishes. The Taylor and HeunD calculations share the endpoint series and angular condition but use different radial propagation representations. Their agreement therefore checks the propagation and convergence of the connection calculation; it is not a fully independent verification of the shared ingredients.

\subsection{Parent continuation and endpoint checks}\label{app:continuation-details}
A Schwarzschild parent is retained by solving the radial condition $\mathcal W_{\mathfrak b}(\omega,\lambda;a)=0$ together with angular regularity $W_\theta(\omega,\lambda;a)=0$. Differentiating the two conditions along a simple root gives
\begin{subequations}\label{eq:parent-transport}
\begin{equation}\label{eq:parent-transport-derivative}
 \frac{{\rm d}\omega_{\hat n}}{{\rm d}a}=
 -\frac{\mathcal W_{\mathfrak b,a}W_{\theta,\lambda}
             -\mathcal W_{\mathfrak b,\lambda}W_{\theta,a}}
           {\mathcal W_{\mathfrak b,\omega}W_{\theta,\lambda}
             -\mathcal W_{\mathfrak b,\lambda}W_{\theta,\omega}},
\end{equation}
\begin{equation}\label{eq:parent-transport-initial}
 \omega_{\hat n}(0)=\omega_{\hat n}^{\rm Sch},
\end{equation}
\end{subequations}
where comma subscripts denote partial derivatives at fixed values of the other arguments. The formula follows by differentiating both equations with respect to $a$ and eliminating ${\rm d}\lambda/{\rm d}a$. The joint Jacobian in the denominator must be nonzero; a multiple-root cluster requires simultaneous continuation. One can use the derivative to predict the frequency at a nearby spin and then solve both conditions there to correct the frequency and separation constant. The angular branch and frequency logarithms are transported along the path, including cut crossings. The initial value $\omega_{\hat n}^{\rm Sch}$ fixes the parent, while an independently solved extremal connection zero fixes the endpoint.

For the $(2,-2)$ branches in Fig.~\ref{fig:parent-identification}, separately solved HeunC roots at $a=0.999999$ anchor the backward continuations to Table~\ref{tab:cook-l2-parent-candidates}. Reversing the continuation and halving the spin steps leave the parent identifications unchanged. The added $\hat n=9,10$ branches and the high-spin segments in Fig.~\ref{fig:transition-l2m1} are obtained from the coupled scattering and angular conditions, with representative frequencies independently reproduced using HeunC.

The $(3,3)$ association requires resolving the nearby $\hat n=33$ and $34$ branches. At $a/M=0.9972$ their separately continued frequencies are
\begin{subequations}\label{eq:l3m3-parent-separation}
\begin{qnmvalues}
\begin{align}
 \omega_{\hat n=33}&\simeq1.376061009349-1.143566555170i,\\
 \omega_{\hat n=34}&\simeq1.366381519327-1.157098835593i.
\end{align}
\end{qnmvalues}
\end{subequations}
Paired forward and backward continuations remain separated and return to independently solved moderate-spin anchors. The $\hat n=34$ branch reaches \mbox{$\omega=1.3634408437-1.1524614727i$} at $a/M=0.9999$; extrapolations approach the independent extremal DM in Table~\ref{tab:extremal-validation} within $2.5\times10^{-11}$. At $a/M=0.99999999999$, the neighboring $\hat n=33,35$ branches instead have
\begin{subequations}\label{eq:l3-high-parent-zdm-rates}
\begin{qnmvalues}
\begin{align}
 \frac{\Delta\omega_{33}}{\kappa}&\simeq-2.7936350-33.4999516i,\\
 \frac{\Delta\omega_{35}}{\kappa}&\simeq-2.7936359-34.4999469i,
\end{align}
\end{qnmvalues}
\end{subequations}
where $\Delta\omega=\omega-m/2$. These rates agree with the leading ZDM values $-\delta_{\rm SR}-i(j+1/2)$ for $j=33,34$, with $\delta_{\rm SR}=2.7936137$. Thus the finite-damping parent $\hat n=34$ lies between two zero-damping parents, as stated in Fig.~\ref{fig:l3-positive-m-continuation-m3}.

The unconventional extremal roots are solved on their stated frequency sheets. Selected roots are checked using independent scattering amplitudes; the additional roots in Eq.~\eqref{eq:additional-negative-heund} are checked by varying precision, endpoint scales, and the matching point in the Taylor continuation. The $(3,-2)$ endpoints in Eq.~\eqref{eq:mminus2-extremal-mst} use the extremal recurrence and independent bilateral $U$-series evaluations; their high-spin samples give $|B^{\mathrm{inc}}/B^{\mathrm{ref}}|<10^{-14}$. Angular eigenvector continuation retains the original branch, including the $\hat n=41$ AS anchor at $a=0$. For the unconventional $(2,-2)$ root, the dagger in Table~\ref{tab:extremal-validation} identifies a numerical $\hat n=8$ association based on the stored trajectory and convergence to the independently solved endpoint in Fig.~\ref{fig:unconventional-endpoint}. The independent sheet check applies at $a=M$; endpoint convergence alone is not a certificate of continuous sheet transport between the stored finite-spin samples.

\section{Conditions for the bounded QNM count}\label{app:holomorphic-radiation}
\subsection{Canonical radial solutions}\label{app:sectorial-states}
The count in Fig.~\ref{fig:connection-count} uses a radial connection whose zeros represent QNMs without additional normalization zeros or poles. At an ordinary matching point, the state $\mathbf v=(\Psi,\partial_r\Psi)^{\mathsf T}$ of a nontrivial solution cannot vanish. A common meromorphic factor in its two components must therefore be removed before forming $\mathcal W_{\mathfrak b}=\det[\mathbf v_{\rm in},\mathbf v_{\rm up}]$. Such factors cancel from the radial Green function and must not be counted as modes~\cite{Kubota:2026hdv,Motohashi:2026mbn}.

The determinant vanishes when the ingoing horizon solution and outgoing infinity solution are linearly dependent; the resulting radial solution obeys both QNM boundary conditions. To count these zeros by the argument principle, one must also show that the chosen angular eigenvalue and the frequency logarithms return to their initial values along the contour, that the connection has no normalization poles in the domain, and that it stays nonzero on the whole boundary. The following construction addresses these conditions in turn. The explicit continuous certificate is for the domain $D_{33}$ and the branch and sheet specified in the main text.

With $z=r-M$ and $\Psi=z^{s+1}R$, the extremal radial equation is
\begin{equation}\label{eq:normal-radial-proof}
\begin{aligned}
 \frac{{\rm d}^2\Psi}{{\rm d}r^2}+Q(r)\Psi&=0,\\
 Q&=\omega^2+\frac{b}{z}+\frac{c}{z^2}+\frac{d}{z^3}+\frac{h^2}{z^4},
\end{aligned}
\end{equation}
where $h=M(2M\omega-m)$, $b=4M\omega^2+2is\omega$, $c=7M^2\omega^2-{}_sA_{\ell m}-s(s+1)$, and $d=(4M\omega-2is)h$. At the horizon, set $u_M=z^{-1}$ and $\Psi={\rm e}^{ihu_M}u_M^{-p_M}f_M$, with $p_M=1-s-2iM\omega$; at infinity, set $u_\infty=z$ and $\Psi={\rm e}^{i\omega u_\infty}u_\infty^{p_\infty}f_\infty$, with $p_\infty=-s+2iM\omega$. The amplitudes $f_e$, $e=M,\infty$, approach one on the selected decay rays and satisfy
\begin{equation}\label{eq:endpoint-volterra-ode}
 f_e''+\left(2i\mathfrak a_e+\frac{q_e}{u}\right)f_e'
 +\left(\frac{C_e}{u^2}+\frac{D_e}{u^3}+\frac{E_e}{u^4}\right)f_e=0,
\end{equation}
where primes denote $u$ derivatives and
\begin{equation}\label{eq:endpoint-volterra-coefficients}
\begin{array}{c|ccccc}
 e&\mathfrak a_e&q_e&C_e&D_e&E_e\\ \hline
 M&h&2-2p_M&c+p_M(p_M-1)&b&\omega^2\\
 \infty&\omega&2p_\infty&c+p_\infty(p_\infty-1)&d&h^2
\end{array}.
\end{equation}
The endpoint coordinates $u_M,u_\infty$ are local variables for this construction; they are distinct from the HeunD coordinate $x$ in the main text. Substituting each endpoint ansatz in the normal radial equation gives the corresponding row of coefficients. Removing the leading exponential and power leaves an amplitude tending to one, fixing its normalization before it is propagated to the matching point.
The coefficients in $f_e\sim\sum_{n\ge0}a_nu^{-n}$ obey
\begin{equation}\label{eq:numerical-endpoint-recurrence}
 a_{n+1}=\frac{[n(n+1-q_e)+C_e]a_n+D_ea_{n-1}+E_ea_{n-2}}
 {2i\mathfrak a_e(n+1)},
\end{equation}
with $a_0=1$, $a_{-1}=a_{-2}=0$, and $n\ge0$. These are the endpoint series used in Eq.~\eqref{eq:numerical-endpoint-series}. For $n\ge0$, the explicit recurrence denominator vanishes only at the excluded critical points $\omega=0$ and $2M\omega=m$; the integer-index factor $n+1$ never vanishes. Analyticity of the selected angular eigenvalue is a separate condition, checked below for the counting domain.

Write an endpoint ray as $u=\rho{\rm e}^{i\theta_e}$, where $\rho>0$ is its radial coordinate and $\theta_e$ its fixed phase. The infima and suprema below are over a compact noncritical parameter chart. Choose
\begin{equation}\label{eq:uniform-endpoint-decay}
 \eta_e=\inf[-\Re(2i\mathfrak a_e{\rm e}^{i\theta_e})]>0.
\end{equation}
Twice integrating Eq.~\eqref{eq:endpoint-volterra-ode} with $f_e\to1$ gives a Volterra equation $f_e=1-K_ef_e$. With $Q_e(u)=C_eu^{-2}+D_eu^{-3}+E_eu^{-4}$, its integral operator is
\[
 \begin{aligned}
 (K_ef)(u)={}&\int_u^\infty {\rm e}^{-2i\mathfrak a_e v}v^{-q_e}\\
 &\quad\times\left[\int_v^\infty {\rm e}^{2i\mathfrak a_e w}
 w^{q_e}Q_e(w)f(w)\,{\rm d}w\right]{\rm d}v .
 \end{aligned}
\]
Both integrals follow the selected ray, and the powers use its continued logarithm. If $B_q=\sup\max(0,\Re q_e)$ and $B_C,B_D,B_E$ bound the three coefficients, then, for $\rho\ge R_e\ge2B_q/\eta_e$,
\begin{equation}\label{eq:volterra-contraction}
 \|K_e\|\le\frac{2}{\eta_e}
 \left(\frac{B_C}{R_e}+\frac{B_D}{2R_e^2}+\frac{B_E}{3R_e^3}\right).
\end{equation}
Here $R_e$ is the minimum ray radius and $B_C,B_D,B_E$ bound the moduli of $C_e,D_e,E_e$. The norm $\|K_e\|$ is induced by $\|f\|_\infty=\sup_{\rho\ge R_e}|f(\rho{\rm e}^{i\theta_e})|$. A sufficiently large $R_e$ makes this a contraction and defines a nonzero solution holomorphic in the parameters. Propagation through ordinary points preserves these properties. The oppositely directed wave solutions give the endpoint basis determinants
\begin{subequations}\label{eq:noncritical-endpoint-basis-determinants}
\begin{align}
 \lim_{r\to M}z^{2(s+1)}W_r[R_{\rm in}^M,R_{\rm up}^M]&=2ih,\\
 \lim_{r\to\infty}z^{2(s+1)}W_r[R_{\rm in}^\infty,R_{\rm up}^\infty]&=2i\omega.
\end{align}
\end{subequations}
They are nonzero away from the critical frequencies, so the continued sectorial bases are related by holomorphic invertible transformations. The canonical radial connection therefore has no finite normalization poles on the noncritical Riemann surface. Its zeros count QNMs, including multiplicity, when the angular branch and frequency logarithms close around the contour.

For an endpoint truncation retaining $N_{\rm end}$ series terms, write $f_e^{(N_{\rm end})}=\sum_{n=0}^{N_{\rm end}-1}a_nu^{-n}$. The recurrence leaves a residual $\mathcal L_ef_e^{(N_{\rm end})}=\sum_{j=1}^3g_{N_{\rm end}+j}^{\rm res}u^{-N_{\rm end}-j}$, where $\mathcal L_e$ is the operator in Eq.~\eqref{eq:endpoint-volterra-ode}. In the norm $\|f\|_{N_{\rm end}}=\sup_{\rho\ge R_e}\rho^{N_{\rm end}}|f(\rho{\rm e}^{i\theta_e})|$, take $N_{\rm end}\ge\Re q_e$ and
\begin{equation}\label{eq:weighted-endpoint-contraction}
\begin{gathered}
 \kappa_{N_{\rm end}}=\frac{1}{\eta_e}\sum_{j=1}^3\frac{B_j}{(N_{\rm end}+j)R_e^j}<1,\\
 (B_1,B_2,B_3)=(|C_e|,|D_e|,|E_e|).
\end{gathered}
\end{equation}
The same integral estimate bounds the amplitude and derivative errors at $u=R_e{\rm e}^{i\theta_e}$:
\begin{subequations}\label{eq:weighted-endpoint-errors}
\begin{equation}\label{eq:weighted-endpoint-error}
\begin{aligned}
 |f_e-f_e^{(N_{\rm end})}|&\le\varepsilon_f,\\
 \varepsilon_f&=\frac{1}{\eta_e(1-\kappa_{N_{\rm end}})}\\
 &\quad\times\sum_{j=1}^3\frac{|g_{N_{\rm end}+j}^{\rm res}|}{(N_{\rm end}+j-1)R_e^{N_{\rm end}+j-1}},
\end{aligned}
\end{equation}
\begin{equation}\label{eq:weighted-endpoint-derivative-error}
\begin{aligned}
 &|\partial_u(f_e-f_e^{(N_{\rm end})})|\\
 &\quad\le\frac{1}{\eta_e}\sum_{j=1}^3\left(\frac{B_j\varepsilon_f}{R_e^{j+1}}+\frac{|g_{N_{\rm end}+j}^{\rm res}|}{R_e^{N_{\rm end}+j}}\right).
\end{aligned}
\end{equation}
\end{subequations}
The factor $(1-\kappa_{N_{\rm end}})^{-1}$ accounts for the effect of the unknown remainder in the integral equation; the residual coefficients alone would not give the full error. The gauge and chain rule convert these bounds to errors in both state components. Angular uncertainty and arithmetic errors are included separately in the connection enclosure below.

\subsection{Isolation of the angular branch}\label{app:bounded-count-details}
For the rectangle $D_{33}$ in Fig.~\ref{fig:connection-count-domain}, the branch is selected by the spherical value $ {}_{-2}A_{33}(0)=10$. Its isolation is verified throughout the rectangle by enclosing a single angular eigenvalue on overlapping spectral circles. The spherical-basis core and its infinite-tail Schur bound are treated together. Let $F$ be the finite-core angular matrix shifted by a trial angular eigenvalue on a spectral circle, $G$ an approximate inverse of $F$, and $I$ the identity matrix. A verified residual $\|I-GF\|_F\le\varepsilon_G<1$ gives
\begin{equation}\label{eq:angular-inverse-certificate}
 \sigma_{\min}(F)\ge\frac{1-\varepsilon_G}{\|G\|_F},
\end{equation}
where $\sigma_{\min}$ is the smallest singular value and $\|\cdot\|_F$ the Frobenius norm. After including frequency-cell variations, complete spectral-circle arcs, and the infinite angular tail, the minimum verified margin is greater than $0.1810$. Neighboring circles enclose the same rank-one projection, so they define one analytic branch throughout $D_{33}$.

In this verification, the finite core retains spherical degrees $3\le\ell'\le42$, and the rectangle is covered by 3645 frequency cells. For each cell, angular spectral circles of radii $2$ and $3$ are checked at 128 nodes each; interval bounds cover the arcs between nodes as well as the variation across the frequency cell. The infinite-tail bound is included in the same margin. Thus the positive margin concerns the continuous cover, rather than only the sampled matrix eigenvalues. The overlapping circles and the spherical anchor identify which of the isolated eigenvalues is followed from cell to cell.

Enlarging each frequency cell by $0.01$ retains a positive margin greater than $0.0744$. This provides complex neighborhoods for the boundary estimates; for $A={}_{-2}A_{33}(M\omega)$, Cauchy's bounds yield
\begin{equation}\label{eq:l3-angular-derivative-bounds}
 \left|\frac{{\rm d}A}{{\rm d}(M\omega)}\right|\le200,\qquad
 \left|\frac{{\rm d}^2A}{{\rm d}(M\omega)^2}\right|\le40000.
\end{equation}
Thus no angular interchange occurs in this rectangle. This isolation condition is essential: a contour that exchanges angular eigenvalues cannot define a count for one fixed angular branch.

For $\ell=m=2$, the sampled connection winding is also one on the three nested rectangles $(0.45,0.56)\times(-0.76,-0.65)$, $(0.15,0.90)\times(-1.50,-0.15)$, and $(0.05,0.98)\times(-4,-0.03)$, where the factors give the real and imaginary frequency intervals. Each contains the DM in Fig.~\ref{fig:transition-dm}. Refinement and parameter variations preserve these numerical windings; they are supporting convergence checks, without the continuous interior and boundary verification established here for $D_{33}$.

\subsection{Continuous boundary control}\label{app:absolute-connection-errors}
To turn the phase curve in Fig.~\ref{fig:connection-count-phase} into a QNM count, the connection must remain nonzero between the computed boundary points. At each point, interval evaluation encloses the regular angular eigenvalue, the endpoint residuals above, and propagation of both radial state components. For a Taylor step $\Delta r$ about an ordinary point $r_0$, the disk $|r-r_0|\le8|\Delta r|$ excludes $r=M$. Choose a positive state-scaling factor $w$ with $w^2\ge\sup|Q|$ on this disk. In the norm of $(\Psi,\partial_r\Psi/w)$, the degree-$n_T$ transfer tails satisfy
\begin{subequations}\label{eq:ordinary-transfer-tails}
\begin{align}
 \varepsilon_\Psi&\le {\rm e}^{8|\Delta r|w}
 \frac{8^{-n_T-1}}{1-1/8},\\
 \varepsilon_{\partial_r\Psi}&\le w{\rm e}^{8|\Delta r|w}
 \frac{8^{-n_T}}{1-1/8}.
\end{align}
\end{subequations}
These bounds apply to the initial column $(1,0)$; division by $w$ gives the bounds for $(0,1)$. Arithmetic enclosures are retained for both columns before propagation. At 180-digit interval precision, endpoint scale 100 and $n_T=110$ give nonzero connection enclosures at all 259 boundary samples. The largest absolute and relative connection error bounds are $1.580\times10^{-53}$ and $4.561\times10^{-51}$, respectively. With these point errors included, the enclosed polygon has winding one, with consecutive phase increments below $0.929$ radians. These pointwise results alone do not exclude an unresolved phase turn between samples; the continuous cover below is needed for that purpose.

Continuous control uses overlapping frequency disks on the same angular branch and Riemann sheet. On a disk of radius $\rho_i$ centered at $\omega_i$, approximate the connection by a degree-$J_i$ polynomial $P_i$ in the scaled frequency, with coefficients $c_{i,q}$:
\begin{subequations}\label{eq:boundary-parameter-enclosure}
\begin{align}
 P_i(\omega)&=\sum_{q=0}^{J_i}c_{i,q}
 \left(\frac{\omega-\omega_i}{\rho_i}\right)^q,\\
 \sup_{|\omega-\omega_i|\le\rho_i}|\mathcal W_{\mathfrak b}-P_i|&\le\varepsilon_i^T.
\end{align}
\end{subequations}
Here $\varepsilon_i^T$ includes the angular, endpoint, transfer, and arithmetic errors. The sufficient condition
\begin{equation}\label{eq:boundary-polynomial-certificate}
 \sum_{q=1}^{J_i}|c_{i,q}|+\varepsilon_i^T<|c_{i,0}|
\end{equation}
places the complete connection image in a disk excluding zero. Uniform errors on a guard disk of radius $2\rho_i$, followed by Cauchy's bounds, control both the retained coefficients and omitted frequency powers. Frequency-scaled endpoint coordinates keep the leading exponentials fixed across each disk; the physical powers and their continued logarithms remain included. All coordinate changes and transfer paths avoid the horizon and both critical frequencies.

More explicitly, for a disk centered at $\omega_i$ the scaled radial coordinates satisfy
\[
 \begin{gathered}
 r-M=\vartheta_e(\omega)(\widetilde r_e-M),\\
 \vartheta_M=\frac{2M\omega-m}{2M\omega_i-m},\quad
 \vartheta_\infty=\frac{\omega_i}{\omega}.
 \end{gathered}
\]
Consequently the leading horizon and infinity exponentials in the scaled equations have their values at $\omega_i$ throughout the disk. The frequency dependence of the power factors is retained, and the scaled derivative is converted back to the physical radial derivative before a short ordinary-point propagation to a common matching radius. This rescaling makes a uniform enclosure less sensitive to rapid exponential variation; it does not replace the original boundary conditions.

Where useful, the same test is applied to $ {\rm e}^{-L_i(\omega)}\mathcal W_{\mathfrak b}$, with a fixed polynomial $L_i(\omega_i)=0$. This entire nonzero factor changes no zeros. If its enclosed image obeys $|{\rm e}^{-L_i}\mathcal W_{\mathfrak b}-c_{i,0}|\le V_i<|c_{i,0}|$, then the phase increment along a covered arc from $\omega_a$ to $\omega_b$ satisfies
\begin{subequations}\label{eq:normalized-phase-transport}
\begin{align}
 \Delta\arg\mathcal W_{\mathfrak b}
 &=\Im[L_i(\omega_b)-L_i(\omega_a)]
   +\epsilon_{i,b}-\epsilon_{i,a},\\
 \max\{|\epsilon_{i,a}|,|\epsilon_{i,b}|\}
 &\le\arcsin\frac{V_i}{|c_{i,0}|}.
\end{align}
\end{subequations}
Here $\epsilon_{i,a}$ and $\epsilon_{i,b}$ are the phases of the normalized connection relative to $c_{i,0}$ at $\omega_a$ and $\omega_b$, respectively; each phase branch is chosen within the image disk that excludes zero. The phase intervals on neighboring disks determine the integer multiple of $2\pi$ relative to each polygon chord. In the verified cover of $\partial D_{33}$, that integer is zero on every edge. The cover has clearance greater than $3.9\times10^{-4}$ and remains valid when the stored vertices are projected onto the exact decimal rectangle. Thus the continuous connection and its numerical polygon have the same winding, establishing Eq.~\eqref{eq:certified-D33-count}. Together with angular isolation and the absence of normalization poles, this proves that $D_{33}$ contains exactly the one simple QNM shown in Fig.~\ref{fig:connection-count-domain}. The statement applies to this domain, angular branch, and Riemann sheet.

The boundary cover uses frequency-polynomial degree $J_i=128$, 250-digit ball arithmetic, endpoint scale 60, and ordinary-point propagation degree 110. Its disks overlap along all 259 edges; their radii are $0.012$ or $0.024$ in $M\omega$. The error enclosure includes the angular, endpoint, transfer, and frequency-series remainders. For each edge, the phase intervals rule out any extra integer turn relative to its polygon chord. This is the additional step that promotes the sampled winding to a count for the continuous boundary.

\section{Uniform matching near the frequency branch points}\label{app:uniform-critical}
This appendix supplies the local analysis used in Sec.~\ref{sec:critical}. At a branch point the generic radial representation degenerates, so a sequence approaching that point does not by itself determine the zeros of the exact extremal connection. We first isolate the angular eigenvalue and match the two radial endpoint solutions near the superradiant bound. The resulting limiting connection identifies the possible zeros on each continued frequency sheet; a uniform remainder then relates those model zeros to the exact connection. We next give quantitative conventional-sheet exclusions for the two diagonal sectors, discuss the other possible critical exponents, and finally treat the separate branch point at $\omega=0$. All exclusions below concern punctured neighborhoods on the specified branches and sheets.
\subsection{Angular isolation and matching at the superradiant bound}\label{app:critical-angular-isolation}
We first isolate the angular eigenvalue so that radial matching follows the same $(\ell,m)$ mode as the frequency varies. Let $\mathcal A(a\omega)$ denote the angular operator in Eq.~\eqref{eq:angular}, with eigenvalue $ {}_sA_{\ell m}(a\omega)$. At fixed $s,m$, $X$ is the multiplication operator $(Xf)(\Theta)=\Theta f(\Theta)$, where $\Theta=\cos\theta$, so $(X^2f)(\Theta)=\Theta^2f(\Theta)$. Matrix elements are evaluated in an orthonormal basis of spin-weighted spherical harmonics at fixed $s,m$. Its norm induced by the angular $L^2$ inner product obeys $\|X\|\le1$. At the real critical frequency, the self-adjoint operator $\mathcal A_*=\mathcal A(m/2)$ therefore satisfies $\mathcal A_*=\mathcal A(0)-\frac{m^2}{4}X^2+smX$. Let $P_{\ell_{\max}}$ be the orthogonal projection onto spherical-basis modes with angular degree $\ell'\le \ell_{\max}$. The retained block is $P_{\ell_{\max}}\mathcal A_*P_{\ell_{\max}}$, the coupling block is $B=P_{\ell_{\max}}\mathcal A_*(1-P_{\ell_{\max}})$, and the omitted tail is $\mathcal T=(1-P_{\ell_{\max}})\mathcal A_*(1-P_{\ell_{\max}})$, with $1$ denoting the identity. Here $\ell_{\max}$ is the angular basis cutoff. The retained block contains $P_{\ell_{\max}}X^2P_{\ell_{\max}}$: multiplying twice before projection includes components that leave the retained basis and return, which squaring $P_{\ell_{\max}}XP_{\ell_{\max}}$ would omit. For gravitational perturbations, the tail eigenvalues have the lower bound
\begin{equation}\label{eq:critical-angular-tail}
 \mathcal T\ge d_{\ell_{\max}},\qquad
 d_{\ell_{\max}}=(\ell_{\max}+1)(\ell_{\max}+2)-2-\frac{m^2}{4}-2m
\end{equation}
where $d_{\ell_{\max}}$ is a lower bound for the tail spectrum. For a cutoff with $d_{\ell_{\max}}>40$ and a real trial eigenvalue below $40$, the positive Schur correction from the tail is bounded by $\eta_{\ell_{\max}}=\frac{\|B\|_{\rm F}^2}{d_{\ell_{\max}}-40}$, with $\|\cdot\|_{\rm F}$ denoting the Frobenius matrix norm. If $v_\ell$ is the corresponding eigenvalue of the retained block, min--max and Schur inertia give $v_\ell-\eta_{\ell_{\max}}\le {}_{-2}A_{\ell m}(\frac{m}{2})\le v_\ell$ whenever the enclosed eigenvalues lie below $40$. Thus $\eta_{\ell_{\max}}$ bounds the eigenvalue shift caused by the omitted basis modes.

The estimate follows by eliminating the tail component of an eigenvector: at a real trial angular eigenvalue $\zeta_A<40$, the retained component obeys the effective equation with operator $P_{\ell_{\max}}\mathcal A_*P_{\ell_{\max}}-B(\mathcal T-\zeta_A)^{-1}B^\dagger$. Since $\mathcal T-\zeta_A$ is positive and its inverse has norm at most $(d_{\ell_{\max}}-40)^{-1}$, the subtracted correction is positive and bounded by $\eta_{\ell_{\max}}$. This explains both the direction and size of the enclosure rather than treating the finite matrix eigenvalues as exact.

For each fixed gravitational sector $(\ell,m)$ with $m>0$, $\mathcal A_*$ has compact resolvent: its difference from the spherical angular operator is a bounded real multiplication potential. The regular boundary condition at either pole selects one local solution, with the logarithmic solution excluded when the endpoint exponent vanishes. Thus two regular eigenfunctions with the same eigenvalue have a vanishing weighted Wronskian and are proportional, so the eigenvalue is simple and isolated. The bounded perturbation $\mathcal A(a\omega)-\mathcal A_*$ is analytic in $a\omega$; a spectral circle inside the neighboring gap therefore retains a rank-one Riesz projection for sufficiently small $|a\omega-m/2|$. This gives a locally analytic angular branch for each fixed sector, with a neighborhood that can depend on $\ell,m$. The following calculation gives explicit bounds for $\ell=2,3$.

Interval enclosure of the core eigenvalues, including the infinite tail, gives a nearest-neighbor gap greater than $6.39$ for every $2\le\ell\le3$ and $1\le m\le\ell$. The calculation uses $\ell_{\max}=30$. For extremal Kerr black holes the complex perturbation satisfies
\begin{equation}\label{eq:critical-angular-complex-bound}
 \|\mathcal A(M\omega)-\mathcal A_*\|
 \le\frac{|k|}{2}\left(m+4+\frac{|k|}{2}\right),
 \qquad k=2M\omega-m.
\end{equation}
Here $k$ is the dimensionless detuning from the superradiant bound. On a circle of radius $3$ in the angular-eigenvalue plane, centered at the critical eigenvalue, the unperturbed resolvent (the inverse of the shifted angular operator) has norm at most $\frac{1}{3}$. For $|k|\le0.1$, the perturbation norm is at most $0.4025$. The resolvent Neumann series therefore preserves the rank-one spectral projection onto the eigenspace inside this circle. Thus exactly one simple angular eigenvalue remains isolated throughout this frequency neighborhood.

We now consider radial matching for $m>0$ and a simple critical angular eigenvalue with real $\delta_{\rm SR}>0$. Use $z=r-M$ for the local radial offset and $h=Mk$. The coefficient $c$ is the radial coefficient in Eq.~\eqref{eq:normal-radial-proof}, and $c_*=c(\omega_{\rm SR})$ denotes its critical value. Analyticity of the angular eigenvalue gives $c-c_*=O(k)$, while Eq.~\eqref{eq:deltasr} gives $c_*-\frac14=\delta_{\rm SR}^2$. The bounds below are taken on a fixed bounded interval of the continuously tracked argument of $k$; their constants need not be uniform over infinitely many Riemann sheets.

For the normal-form radial function $\Psi=z^{s+1}R$ in Eq.~\eqref{eq:numerical-radial-state}, the limiting outer and scaled inner solutions are
\begin{subequations}\label{eq:critical-whittaker-pair}
\begin{align}
 \Psi_{\mathrm{up}}^{\infty,0}
 &=(-2i\omega_{\rm SR})^{s-im}W_{-s+im,i\delta_{\rm SR}}(-2i\omega_{\rm SR}z),\\
 \Psi_{\mathrm{in}}^{M,0}
 &=(-2ih)^{-s-im}zW_{s+im,i\delta_{\rm SR}}\!\left(\frac{-2ih}{z}\right).
\end{align}
\end{subequations}
Here $W$ denotes the Whittaker $W$ function, and the superscript $0$ labels a limiting radial equation. These are the near and far limits underlying the matching construction~\cite{Gralla:2017lto,Casals:2018eev}. The displayed factors fix their critical power normalizations. All powers and special functions follow the selected continuation path.

We use the logarithmic state $(Y,{\rm d}Y/{\rm d}\xi)^{\mathsf T}$, with $Y=z^{-1/2}\Psi$ and $\xi=\log(z/M)$. At a fixed outer radius the outgoing state differs from its Whittaker limit by $O(k)$. The dimensionless near-horizon coordinate $u=z/h$ resolves the region $z=O(h)$; taking $k\to0$ at fixed $u$ gives the same estimate for the horizon state. Matching at $|z|/M=\sqrt{|k|}$, variation of constants in the logarithmic coordinate bounds the integrated perturbation by $O(\sqrt{|k|}+|k\log|k||)$. The comparison fundamental matrices and their inverses are bounded on each fixed finite interval of the continued argument because $\delta_{\rm SR}$ is real and nonzero.

The choice of matching point makes the overlap explicit: as $k\to0$, both $|z|/M$ and $|h/z|$ are $O(\sqrt{|k|})$. Thus the outer solution is evaluated close to the horizon while the inner solution is evaluated far outside its scaled horizon region. The two power solutions at the match are independent because their logarithmic derivatives tend to $\pm i\delta_{\rm SR}$. The connection is obtained by taking the determinant of the two endpoint states in this common basis.

The two coefficients needed to normalize the determinant are
\begin{subequations}\label{eq:critical-prefactor-coefficients}
\begin{align}
 a_{\mathrm{in},-}
 &=\frac{\Gamma(-2i\delta_{\rm SR})}{\Gamma(\frac{1}{2}-i\delta_{\rm SR}-s-im)}
 (-2ih)^{\frac{1}{2}+i\delta_{\rm SR}-s-im},\\
 a_{\mathrm{up},+}
 &=\frac{\Gamma(-2i\delta_{\rm SR})}{\Gamma(\frac{1}{2}-i\delta_{\rm SR}+s-im)}
 (-2i\omega_{\rm SR})^{\frac{1}{2}+i\delta_{\rm SR}+s-im}.
\end{align}
\end{subequations}
They multiply $z^{\frac{1}{2}-i\delta_{\rm SR}}$ and $z^{\frac{1}{2}+i\delta_{\rm SR}}$, respectively. With
\begin{subequations}\label{eq:critical-prefactor}
\begin{align}
 L_{\mathfrak b}^{\rm crit}&=\operatorname{Log}_{\mathfrak b}(-2ik)
 +\operatorname{Log}(-im),\\
 P_{\mathfrak b}&=2i\delta_{\rm SR}\,a_{\mathrm{in},-}a_{\mathrm{up},+}
 {\rm e}^{-2i\delta_{\rm SR} L_{\mathfrak b}^{\rm crit}}.
\end{align}
\end{subequations}
The superscript ${\rm crit}$ indicates that the second logarithm in $L_{\mathfrak b}$ is evaluated at the critical frequency; $P_{\mathfrak b}$ removes the explicit endpoint powers from the Wronskian. This factor is nonzero and holomorphic on the punctured logarithmic sector, provided the displayed gamma coefficients do not vanish. This condition holds for the gravitational diagonal sectors studied here. Expanding the two Whittaker functions gives
\begin{align}\label{eq:critical-gamma-coefficient}
 C_{\ell m}={}&
 \left[\frac{\Gamma(2i\delta_{\rm SR})}
 {\Gamma(-2i\delta_{\rm SR})}\right]^2
 \nonumber\\
 &\times\prod_{\sigma=\pm1}
 \frac{\Gamma(\frac{1}{2}-i\delta_{\rm SR}-im+\sigma s)}
 {\Gamma(\frac{1}{2}+i\delta_{\rm SR}-im+\sigma s)}.
\end{align}
The state estimates then give
\begin{subequations}\label{eq:uniform-critical-connection}
\begin{align}
 \frac{\mathcal W_{\mathfrak b}}{P_{\mathfrak b}}
 &={\rm e}^{2i\delta_{\rm SR} L_{\mathfrak b}^{\rm crit}}-C_{\ell m}
 +\varepsilon_{\mathfrak b}(k),\\
 |\varepsilon_{\mathfrak b}(k)|
 &\le K_{\mathfrak b}\big[\sqrt{|k|}+|k\log|k||\big].
\end{align}
\end{subequations}
Here $\varepsilon_{\mathfrak b}(k)$ is the matching remainder, and $K_{\mathfrak b}$ is a positive bound independent of sufficiently small $|k|$. This bound is uniform for sufficiently small $|k|$ on the chosen finite argument interval; it requires real $\delta_{\rm SR}>0$. Replacing $L_{\mathfrak b}^{\rm crit}$ by $L_{\mathfrak b}$ contributes only $O(k)$.

Label the zeros so that $k_N\to0$ as $N\to\infty$. Let $k_N^0$ be the zeros obtained by omitting $\varepsilon_{\mathfrak b}$ from Eq.~\eqref{eq:uniform-critical-connection}, and $k_N$ the corresponding exact Wronskian zeros. Their asymptotic geometric spacing is $q_{\rm geo}={\rm e}^{-\frac{\pi}{\delta_{\rm SR}}}$. Choose a fixed relative radius $0<\rho_{\rm rel}<\frac{1-q_{\rm geo}}{1+q_{\rm geo}}$, small enough that the circles remain in the specified angular interval. On $|k-k_N^0|=\rho_{\rm rel}|k_N^0|$, the model equals $C_{\ell m}[(\frac{k}{k_N^0})^{2i\delta_{\rm SR}}-1]$ and has a positive minimum independent of $N$. Equation~\eqref{eq:uniform-critical-connection} makes the error smaller than this minimum for all sufficiently small $|k_N^0|$. Rouch\'e's theorem then gives exactly one simple zero in each circle. In particular,
\begin{equation}\label{eq:exact-critical-ladder}
 k_N=k_N^0+O(|k_N^0|^{\frac{3}{2}}),\qquad
 \frac{|k_{N+1}|}{|k_N|}\longrightarrow q_{\rm geo}.
\end{equation}
On the remaining parts of each scaled annulus the model has a uniform positive lower bound, excluding additional sufficiently small zeros in that sector. The resulting sequence is countably infinite on a single fixed unconventional horizon-frequency sheet and accumulates at the superradiant branch point. Its continued argument stays in the chosen bounded interval as $N$ increases; only the modulus shrinks. Thus uniformity on this finite argument interval suffices for the infinite sequence, without a bound uniform across infinitely many sheets. These zeros are poles of the continued radial Green function, since the noncritical canonical states are primitive. The index $N$ in Eq.~\eqref{eq:exact-critical-ladder} orders this local accumulation sequence and is distinct from a Schwarzschild parent overtone.

For fixed sufficiently small $\epsilon_0$, the unconventional Riemann sheet count in a shrinking critical annulus therefore satisfies
\begin{equation}\label{eq:critical-annulus-density}
 \mathcal N_{\mathfrak b}^{\rm crit}(\epsilon,\epsilon_0)
 =\frac{\delta_{\rm SR}}{\pi}\log\frac{\epsilon_0}{\epsilon}+O(1),
 \qquad\epsilon\to0,
\end{equation}
where $\mathcal N_{\mathfrak b}^{\rm crit}$ counts the continued Green function poles, including multiplicity, in $\epsilon<|\omega-\omega_{\rm SR}|<\epsilon_0$. The positive numbers $\epsilon$ and $\epsilon_0$ are the inner and outer frequency radii; $O(1)$ remains bounded as $\epsilon\to0$ at fixed $\epsilon_0$. These poles are not counted as DM parents on the conventional Riemann sheet.

On the conventional Riemann sheet, the positive model gap ${\rm e}^{-\pi\delta_{\rm SR}}-|C_{\ell m}|$ instead excludes zeros sufficiently close to the critical frequency. This gap is positive for the two diagonal sectors in Table~\ref{tab:critical-ray-arguments}; the following bounds make their exclusion radii explicit.

\subsection{Quantitative bounds for zero-free regions on the conventional Riemann sheet}\label{app:explicit-conventional-sheet-collar}
The preceding matching argument shows that a sufficiently small zero-free neighborhood exists, but its unspecified remainder constant does not give a numerical radius. Here we use the exact condition that the incoming amplitude vanishes and enclose its terms throughout a disk. Angular isolation fixes the branch, and the continued-fraction enclosures control the renormalized angular momentum and matching sums. We then bound the modulus of the two sides in opposite directions; a strict separation excludes an incoming-amplitude zero throughout the stated region.
The extremal MST matching coefficients~\cite{Casals:2018eev} give an exact quantitative exclusion near the superradiant bound. Write $\widetilde\omega=2M\omega=m+k$ and set $
\nu_0=-\frac12+i\delta_{\rm SR}$, where $\delta_{\rm SR}>0$. The angular eigenvalue belongs to the isolated rank-one branch on $|k|\le0.1$ established above. Here $\widetilde\omega$ is the dimensionless frequency, $\nu$ is the renormalized angular momentum of Appendix~\ref{app:mst}, and $\nu_0$ is its critical value. The coefficients $a_n^\nu$ obey Eq.~\eqref{eq:mst-recurrence}, with $a_0^\nu=1$; $n$ labels series terms. We bound both sides of the incoming-amplitude condition to show that they cannot coincide.

At matching index zero, divide each matching sum by its Gamma factor at index zero. The resulting sums are
\begin{subequations}\label{eq:mst-normalized-matching-sums}
\begin{align}
 \mathscr C_\nu&=\sum_{n=0}^\infty c_n^\nu a_n^\nu,\\
 \mathscr D_\nu&=\sum_{n=0}^\infty
 \frac{(-2\nu-1)_n}{n!}a_{-n}^\nu,\\
 c_n^\nu&=\frac{(2\nu+1)_n}{n!}
 \prod_{\sigma=\pm1}
 \frac{(\nu+1+\sigma\chi_{\sigma s})_n}
      {(\nu+1-\sigma\chi_{\sigma s})_n}.
\end{align}
\end{subequations}
Here $c_n^\nu$ are matching weights and $\mathscr C_\nu,\mathscr D_\nu$ are normalized matching sums. The auxiliary parameters $\chi_{\sigma s}=\sigma s-i\widetilde\omega$, with $\sigma=\pm1$, are distinct from the logarithmic coordinate $\xi$ above. The Pochhammer symbol is $(a)_n=\Gamma(a+n)/\Gamma(a)$, with $(a)_0=1$. The partner $\widetilde\nu=-\nu-1$ has $a_n^{\widetilde\nu}=a_{-n}^\nu$. Gamma shift identities establish Eq.~\eqref{eq:mst-normalized-matching-sums} term by term, rather than as a small-$k$ approximation. Define
\begin{equation}\label{eq:mst-critical-gamma-factor}
 \mathcal G(\nu,\widetilde\omega)=
 \left[\frac{\Gamma(2\nu+1)}{\Gamma(-2\nu-1)}\right]^2
 \prod_{\sigma=\pm1}
 \frac{\Gamma(-\nu+\chi_{\sigma s})}
      {\Gamma(\nu+1+\chi_{\sigma s})}.
\end{equation}
The exact condition for the incoming amplitude is then
\begin{equation}\label{eq:mst-exact-critical-condition}
 {\rm e}^{(2\nu+1)L_{\mathfrak b}}
 =\mathcal G(\nu,\widetilde\omega)
 \frac{\mathscr C_
\nu\mathscr D_{\widetilde\nu}}
      {\mathscr D_
\nu\mathscr C_{\widetilde\nu}},
\end{equation}
where $L_{\mathfrak b}$ is the same continued logarithm as in Eq.~\eqref{eq:critical-sheet-model}. Indeed,
$L_{\mathfrak b}=\operatorname{Log}_{\mathfrak b}k+\operatorname{Log}(2\widetilde\omega)-i\pi$ here. With these logarithms the unsimplified incidence factor of the MST connection is $-1$. Mixing a principal-power piecewise factor with these continued logarithms would change the condition. At $k=0$, $\mathscr C_\nu=\mathscr D_\nu=1$ and $\mathcal G(
\nu_0,m)=C_{\ell m}$, recovering Eq.~\eqref{eq:critical-sheet-model}.

We enclose the infinite continued fractions of Eq.~\eqref{eq:mst-recurrence}, including their $\nu$ derivatives. For auxiliary parameter disks $|k|\le R_k$ and $|\nu-\nu_0|\le r_\nu$, the minimal-ratio tails have the invariant bound $1/64$ from depth $48$ onward. Uniform contraction encloses the infinite tails analytically. Backward differentiation of the exact ratios then encloses the closure derivative. The map $
\nu\mapsto\nu-F_{\rm CF}(\nu,k)/(2
\nu_0+1)$, where $F_{\rm CF}(\nu,k)$ is the residual of the $n=0$ recurrence after matching the minimal sequences from both index directions, contracts and maps the closed complex disk strictly into itself. The angular perturbation bound above includes the infinite angular problem. Uniqueness and holomorphic dependence give
\begin{equation}\label{eq:mst-critical-nu-schwarz}
 |\nu(k)-
\nu_0|\le B_\nu|k|,\qquad |k|\le R_k,
\end{equation}
by Schwarz's lemma, with $B_\nu=1.796$ for $(2,2)$ and $2.206$ for $(3,3)$. The certified contraction factors are below $0.077$ and $0.186$ for the two diagonal sectors in Table~\ref{tab:critical-ray-arguments}. The radii $R_k$ and $r_\nu$ bound the auxiliary detuning and renormalized-angular-momentum disks, respectively; $B_\nu$ bounds the change in $\nu$ per unit $|k|$.

The infinite-tail enclosure can be checked from the continued fraction (CF) recurrence coefficients as follows. Let $L_{\rm CF}$ be the truncation depth from which the two recurrence tails are bounded, and $h_{\rm CF}=1/64$ the invariant bound on the corresponding minimal-solution coefficient ratios. Set $V=|\nu_0|+2r_\nu+1$, $d_{\rm CF}=L_{\rm CF}-V-1$, $e_k=(m+R_k)R_k$, and $B_A=R_k(m+4+R_k/2)/2$. For $|n|\ge L_{\rm CF}$, the off-diagonal coefficients are bounded by $T_{\rm CF}$ and the diagonal coefficient from below by $B_{\rm CF}$, where
\begin{align}
 T_{\rm CF}&=e_k\left(\frac{L_{\rm CF}+V+1+|s|+m+R_k}{d_{\rm CF}}\right)^2,\nonumber\\
 B_{\rm CF}&=d_{\rm CF}^2-\delta_{\rm SR}^2-\frac14-B_A-\frac{7R_k(2m+R_k)}4\nonumber\\
 &\quad-\frac{e_k[s^2+(m+R_k)^2]}{d_{\rm CF}^2}.
\end{align}
The inequalities $T_{\rm CF}/(B_{\rm CF}-T_{\rm CF}h_{\rm CF})<h_{\rm CF}$ and $T_{\rm CF}^2/(B_{\rm CF}-T_{\rm CF}h_{\rm CF})^2<1$ make the tail disk invariant and the tail map contractive. The tail bound holds on the auxiliary $\nu$ disk enlarged by one complex unit, so Cauchy's inequality also bounds its $\nu$ derivative by $h_{\rm CF}$ on the original disk. The finite backward recurrence is therefore initialized with an enclosure of the infinite tail. We use $(R_k,r_\nu)=(0.025,0.060)$ for $(2,2)$ and $(0.050,0.120)$ for $(3,3)$. Two configurations, with 180-digit intervals and $L_{\rm CF}=48$, and 220-digit intervals and $L_{\rm CF}=64$, satisfy the enclosure, contraction, normalization, and modulus-separation tests.

On a detuning disk $0<|k|\le\rho<R_k$, positive term majorants and geometric infinite tails bound the matching sums. All four matching sums have deviations from one below one. The normalized bilateral sums entering Eq.~\eqref{eq:mst-normalized-bases} obey the same nonvanishing test. Moreover, $B_
\nu\rho+\rho<\frac12$ and $B_
\nu\rho<\delta_{\rm SR}$ separate the relevant Gamma arguments from their poles and the sine factors from their zeros. Hence Eq.~\eqref{eq:mst-exact-critical-condition} is equivalent to vanishing incoming amplitude throughout the stated punctured disk. Analytic continuation gives the same condition on the complete retained argument sector.

For the Gamma factor, bound the logarithmic derivatives by $L_
\nu\ge|\partial_
\nu\log\mathcal G|$ and $L_{\widetilde\omega}\ge|\partial_{\widetilde\omega}\log\mathcal G|$ on the full parameter enclosure. The bounds follow by enclosing the digamma functions away from their poles. Consequently,
\begin{equation}\label{eq:mst-gamma-disk-bound}
 |\mathcal G|\le |C_{\ell m}|
 {\rm e}^{(L_\nu B_\nu+L_{\widetilde\omega})\rho}.
\end{equation}
Multiplication by the matching sum bounds gives a uniform upper bound $C_{\max}$ on the right-hand side of Eq.~\eqref{eq:mst-exact-critical-condition}.

For every $0<|k|\le\rho$ and $-\frac\pi2\le\arg_{\mathfrak b}k\le\frac{3\pi}2$, the left-hand side satisfies
\begin{subequations}\label{eq:mst-conventional-sheet-disk-separation}
\begin{align}
 \left|{\rm e}^{(2\nu+1)L_{\mathfrak b}}\right|
 &\ge {\rm e}^{-\pi\delta_{\rm SR}-E_\rho},\\
 E_\rho&=2B_
\nu\rho\left[-\log\rho+\log(2m)+\frac{3\pi}2\right]
\nonumber\\
 &\quad+\frac{(2\delta_{\rm SR}+2B_
\nu\rho)\rho}{m-\rho}.
\end{align}
\end{subequations}
The nonnegative quantity $E_\rho$ bounds the correction to the modulus of the critical exponential. Here $\rho<{\rm e}^{-1}$, so $|k||\log|k||$ increases with $|k|$ over the required interval; the frequency-logarithm correction is bounded by $|k|/(m-|k|)$. These inequalities control every intermediate radius and argument, including both limiting local cut banks.

At the stated radii, the ratios $C_{\max}{\rm e}^{\pi\delta_{\rm SR}+E_\rho}$ are below $0.018$ for $(2,2)$ and $0.046$ for $(3,3)$, both below one. The two sides of Eq.~\eqref{eq:mst-exact-critical-condition} therefore cannot coincide. Restoring units, Table~\ref{tab:critical-ray-arguments} proves
\begin{equation}\label{eq:mst-explicit-conventional-sheet-disks}
 0<\left|\omega-\omega_{\rm SR}\right|
 \le\frac{\rho}{2M},\qquad\ell=m=2,3,
\end{equation}
with the sector radii in the table and common radius $0.018/(2M)$. The numbers $0.018$ and $0.046$ in the preceding modulus comparison are upper bounds on a ratio, whereas the sector radii in $k$ are $0.018$ and $0.040$. These disks are local results. In particular, the $(3,3)$ disk does not reach the bounded counting domain $D_{33}$: its nearest point has $|k|=\sqrt{0.0052}>0.0721$. The separate certificates therefore do not count zeros in the intervening region.

\subsection{Nonoscillatory and repeated critical exponents}\label{app:negative-critical}
The oscillatory powers used above require real positive $\delta_{\rm SR}$. We now examine the other possibilities separately: when $\delta_{\rm SR}^2<0$ the two powers have unequal real exponents, while at $\delta_{\rm SR}=0$ they coalesce and a logarithmic solution replaces one power. These cases determine whether the preceding sheet-dependent accumulation mechanism persists.
For $m>0$ and $\delta_{\rm SR}^2<0$, write $\mu=\sqrt{-\delta_{\rm SR}^2}>0$. Here $\mu$ is the real power index of the comparison solutions. Their propagators act on the logarithmic state $(Y,{\rm d}Y/{\rm d}\xi)^{\mathsf T}$, with a fixed vector norm and its induced matrix norm. The constant $K$ is a finite positive bound on the chosen angular branch and finite argument interval, independent of sufficiently small $|k|$. Their logarithmic propagators satisfy weighted bounds: inward outer propagation from $|z_2|$ to $|z_1|<|z_2|$ is bounded by $K|\frac{z_2}{z_1}|^\mu$, and outward inner propagation from $|u_1|$ to $|u_2|>|u_1|$ by $K|\frac{u_2}{u_1}|^\mu$. To see this, factor each endpoint fundamental matrix into a bounded invertible matrix and the diagonal powers. When the exponent difference is an integer, the subdominant logarithm enters the bounded factor as $z^{2\mu}\log z$ or $u^{-2\mu}\log u$; it does not change the dominant bound.

Multiplying the outer state norm by $|z|^\mu$ and the inner norm by $|u|^{-\mu}$ cancels these growth factors in the variation-of-constants integrals. The weighted variation-of-constants estimate then gives the same vanishing relative state error. At the matching point, the dominant outer and inner directions are $(1,-\mu)^{\mathsf T}$ and $(1,\mu)^{\mathsf T}$. Their Wronskian, ordered as horizon and infinity states, is $-2\mu$. The coefficients in Eq.~\eqref{eq:critical-prefactor-coefficients}, evaluated at $\delta_{\rm SR}=i\mu$, are precisely their dominant coefficients. Their numerators are $\Gamma(2\mu)$ and the arguments of their denominator Gamma functions have nonzero imaginary part $-m$, so neither vanishes. Consequently,
\begin{equation}\label{eq:negative-critical-connection}
 \frac{\mathcal W_{\mathfrak b}}{-2\mu\,a_{\mathrm{in},-}a_{\mathrm{up},+}}
 =1+o(1),\qquad k\to0.
\end{equation}
The error is uniform on a fixed finite logarithmic sector and can be bounded by a constant times $\sqrt{|k|}+|k\log|k||+|k|^\mu(1+|\log|k||)$. Thus each such sheet has a sufficiently small zero-free punctured critical neighborhood. This statement includes the remaining corotating sectors with $2\le\ell\le3$; it does not cover $\delta_{\rm SR}=0$.

The case $\delta_{\rm SR}=0$ uses the logarithmic Whittaker expansion. The coefficients of the logarithmic terms in the inner and outer comparison solutions are, respectively,
\begin{subequations}\label{eq:critical-log-coefficients}
\begin{align}
 \widetilde a_{\mathrm{in}}
 &=\frac{(-2ih)^{\frac{1}{2}-s-im}}{\Gamma(\frac{1}{2}-s-im)},\\
 \widetilde a_{\mathrm{up}}
 &=\frac{(-2i\omega_{\rm SR})^{\frac{1}{2}+s-im}}{\Gamma(\frac{1}{2}+s-im)}.
\end{align}
\end{subequations}
The inner and outer logarithmic factors are $\log z-\operatorname{Log}_{\mathfrak b}(-2ih)-\psi^{(0)}(\frac{1}{2}-s-im)-2\gamma_{\rm Euler}$ and $-\log z-\operatorname{Log}(-2i\omega_{\rm SR})-\psi^{(0)}(\frac{1}{2}+s-im)-2\gamma_{\rm Euler}$, respectively. Here $\log z$ follows the radial matching path, $\psi^{(0)}(\zeta)=\Gamma'(\zeta)/\Gamma(\zeta)$ is the digamma function, and $\gamma_{\rm Euler}$ is Euler's constant. Their determinant gives
\begin{align}\label{eq:critical-repeated-exponent}
 \frac{\mathcal W_{\mathfrak b}}{\widetilde a_{\mathrm{in}}\widetilde a_{\mathrm{up}}}
 ={}&L_{\mathfrak b}^{\rm crit}+\psi^{(0)}(\frac{1}{2}-s-im)
\nonumber\\
 &+\psi^{(0)}(\frac{1}{2}+s-im)+4\gamma_{\rm Euler}+o(1).
\end{align}
The comparison fundamental matrices and their inverses now grow at most logarithmically. At $\frac{|z|}{M}=\sqrt{|k|}$, multiplying the perturbation integrals by their logarithmic factors bounds the determinant error, for example, by $K\sqrt{|k|}(1+|\log|k||)^6$. This tends to zero. Since $L_{\mathfrak b}^{\rm crit}$ diverges on a fixed finite sector and both gamma coefficients are nonzero for $m>0$, this case also has a zero-free punctured critical neighborhood.

For every fixed gravitational corotating sector, integration by parts in the regular angular equation at $a\omega=m/2$ gives
\begin{equation}\label{eq:critical-angular-variational-bound}
 {}_{-2}A_{\ell m}(m/2)\ge-\frac{m^2}{4}-2m+2.
\end{equation}
Regularity makes the endpoint integration terms vanish. Combining this with Eq.~\eqref{eq:deltasr} yields
\begin{equation}\label{eq:critical-gravitational-delta-bound}
 \delta_{\rm SR}^2\le2m^2+2m-\frac{17}{4}<\frac{9m^2}{4}.
\end{equation}
The last difference is $\frac{(m-4)^2+1}{4}>0$. Thus a real positive exponent has $\delta_{\rm SR}<\frac{3m}{2}$. For integer $m\ge1$, this implies
$ {\rm e}^{2\pi m}>{\rm e}^{\pi\delta_{\rm SR}}+1+{\rm e}^{-\pi\delta_{\rm SR}}$, which is equivalent to the positive gap on the conventional Riemann sheet $|C_{\ell m}|<{\rm e}^{-\pi\delta_{\rm SR}}$. The required model ray is therefore in $(\frac{3\pi}{2},2\pi)$, on the adjacent horizon sheet.

Together the real, imaginary, and repeated exponent cases establish a zero-free punctured neighborhood of the critical frequency on the conventional Riemann sheet for each fixed gravitational $(\ell,m)$ with $m>0$. In sufficiently small annuli,
\begin{subequations}\label{eq:general-critical-local-counts}
\begin{align}
 \mathcal N_{\rm ret}^{\rm crit}(\epsilon,\epsilon_0)&=0,\\
 \mathcal N_{\rm adj}^{\rm crit}(\epsilon,\epsilon_0)
 &=\frac{\delta_{\rm SR}}{\pi}\log\frac{\epsilon_0}{\epsilon}+O(1),
 \quad\delta_{\rm SR}^2>0.
\end{align}
\end{subequations}
The subscripts $ {\rm ret}$ and $ {\rm adj}$ denote the conventional sheet selected from the upper half-plane and the adjacent horizon-frequency sheet, respectively. These quantities count zeros, including multiplicity, in the annulus $\epsilon<|\omega-\omega_{\rm SR}|<\epsilon_0$. For $\delta_{\rm SR}^2\le0$, every fixed finite horizon sheet instead has a zero-free punctured neighborhood of the critical frequency. The radii and constants depend on the chosen angular branch and continuation.

\subsection{The punctured zero-frequency neighborhood}\label{app:origin-collar}
The origin requires a separate limit because the asymptotic behavior at radial infinity changes as $\omega\to0$. We compare the exact endpoint states with their static counterparts and normalize their Wronskian by its nonzero leading factor. If the normalized Wronskian stays away from zero, the punctured neighborhood contains no QNM zeros on the specified sheets. The cases $m\ne0$ and $m=0$ differ because, in the latter, the superradiant and zero-frequency branch points coincide.
Fix a local angular branch with $ {}_sA_{\ell m}(0)=\ell(\ell+1)-s(s+1)$ and $\ell\ge|s|$. The radial inversion $v=\frac{1}{z}$, $\Phi(v)=v\Psi(\frac{1}{v})$ gives exactly
\begin{align}\label{eq:origin-dual-normal}
 \frac{{\rm d}^2\Phi}{{\rm d}v^2}+\Big[&h^2+\frac{d}{v}+\frac{c}{v^2}
\nonumber\\
 &+\frac{b}{v^3}+\frac{\omega^2}{v^4}\Big]\Phi=0.
\end{align}
Here $v$ is the inverted radial coordinate and $\Phi$ the rescaled normal-form solution; $b,c,d,h$ retain their definitions in Eq.~\eqref{eq:normal-radial-proof}. The original horizon ingoing state becomes the dual infinity outgoing state, and the original infinity outgoing state becomes the dual origin ingoing state. The coordinate reversal and the exchange of state order cancel in the Wronskian.

For $m\ne 0$, $h\to-Mm\ne 0$ as $\omega\to0$. The dual radial state vector at infinity is therefore nondegenerate. Scaling the dual origin before taking the limit gives the same weighted matching argument as above, now with the static power index $\mu_0=\ell+\frac{1}{2}$ and limiting Whittaker parameters $s$ and $-s$. The dominant gamma coefficients are $\frac{\Gamma(2\ell+1)}{\Gamma(\ell+1-s)}$ and $\frac{\Gamma(2\ell+1)}{\Gamma(\ell+1+s)}$. Both are nonzero. The two dominant state directions again have a nonzero determinant.

When $m=0$, $h=2M^2\omega$ and both endpoints degenerate. Use the dimensionless endpoint coordinates $v_M=h/z$ and $v_\infty=\omega z$ for the horizon and infinity, separately removing their nonzero frequency-dependent normalization factors. The leading exponentials become $ {\rm e}^{iv_M}$ and $ {\rm e}^{iv_\infty}$. Match at fixed $z=M$, where both scaled radii shrink in proportion to $|M\omega|$. Relative to the corresponding Whittaker equations, the weighted logarithmic perturbation integrals are $O(|M\omega|+|M\omega\log(M\omega)|)$. The dominant states therefore have vanishing relative errors. Subdominant resonant logarithms do not alter this limit. Since the two frequency branch points coincide, their logarithms must be continued together.

In both cases the leading determinant is
\begin{subequations}\label{eq:origin-leading-connection}
\begin{align}
 P_0={}&-\frac{(2\ell+1)\Gamma(2\ell+1)^2}
 {\Gamma(\ell+1-s)\Gamma(\ell+1+s)}
\nonumber\\
 &\times(-2ih)^{-\ell-s}(-2i\omega)^{s-\ell},\\
 \frac{\mathcal W_{\mathfrak b}}{P_0}&=1+o(1),\qquad \omega\to0.
\end{align}
\end{subequations}
This estimate is uniform on each fixed finite logarithmic sector and excludes zeros near the origin, including for $m=0$. It is not uniform in $\ell$ or over infinitely many sheets. The following calculation bounds the size of this punctured neighborhood. The singular power $P_0$ at the branch point is distinct from a finite noncritical normalization pole.

For $\ell=2,3$, consider the punctured frequency disk $0<|\omega|\le\rho_0$ in the adopted units, with $\rho_0=10^{-12}$. Write $A={}_{-2}A_{\ell m}(\omega)$ for the continued angular eigenvalue. The angular perturbation bound $|A-[\ell(\ell+1)-2]|\le\rho_0^2+4\rho_0$ preserves the simple angular branch. The static dominant solutions are finite polynomials multiplying the endpoint exponentials. In the positive-$m$ horizon, scaled infinity, and scaled axisymmetric horizon constructions, respectively, they are
\begin{align}
 f_H(z)&=z^3{\rm e}^{-im/z}\sum_{j=0}^{\ell-2}a_jz^j,\nonumber\\
 g_I(v_\infty)&={\rm e}^{iv_\infty}v_\infty^2\sum_{j=0}^{\ell+2}b_jv_\infty^{-j},\nonumber\\
 g_0(v_M)&={\rm e}^{iv_M}v_M^{-2}\sum_{j=0}^{\ell-2}d_jv_M^{-j},
\end{align}
where $v_\infty=\omega z$ is used for every $m$, while $v_M=h/z$ is used for the axisymmetric horizon. The coefficients are $a_j=\binom{\ell-2}{j}(\ell+3)_j/(2im)^j$, $b_j=\binom{\ell+2}{j}(\ell-1)_j/(-2i)^j$, and $d_j=\binom{\ell-2}{j}(\ell+3)_j/(-2i)^j$. Complementary solutions that are regular in the corresponding comparison problem are
\begin{align}
 s_H(z)&=z^{-\ell}{\rm e}^{-im/z}\,{}_1F_1(\ell+3;2\ell+2;2im/z),\nonumber\\
 r_I(v_\infty)&=v_\infty^{\ell+1}{\rm e}^{iv_\infty}
 {}_1F_1(\ell-1;2\ell+2;-2iv_\infty),\nonumber\\
 s_0(v_M)&=v_M^{\ell+1}{\rm e}^{iv_M}
 {}_1F_1(\ell+3;2\ell+2;-2iv_M).
\end{align}
Taking the Wronskians in the order displayed, with derivatives in the argument of each pair, gives
\[
\begin{aligned}
 W(f_H,s_H)&=-(2\ell+1)a_{\ell-2},\\
 W(g_I,r_I)&=(2\ell+1)b_{\ell+2},\\
 W(g_0,s_0)&=(2\ell+1)d_{\ell-2}.
\end{aligned}
\]
 These nonzero Wronskians enter the Green kernels built from each dominant and regular pair. For each endpoint construction $e$, let $E_e^{\rm stat}$ bound each anchor-state component error, $B_e^{\rm stat}$ the propagated weighted-state error, and $C_e^{\rm stat}$ the static comparison-state norm. Variation of constants gives
\begin{equation}\label{eq:origin-volterra-error}
\begin{aligned}
 B_e^{\rm stat}&\le\frac{2K_e^{\rm stat} E_e^{\rm stat}+q_e^{\rm stat} C_e^{\rm stat}}{1-q_e^{\rm stat}},\\
 q_e^{\rm stat}&=K_e^{\rm stat}\int|\zeta\Delta Q_e(\zeta)|\,|d\zeta|<1.
\end{aligned}
\end{equation}
Here $e$ labels the horizon, infinity, or scaled axisymmetric horizon construction. The scalar $K_e^{\rm stat}$ bounds the weighted comparison Green kernel, and $\Delta Q_e$ is the exact potential minus its static comparison potential in the corresponding endpoint coordinate. The integral runs along the chosen endpoint path from the finite anchor toward the matching region, with integration variable $\zeta$; $q_e^{\rm stat}<1$ is the resulting contraction bound. Frequency-disk enclosures of the endpoint states include the radial residual, angular error, and propagation remainder. Two configurations use endpoint scale, Taylor degree, and decimal interval precision $(10,110,180)$ and $(12,130,220)$, retaining ten endpoint terms. Rounding upward over both configurations, the anchor bounds $(E_H^{\rm stat},E_0^{\rm stat},E_I^{\rm stat})$ are $(3.49\times10^{-8},6.42\times10^{-9},1.17\times10^{-8})$ for $\ell=2$ and $(6.08\times10^{-7},1.32\times10^{-7},1.04\times10^{-6})$ for $\ell=3$. Here $H$ is the $m>0$ horizon construction, $0$ the axisymmetric horizon construction, and $I$ infinity; $E_H^{\rm stat}$ is maximized over $1\le m\le\ell$ and $E_I^{\rm stat}$ covers $|m|\le\ell$. Thus every required state has $E_e^{\rm stat}<1.05\times10^{-6}$.

Matching at $z=|\omega|^{-1/2}$ for $m\ne0$ and at $z=1$ for $m=0$ controls the state errors. Let $e_H,e_I$ bound the resulting component errors relative to the dominant logarithmic-state directions $(1,\ell+1)^{\mathsf T}$ and $(1,-\ell)^{\mathsf T}$. Expanding their determinant shows that its relative error is bounded by
\begin{equation}
 d_{\ell m}=\frac{(\ell+1)e_H+(\ell+2)e_I+2e_He_I}{2\ell+1}.
\end{equation}
If $f_{\ell m}$ bounds the additional discrepancy in the retained frequency powers, the complete normalized error is at most $f_{\ell m}+(1+f_{\ell m})d_{\ell m}$. Including these powers and the static polynomial tails yields the uniform determinant estimate
\begin{align}\label{eq:explicit-origin-collar-error}
 \left|\frac{\mathcal W_{\mathfrak b}}{P_0}-1\right|
 &<0.080,
\nonumber\\
 0<|M\omega|&\le10^{-12},\qquad
 -\frac{5\pi}{2}\le\arg_{\mathfrak b}\omega\le\frac{7\pi}{2},
\nonumber\\
 \ell&=2,3,\quad |m|\le\ell.
\end{align}
The bound is uniform at every smaller radius and includes the local cut banks. Negative $m$ follows from $(\omega,m)\mapsto(-\omega^*,-m)$. Thus the stated punctured disk is zero-free on the conventional origin chart and the two charts reached by one full turn about $\omega=0$ in either direction; for $m=0$, both frequency logarithms are continued together.

\bibliography{ExtremeKerrQNMRefs}
\end{document}